\documentclass[11pt,twoside]{article}

\usepackage{amsfonts}
\usepackage{amsmath}
\usepackage{amsthm}
\usepackage{amscd}
\usepackage{cite}
\usepackage{eucal}
\usepackage{graphicx}
\usepackage{a4}
\usepackage{booktabs}
\usepackage{microtype}
\usepackage{times}
\usepackage{float}

\input{macro.lib}

\renewcommand{\appendix}{%
   \renewcommand{\section}{
        \secdef\Appendix\sAppendix}%
   \setcounter{section}{0}%
   \renewcommand{\thesection}{\Alph{section}}%
   \renewcommand{\theequation}{\thesection.\arabic{equation}}%
}

\newcommand{\Appendix}[2][?]{%
     \refstepcounter{section}%
     \setcounter{equation}{0}%
     \addcontentsline{toc}{appendix}%
          {\protect\numberline{\appendixname~\thesection} #1}%
     \vspace{\baselineskip}%
     {\noindent\large\bfseries\appendixname\ \thesection: #2\par}%
     \sectionmark{#1}\vspace{\baselineskip}}

\newcommand{\sAppendix}[1]{%
     {\noindent\large\bfseries\appendixname\:: #1\par}%
     \sectionmark{#1}\vspace{\baselineskip}}

\allowdisplaybreaks

\begin{document}

\thispagestyle{empty}

\begin{center}

{\Large {\bf Thermal form factor approach\\ to the ground-state
correlation functions of the XXZ chain\\ in the antiferromagnetic
massive regime\\}}

\vspace{7mm}

{\large
Maxime Dugave,\footnote{e-mail: dugave@uni-wuppertal.de}
Frank G\"{o}hmann\footnote{e-mail: goehmann@uni-wuppertal.de}}%
\\[1ex]
Fakult\"at f\"ur Mathematik und Naturwissenschaften,\\
Bergische Universit\"at Wuppertal,
42097 Wuppertal, Germany\\[2.5ex]
{\large Karol K. Kozlowski\footnote{e-mail: karol.kozlowski@ens-lyon.fr}}%
\\[1ex]
Univ Lyon, ENS de Lyon, Univ Claude Bernard,  CNRS, Laboratoire de Physique, F-69342 Lyon, France\\[2.5ex]
{\large Junji Suzuki\footnote{e-mail: sjsuzuk@ipc.shizuoka.ac.jp}}%
\\[1ex]
Department of Physics, Faculty of Science, Shizuoka University,\\
Ohya 836, Suruga, Shizuoka, Japan

\vspace{15mm}
{\it Dedicated to the memory of Petr Petrovich Kulish}
\vspace{15mm}

{\large {\bf Abstract}}

\end{center}

\begin{list}{}{\addtolength{\rightmargin}{9mm}
               \addtolength{\topsep}{-5mm}}
\item
We use the form factors of the quantum transfer matrix in the
zero-temperature limit in order to study the two-point ground-state
correlation functions of the XXZ chain in the antiferromagnetic massive
regime. We obtain novel form factor series representations of
the correlation functions which differ from those derived
either from the q-vertex-operator approach or from the algebraic
Bethe Ansatz approach to the usual transfer matrix. We advocate
that our novel representations are numerically
more efficient and allow for a straightforward calculation of
the large-distance asymptotic behaviour of the two-point
functions. Keeping control over the temperature corrections
to the two-point functions we see that these are of order
$T^\infty$ in the whole antiferromagnetic massive regime. The
isotropic limit of our result yields a novel form factor series
representation for the two-point correlation functions of
the XXX chain at zero magnetic field.
\end{list}

\clearpage

\section{Introduction}
Two-point correlation functions of the Heisenberg XXZ chain can be
studied by means of form factor expansions. These have turned out
to be particularly useful for extracting the large-distance
asymptotics \cite{KMS11a,KKMST11b,DGK13a,DGK14a,DGKS15a} and are
currently the only efficient means to study time dependent correlation
functions analytically \cite{BKM98,CaHa06,KKMST12,DGKS16a}.
We distinguish usual form factors from thermal form factors. Usually
form factors are understood as matrix elements of local operators
between the ground state and excited states of a given Hamiltonian
or transfer matrix. The usual form factors of the XXZ chain were
studied in \cite{JiMi95,IKMT99,KMT99a,KKMST09b,KKMST11a,DGKS15a}.
Finite-temperature static correlation functions can also be expanded
in terms of matrix elements of certain non-local operators between
the dominant state and excited states of the quantum transfer matrix
\cite{DGK13a,DGK14a}. In order to distinguish them from the usual
form factors we have introduced the term `thermal form factors'
in \cite{DGK13a}. Thermal form factor expansions are particularly
convenient for studying the large-distance asymptotics of thermal
correlation functions, since the latter is determined by a few
terms of the series as long as the temperature remains strictly
finite. In the zero-temperature limit infinitely many terms have
to be taken into account. Then thermal form factor expansions generate
different but equivalent expansions of the zero-temperature static two-point functions.
These are the subject of this work.

Form factor densities of the XXZ chain were first obtained within the
q-vertex operator approach \cite{JiMi95}. This elegant method is
designed to work directly for the infinite chain and applies
only to the antiferromagnetic massive ground state regime. Still,
unlike other methods, it also works for the fully anisotropic XYZ
chain \cite{Lashkevich02}. An alternative approach to the
calculation of form factors, at least of the XXZ chain, is
the algebraic Bethe Ansatz approach. Combining what is called
`the solution of the quantum inverse problem' with a remarkable
scalar product formula \cite{Slavnov89} due to N.~Slavnov,
the form factors of the finite-length XXZ chain were expressed
by certain determinants in \cite{KMT99a}. Although the determinants
cannot be computed explicitly, they can be evaluated numerically
by solving the underlying Bethe Ansatz equations on a computer.
This approach turned out to be efficient for the calculation of
experimentally relevant correlation functions \cite{BKM03,SST04,%
CaMa05,PSCHMWA07}.

\begin{figure}
\begin{center}
\includegraphics[width=.8\textwidth,angle=0,clip=true]{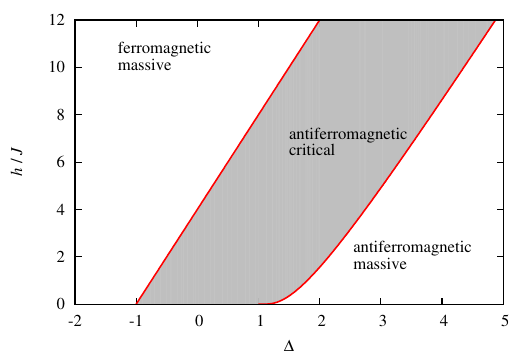}
\caption{\label{fig:phasediagram} The ground state phase diagram of
the XXZ chain in the $\D$-$h$ plane. The ferromagnetic massive regime
and the antiferromagnetic critical regime are separated by the upper
critical field $h_u$ (left red line). The lower critical field $h_\ell$
(right red line) marks the border between the antiferromagnetic critical
and the antiferromagnetic massive regime.
}
\end{center}
\end{figure}

The form factor formulae for finite spin chains of length
$L$ obtained within the algebraic Bethe Ansatz approach hold
for arbitrary values of the parameters of the model, which are
the strength of the external magnetic field $h$ and the 
anisotropy parameter $\D$. In the thermodynamic limit, $L
\rightarrow \infty$, the $\D$-$h$ parameter plane is separated
into three different regimes (or ground state phases) depicted in
Figure~\ref{fig:phasediagram}. Like all other quantities which
can be calculated by means of the algebraic Bethe Ansatz, when $L$
is sent to infinity, the form factors become functionals of a
few basic functions such as the dressed energy, dressed momentum
and dressed phase. It is not easy to actually perform this limit,
and the situation is different in the different ground state
regimes sketched in Figure~\ref{fig:phasediagram}. In the
antiferromagnetic critical regime for $h > 0$ the ground state
has finite magnetization and the low-lying excitations are of
particle-hole type, much like in the case of free Fermions. The
corresponding Bethe roots are real. For this case the thermodynamic
limit of the particle-hole form factors was analysed in \cite{KKMST09b,%
KKMST11a}. A formula for the summation of all particle-hole form factors
was obtained in \cite{KKMST11b}. This formula made it possible to
determine the large-distance asymptotics of the two-point functions
including the non-universal amplitudes.

At zero magnetic field and also in the whole antiferromagnetic massive regime
the ground state magnetization of the finite-size system vanishes. The
lowest-lying excitations above the massive ground state involve non-real Bethe
roots. This makes the analysis of the thermodynamic limit of certain determinants
that are part of the Bethe Ansatz form factors more involved as these
determinants are rather singular in this limit. In the antiferromagnetic
massive regime there is one state which is almost degenerate with the ground
state 
(a `pseudo ground state'). It is the only other state with
only real Bethe roots. The matrix element of $\s^z$ between this state and the
ground state is, in a sense, the simplest non-vanishing form factor. Its
thermodynamic limit was obtained in \cite{Baxter73a,Baxter76a,IKMT99}.
The thermodynamic limit of \emph{all} other form factors of $\s^z$ in
the antiferromagnetic massive regime was obtained in \cite{DGKS15a}.
The corresponding form factor series represents the longitudinal two-point
functions not only asymptotically but at all distances.

The critical regime at $h = 0$ is, so far, the least well understood
from a Bethe Ansatz perspective. In particular, the calculation of
the thermodynamic limit of Bethe Ansatz form factors is still open
in this regime. Note, however, that in this regime rather explicit
results for the correlation amplitudes, that determine the leading
large-distance asymptotic decay of the two-point functions, were
obtained \cite{Lukyanov99,LuTe03} from a clever combination of
perturbation theory applied to the Gaussian conformal field theory
\cite{Lukyanov98} with results from the $q$-vertex operator
approach to the form factors of the XYZ model \cite{Lashkevich02}.

The amplitude densities in the form factor series for the longitudinal
correlation functions obtained in \cite{DGKS15a} involve higher
dimensional residues originating from the fact that one has to sum
up the contributions of the non-real Bethe roots whose \textit{loci}
are constrained by the higher-level Bethe equations. These higher
dimensional residues make a numerical analysis beyond the so-called
two-spinon contribution hard. As we shall see below, the form factor
series obtained from the zero temperature limit of a thermal form factor
expansion is free of this difficulty. The reason for this stems from the 
different role played by the magnetic field for the usual transfer matrix
and for the quantum transfer matrix. As opposed to the Bethe Ansatz
equations of the usual transfer matrix, the Bethe Ansatz equations
and hence the eigenvalues and eigenvectors of the quantum transfer
matrix depend parametrically on the magnetic field, even in
the antiferromagnetic massive regime. As we shall see below, in 
the zero-temperature limit, although the individual terms in the
form factors expansion of the two-point functions do depend on the
magnetic field, their sum does not. In such a way, one does recover that
the zero-temperature correlation functions are field independent. In the
presence of a magnetic field the Bethe root patterns for \emph{all}
low-temperature `excitations' of the quantum transfer matrix are
of particle-hole type \cite{DGKS15b}. In the zero-temperature limit
the particle and hole roots become unconstrained and densely fill two
curve segments in the complex plane. No higher-level Bethe Ansatz equations
have to be satisfied. Using this fact we shall obtain rather explicit
expressions for form factor densities in the limit which are different
from those obtained for the usual transfer matrix \cite{DGKS15a} and
also from those obtained within the vertex operator approach
\cite{JiMi95,Lashkevich02}.

The paper is organized as follows. In the remainder of this introduction
we recall the definition of the model, the form factor series expansions
of the two-point functions and our recent results \cite{DGKS15b} for the
spectrum and Bethe root patterns of the quantum transfer matrix in the
antiferromagnetic massive regime at finite magnetic field $h$. Then in
Section~\ref{sec:amplitudes} we present our results for the amplitudes
in the form factor series in the low-temperature limit. The amplitudes
can be decomposed in three factors, a universal part, a determinant
part and a factoring part, which will be treated separately. In
Section~\ref{sec:series} we show how the form factor series can be
written as series of multiple integrals corresponding to integration
over particle and hole parameters. We compare with previous results
which were interpreted in terms of multi-spinon contributions and we
perform numerical tests against known exact results and purely numerical
calculations in order to assess the efficiency of our novel series
representations. In Section~\ref{sec:isolimit} we perform and discuss
the isotropic limit. We conclude in Section~\ref{sec:conclusions}
with a summary and the discussion of open questions. Almost all
technical details of the calculations are deferred to a series of
appendices.

\subsection{Hamiltonian and correlation functions}
The Hamiltonian defining the spin-$\2$ XXZ chain in a magnetic
field of strength $h$ along the magnetic anisotropy direction is
\begin{equation} \label{ham}
     H = J \sum_{j = - L + 1}^L \Bigl( \s_{j-1}^x \s_j^x + \s_{j-1}^y \s_j^y
                               + \D \bigl( \s_{j-1}^z \s_j^z - 1 \bigr) \Bigr)
         - \frac{h}{2} \sum_{j = - L + 1}^L \s_j^z \epc
\end{equation}
where the $\s_j^\a$ are Pauli matrices $\s^\a$ acting on the $j$th factor of
the tensor-product space of states ${\cal H} = {\mathbb C}^{\otimes 2L}$
of $2L$ spins $\2$. The intrinsic parameters of the model are the strength  $J > 0$ of
the exchange interaction  and the real anisotropy parameter~$\D$.
The exchange interaction merely fixes the energy scale. $\D$ and
$h/J$ are the two physical parameters which determine the ground state
phase diagram \cite{YaYa66d}.
We shall use the standard reparameterization $\D = (q + q^{-1})/2$ with
$q = \re^{- \g}$. In the following we will consider easy-axis anisotropy
corresponding to $\D > 1$. Hence, we assume that $\g > 0$. We shall also
assume that the magnetic field is positive and below the lower critical
field $h_\ell$. The latter as well as several other functions we shall
encounter below are sometimes conveniently expressed in terms of the elliptic
modulus $k = k(q) = \dh_2^2 (0, q)/\dh_3^2 (0, q)$, where the $\dh_j (x, q)$
are Jacobian Theta functions (see \cite{WhWa63ch22}). Denoting the
complementary modulus by $k' = \sqrt{1 - k^2}$ and the complete elliptic
integral of the first kind by $K = K(k)$ we have
\begin{equation}
     h_\ell/J = 2 (q^{-1} - q) \dh_4^2 (0, q)
              = 8 K k' \sh(\p K/K')/\p \epc
\end{equation}
where $K' = K(k')$.

Our goal is to derive efficient series representations for the two-point
correlation functions of the Hamiltonian (\ref{ham}). Correlation functions
of two operators $X$, $Y$ acting on ${\cal H}$ are defined as
\begin{equation}
     \< X Y \> = \frac{\Tr \re^{ - H/T} X Y}{\Tr \re^{ - H/T}} \epc
\end{equation}
where $T$ is the temperature. In this work we shall focus of on the
longitudinal and transversal two-point functions $\<\s_1^z \s_{m+1}^z\>$
and $\<\s_1^- \s_{m+1}^+\>$ in the thermodynamic limit $L \rightarrow
\infty$.

Static temperature dependent correlation functions can be treated most
efficiently within the quantum transfer matrix formalism \cite{GKS04a}
which was originally developed to calculate numerically the free energy
per lattice site of quantum spin systems in the thermodynamic limit
\cite{Suzuki85,SuIn87} and turned out to be compatible with the integrable
structure of vertex models connected with the Yang-Baxter equation
\cite{SAW90,Kluemper92,Kluemper93}. The quantum transfer matrix associated
with a spin model like (\ref{ham}) can be introduced as the column-to-column
transfer matrix of a vertex model on a $2L \times N$ rectangular lattice,
where $2L$ is the number of lattice sites along the spin chain and
$N$ is the number of auxiliary lattice sites in perpendicular direction.
The perpendicular direction may be interpreted as the imaginary time
direction in a path-integral realization of the partition function of
the spin chain. $N$ is called the Trotter number. It was shown in
\cite{Suzuki85} that the `Trotter limit' $N \rightarrow \infty$ of a
single dominant eigenvalue of the quantum transfer matrix determines
the free energy per lattice site of the spin chain in the thermodynamic
limit. In \cite{GKS04a} it was realized that, likewise, the corresponding
dominant eigenvector determines all temperature dependent correlation
functions. They can be written as expectation values of products of
`monodromy matrix elements' and quantum transfer matrices with respect to
the dominant state (see \cite{GKS04a} and Appendix~\ref{app:qtm}).

Expanding those expressions in a basis of eigenvectors of the quantum
transfer matrix we obtained `thermal form factor expansions' in
\cite{DGK13a}. These are series of the form
\begin{equation} \label{ffexp}
     \<\s_1^z \s_{m+1}^z\> - \<\s_1^z\>\<\s_{m+1}^z\>
     = \sum_{n=1}^\infty A_n^{zz} \r_n^m \epc \qd
     \<\s_1^- \s_{m+1}^+\> = \sum_{n=1}^\infty A_n^{-+} \r_n^m \epc
\end{equation}
where the $\r_n$ are ratios of eigenvalues of `excited states' of
the quantum transfer matrix by the dominant eigenvalue in the
Trotter limit and where the amplitudes $A_n^{zz}$ and $A_n^{-+}$
are products of two thermal form factors, each being a normalized matrix
element of an entry of the monodromy matrix taken between the dominant state and an excited
state of the quantum transfer matrix (for more details see \cite{DGK13a}
and Appendix~\ref{app:qtm}). The sums run over all relevant excited
states, i.e.\ over all excited states compatible with the conservation
of the $z$-component of the total spin.

If $T > 0$ the absolute values $|\r_n|$ form a decreasing sequence,
$1 \ge |\r_1| \ge |\r_2| \ge |\r_3| \dots$, with only finitely many
of the $|\r_j|$ equal to $|\r_1|$. These determine the large-$m$
asymptotic behaviour and, together with the corresponding
amplitudes $A_j^{zz}$ or $A_j^{-+}$, can be obtained numerically
from the expressions derived in \cite{DGK13a}. The situation
changes in the limit $T \rightarrow 0+$. In this limit infinitely
many eigenstates of the quantum transfer matrix degenerate and
have to be summed up in order to obtain the asymptotic behaviour
of the two-point functions.  Nevertheless, the situation remains
comfortable, since the eigenvalue ratios and amplitudes simplify
in this limit. An analysis of the $T \rightarrow 0+$ limit of the
two-point functions of the XXZ chain in the critical regime was
carried out in \cite{DGK13a,DGK14a}. Here we perform a similar
analysis for the antiferromagnetic massive regime. We are going
to build on our recent paper \cite{DGKS15b}. Based on a set of
mild assumptions we have classified in that paper all excitations
of the quantum transfer matrix in the low-temperature limit for
the model in the antiferromagnetic massive regime, and we have
calculated the corresponding eigenvalue ratios $\r_n$. This is
equivalent to having determined all correlation lengths $\x_n =
- 1/\ln \r_n$.

\subsection{Low-temperature spectrum of correlation lengths}
For the description of the low-temperature spectrum of correlation
length we have to introduce a number of functions that determine the
physical properties of the XXZ chain at $T = 0+$. These are
the dressed momentum $p$, the dressed energy $\e$ and the
dressed phase~$\ph$. In the antiferromagnetic massive regime
we can express these functions explicitly in terms of known
special functions. We define the dressed momentum as
\begin{equation} \label{ptheta4}
      p(x) = \4 + \frac x {2 \p}
             + \frac 1 {2\p\i} \ln \biggl(
	       \frac{\dh_4 (x + \i \g/2, q^2)}{\dh_4 (x - \i \g/2, q^2)}
	       \biggr)
\end{equation}
and the dressed energy as
\begin{equation} \label{epsdn}
     \e(x) = \frac h 2 - \frac{4 J K \sh (\g)}{\p}
             \dn \biggl( \frac{2 K x}{\p} \bigg| k \biggr) \epc
\end{equation}
where $\dn$ denotes the Jacobian elliptic $\dn$-function. Note that
the dressed energy depends explicitly on the magnetic field $h$.

The Jacobian Theta functions and the Jacobian elliptic functions
are special cases of functions that can be expressed in terms
of (infinite) $q$-multi factorials which, for $|q_j| < 1$ and
$a \in {\mathbb C}$, are defined as
\begin{equation}
     (a;q_1, \dots, q_p) =
        \prod_{n_1, \dots, n_p = 0}^\infty (1 - a q_1^{n_1} \dots q_p^{n_p}) \epp
\end{equation}
We shall make extensive use of $q$-multi factorials below, when
we describe the amplitudes in the form factor expansions of two-point
functions. Here we need them to define the dressed phase,
\begin{equation} \label{dressedphase}
     \ph(x_1, x_2) = \i \Bigl( \frac \p 2 + x_{12} \Bigr)
        + \ln \Biggl\{ \frac{\G_{q^4} \bigl(1 + \frac{\i x_{12}}{2\g}\bigr)
	                     \G_{q^4} \bigl(\2 - \frac{\i x_{12}}{2\g}\bigr)}
		            {\G_{q^4} \bigl(1 - \frac{\i x_{12}}{2\g}\bigr)
			     \G_{q^4} \bigl(\2 + \frac{\i x_{12}}{2\g}\bigr)}
			     \Biggr\} \epc
\end{equation}
where $x_{12} = x_1 - x_2$, $|\Im x_2| < \g$ and where the $q$-Gamma
function $\G_q$ is given in terms of $q$ factorials,
\begin{equation}
     \G_q (x) = \frac{(1 - q)^{1 - x} (q;q)}{(q^x;q)} \epp
\end{equation}

In \cite{DGKS15b} we have conjectured that at temperatures low enough
all excitations of the quantum transfer matrix can be parameterized
by an even number of complex parameters located inside the strip
$|\Im x| < \g/2$. Referring to \cite{DGKS15b} we call the parameters
in the upper half plane particles and denote them by $y_i$, $i = 1,
\dots, n_p$. The parameters in the lower half plane will be called
holes and will be denoted $x_j$, $j = 1, \dots, n_h$. Here $n_p$
and $n_h$ are non-negative integers. Their difference is
\begin{equation}
     n_h - n_p = 2s \epc
\end{equation}
where $s \in {\mathbb Z}$ is the conserved pseudo spin of the quantum
transfer matrix (for a definition see Appendix~\ref{app:qtm}). In
our form factor expansions (\ref{ffexp}) $s$ is fixed and equal
to the spin of the operator\footnote{In analogy with conformal field
theory the spin of an operator is defined by the value it changes
the spin of a state it is acting on.} that stands to the right in
the two-point functions, i.e.\ $s = 0$ for the longitudinal correlation
function $\< \s_1^z \s_{m+1}^z \>$ and $s = 1$ for $\< \s_1^- \s_{m+1}^+ \>$
as $\s_{m+1}^z$ leaves the $z$-component of the total spin unchanged,
while $\s_{m+1}^+$ changes it by $+1$.

Then, up to corrections of the order $T^\infty$, the particles and holes
are determined by the higher-level Bethe Ansatz equations
\begin{subequations}
\label{hlbaes}
\begin{align}
     & \e(y_j) = 2 \p \i T \bigl(\ell_j + F(y_j)\bigr) \epc
			     && j = 1, \dots, n_p \epc \\
     & \e(x_j) = 2 \p \i T \bigl(m_j + F(x_j)\bigr) \epc
			     && j = 1, \dots, n_h \epc
\end{align}
\end{subequations}
where $\ell_j, m_j \in {\mathbb Z}$ and where we assume the
$\ell_j$ and $m_j$ to be mutually distinct. $F$ is the `shift
function' defined by
\begin{equation} \label{defshiftfun}
     2 \p \i F(x) = \i \p k + \a \g + \sum_{\ell=1}^{n_p} \ph(x,y_\ell) 
                                    - \sum_{\ell=1}^{n_h} \ph(x,x_\ell) \epp
\end{equation}
Here $\a $ is an auxiliary twist related to the magnetic field,
which serves as a regularization parameter and will be set equal
to zero at the end of the calculation. The parameter $k \in \{0, 1\}$
distinguishes between two sectors of excitations of the quantum
transfer matrix corresponding to staggered and non-staggered
contributions to the form factor series below.

\begin{figure}
\begin{center}
\includegraphics[width=.70\textwidth]{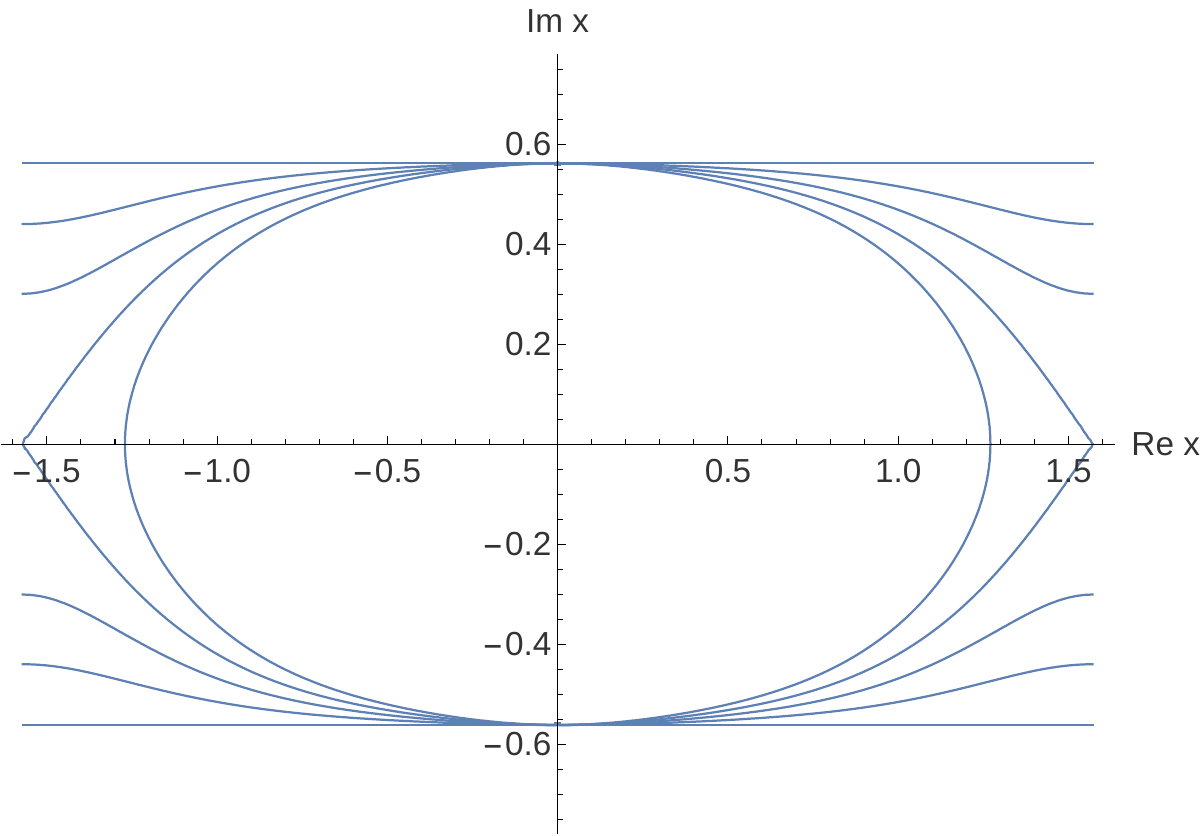}
\end{center}
\caption{\label{fig:regimes} The curves $\Re \e (x) = 0$ for various
values of the magnetic field. Here $\D = 1.7$, $h_\ell/J = 0.76$.
The values of the magnetic field decrease proceeding from the inner
to the outer curve: $h/h_\ell = 1.34, 1, 2/3, 1/3, 0$. In the
critical regime $h_\ell < h < h_u = 4J (1 + \D)$ the curves are
closed. In this regime $\e$ is not given by (\ref{epsdn}), but is
rather defined as a solution of a linear integral equation (see
e.g.\ \cite{DGK14b}). At the lower critical field $h = h_\ell$
the closed curves develop two cusps, and a gap opens for $0 < h
< h_\ell$ which is the parameter regime considered in this work.}
\end{figure}

In the limit $T \rightarrow 0+$ at finite $n_p$ and $n_h$ the higher-%
level Bethe Ansatz equations (\ref{hlbaes}) decouple, $\i \p \ell_j T$
and $\i \p m_j T$ turn into independent continuous variables, and the
particles and holes become free parameters on the curves
\begin{equation} \label{defbpm}
     {\cal B}_\pm = \bigl\{ x \in {\mathbb C} \big|
                            \Re \e (x) = 0, - \p/2 \le \Re x \le \p/2,
			    0 < \pm \Im x < \g \bigr\} \epp
\end{equation}
These curves are shown in Figure~\ref{fig:regimes}. As we can see, the
massive regime is distinguished from the massless regime by the
opening of a `band gap' at the critical field $h_\ell$.

The main result of our work \cite{DGKS15b} was an explicit formula
for all correlation lengths, or rather all eigenvalue ratios, in
the low-temperature regime. At low enough temperatures all excitations
are parameterized by solutions of the higher-level Bethe Ansatz
equations (\ref{hlbaes}). Thus, instead of $\r_n$ we shall rather
write $\r = \r\bigl(\{x_i\}_{i=1}^{n_h}, \{y_j\}_{j=1}^{n_p} |k\bigr)$.
With this change of notation the eigenvalue ratios at finite magnetic
field are expressed as follows \cite{DGKS15b}
\begin{multline} \label{evarat}
     \r \bigl(\{x_i\}_{i=1}^{n_h}, \{y_j\}_{j=1}^{n_p}|k\bigr)
        = (-1)^k \exp \biggl\{
                    2 \p \i \Bigl[ \sum_{j=1}^{n_p} p(y_j)
		                 - \sum_{j=1}^{n_h} p(x_j) \Bigr] \biggr\} \\
               = (-1)^{k}
	\biggl[ \prod_{j=1}^{n_p}
	        \frac{\dh_1(y_j - \i \g/2, q^2)}
		     {\dh_4(y_j - \i \g/2, q^2)} \biggr]
	\biggl[ \prod_{j=1}^{n_h}
	        \frac{\dh_4(x_j - \i \g/2, q^2)}
		     {\dh_1(x_j - \i \g/2, q^2)} \biggr] \epc
\end{multline}
this being valid up to multiplicative corrections of the order
$\bigl(1 + {\cal O} (T^\infty)\bigr)$. Note that the value of the
magnetic field $h$ enters here through the particle and hole
parameters.

In the special case when there are neither particles nor holes,
$n_p = n_h = 0$, the eigenvalue ratios reduce to $\r (\emptyset,
\emptyset |k) = (-1)^k$.  As we have discussed in \cite{DGKS15b}
the value $k = 0$ corresponds to the case when the dominant
state eigenvalue is divided by itself, while $k = 1$ corresponds
to an eigenvalue of an excited state which is almost degenerate
in absolute value with the dominant state, meaning that up to
sign the two eigenvalues differ only by a factor of $\bigl(1 +
{\cal O} (T^\infty)\bigr)$. Using the properties of the dressed
momentum function $p$ in the complex plane, it is not difficult
to see \cite{DGKS15b} that for all other states, for which
$n_p$ or $n_h$ is non-zero, $|\r| < 1$. For small finite
temperature the eigenvalue ratios form a sequence of discrete
values corresponding to a discrete spectrum of correlation
lengths \cite{DGKS15b}. Their numerical values can be easily
obtained from (\ref{hlbaes}), (\ref{evarat}). The largest
correlation length for $s = 0$ corresponds to a single
particle-hole excitation with $n_p = n_h = 1$ and $\ell_1 =
- m_1 = 1$ in (\ref{hlbaes}). The corresponding particle is located at $x$
and the hole at $y$. Larger `quantum numbers'
$\ell_1$, $m_1$ or multiple particle-hole excitations lead to
shorter, subdominant correlation lengths.

As $T \rightarrow 0+$ the eigenvalue ratio corresponding to the
largest correlation length converges to
\begin{multline}
     \r_{\rm max} (h)
        = \lim_{T \rightarrow 0+} \max \bigl|\r (\{x\},\{y\}|0)\bigr| \\
        = \Biggl[
            \sqrt{\frac{1}{k^2} - \biggl( \frac{1}{k^2} - 1 \biggr)
                  \biggl( \frac{h}{h_\ell} \biggr)^2} -
            \sqrt{\biggl( \frac{1}{k^2} - 1 \biggr)
                  \biggl( 1 - \biggl( \frac{h}{h_\ell} \biggr)^2 \biggr)}
            \Biggr]^2 \epc
\end{multline}
(see \cite{DGKS15b}) where $k = k(q)$ is the elliptic modulus. For
$h = 0$ this simplifies to
\begin{equation}
     \r_{\rm max} (0) = \frac{1 - \sqrt{1 - k^2}}{1 + \sqrt{1 - k^2}}
        = k(q^2) \epp
\end{equation}
As we know from the analysis of the form factor series of the
ordinary transfer matrix \cite{DGKS15a} and from classical work on the
eight-vertex model \cite{JKM73}, it is this ratio $\r_{\rm max} (0)$
and not $\r_{\rm max} (h)$ which determines the zero-temperature 
correlation length of the longitudinal two-point functions in the
whole antiferromagnetic massive regime $0 \le h < h_\ell$. We
shall see below that the amplitudes as well depend on the magnetic
field through the positions of the particle and hole parameters.
The summation (or rather integration) over infinitely many almost
degenerate terms in the form factor series causes that the
field dependence of the full correlation functions drops out
in the end.

\section{Zero-temperature limit of the amplitudes in the form-factor
expansion} \label{sec:amplitudes}
The amplitudes in the thermal form factor expansion of the
finite-temperature two-point functions were considered in
\cite{DGK13a}. They are products of two matrix elements of
certain non-local operators between dominant state and excited
states of the quantum transfer matrix in the Trotter limit.
In \cite{DGK13a} it was observed that they can be written
as products of three factors
\begin{equation}
     A_n^{xy} = \lim_{\a \rightarrow 0} U_{n, s} (\a) D_n^{xy} (\a)
                F_n^{xy} (\a)  \epc
\quad \text{with} \; \; xy = zz, -+ \epc 
\end{equation}
 which were called `universal part', `determinant
part' and `factorizing part', respectively. It was conjectured
that this structure holds for arbitrary form factors involving
finite products of local operators at neighbouring sites.
For technical reasons dominant state and excited states were
considered at different magnetic fields $h$ and $h'$ related
by the twist parameter $\a$ introduced in (\ref{defshiftfun}),
\begin{equation}
     \a = \frac{h - h'}{2 \g T} \epp
\end{equation}
This leads to slightly generalized amplitudes which are more
convenient in the intermediate steps of the calculations but
in which the limit $\a \rightarrow 0$ has to be performed
eventually in order to obtain the physically relevant correlation
functions. Using $\a$ has the additional advantage that we may
obtain the amplitudes of the longitudinal correlation functions
as the second derivative with respect to $\a$ of a properly defined
generating function (see Appendix~\ref{app:qtm} for the definition).
\subsection{The universal part}
The universal part originates from products over ratios of
eigenvalues of the quantum transfer matrix evaluated at the
Bethe roots of dominant and excited states \cite{DGK13a}. It
depends on the local operators, whose correlation functions
are considered, only through the spin $s$. In order to express
it in a compact way and to prepare for rewriting sums over particle
and hole parameters as integrals, we introduce two more functions.
One corresponds to the leading low-temperature asymptotics of the
auxiliary function,
\begin{equation}
     \fa_n (x|\k') =
        \fa \bigl(x|\{x_i\}_{i=1}^{n_h}, \{y_j\}_{j=1}^{n_p}|k\bigr) =
        \re^{- \e(x)/T + 2 \p \i F(x)} \epc
\end{equation}
which plays a major role in the analysis of the spectrum of the
quantum transfer matrix (cf.\ \cite{DGKS15b} and Appendix~\ref{app:qtm}).
$\fa_n$ approximates the exact auxiliary function up to multiplicative
$1 + \CO \bigl( T^\infty \bigr)$ corrections and depends on a parameter
$\k'$ referring to the rescaled magnetic field $\k' = - h'/(2 \g T)$.
The other one is a ratio of $q$-gamma and $q$-Barnes functions (see
Appendix~\ref{app:qfunctions}),
\begin{equation}
     \Ps (x) = \prod_{\epsilon=\pm}
               \frac{1}{\G_{q^4} \bigl(\2 -\epsilon \frac{\i x}{2 \g} \bigr)
                        \G_{q^4} \bigl(\epsilon\frac{\i x}{2 \g} \bigr)} \:
               \frac{G_{q^4}^4 \bigl(1 +\epsilon \frac{\i x}{2 \g} \bigr)}
	            {G_{q^4}^4 \bigl(\2 -\epsilon \frac{\i x}{2 \g} \bigr)} \epp
\end{equation}
In terms of these functions we can represent the universal part of the
amplitudes as 
\begin{align} \label{unilowtexpl}
     U_{n, s} (\a) & =
     U \bigl( \{x_i\}_{i=1}^{n_h}, \{y_j\}_{j=1}^{n_p}|k\bigr) \notag \\
        & = q^{\a s + s^2}
	    \biggl[ \frac{2 q^{- \a}}{(1 - q^4) \G_{q^4} (\2) G_{q^4}^4 (\2)}
	            \biggr]^{n_p + n_h} \notag \\[1ex]
        & \qd \times
	    \biggl[ \prod_{j=1}^{n_h}
	            \frac{1 - \re^{- 2 \p \i F(x_j)}}{\fa_n' (x_j|\k')} \biggr]
	    \biggl[ \prod_{j=1}^{n_p}
	            \frac{1 - \re^{- 2 \p \i F(y_j)}}{\fa_n' (y_j|\k')} \biggr]
            \biggl[\prod_{j=1}^{n_h} \prod_{k=1}^{n_p}
	           \re^{\ph(x_j, y_k) - \ph(y_k, x_j)} \biggr] \notag \\[1ex]
        & \qd \times
	  \frac{\Bigl[\prod_{1 \le j < k \le n_h} \Ps (x_{jk}) \Bigr] 
	        \Bigl[\prod_{1 \le j < k \le n_p} \Ps (y_{jk}) \Bigr]}
               {\prod_{j=1}^{n_h} \prod_{k=1}^{n_p} \Ps (x_j - y_k)} \epc
\end{align}
where equality holds up to multiplicative $1 +
\CO \bigl( T^\infty \bigr)$ corrections. Here $x_{jk} = x_j - x_k$,
$y_{jk} = y_j - y_k$, and the prime in $\fa_n'$ denotes the
derivative with respect to the first argument. A derivation
of the formula is presented in Appendix~\ref{app:unipart}.
The advantages of expressing everything in terms of $q$-gamma
and $q$-Barnes functions are first, that the isotropic limit
will be rather obvious in this formulation, and second, that
we can use the known functional equations among these functions
to rewrite the universal part in many useful ways.

\subsection{The determinant part}
The determinant part has its origin in a ratio of two products of
determinants \cite{DGK13a}. We derive its low-temperature limit
in the antiferromagnetic massive regime in Appendix~\ref{app:detpart}.
Here we summarize the result.

We introduce a `weight function'
\begin{equation} \label{defw}
     w(x) = (-1)^k
            \biggl[\prod_{\ell = 1}^{n_h}
	           \frac{\dh_1 (x - x_\ell, q^2)}{\dh_4 (x - x_\ell, q^2)} \biggr]
                   \biggl[\prod_{\ell = 1}^{n_p}
                          \frac{\dh_4 (x - y_\ell, q^2)}{\dh_1 (x - y_\ell, q^2)}
			  \biggr]
\end{equation}
and the `deformed kernel'
\begin{equation} \label{stupidkernel}
     K_\a (x) = \frac{1}{2\p\i}
                \bigl( q^{- \a} \ctg(x - \i \g) - q^\a \ctg(x + \i \g) \bigr) \epp
\end{equation}
The determinant part is parameterized by two kernel functions
$K^\pm$ which are different for the transversal and the
longitudinal case. They can be expressed in terms of $K_\a$
and will be specified below in (\ref{klongi}) and (\ref{ktrans}).
Given the kernel functions we define
\begin{subequations}
\label{vpm}
\begin{align}
     & \frac{v^- (x_j,y)}{2\p\i} = \frac{\res \{ w^{-1} \} (x_j) K^- (x_j, y)}
                          {1 - \re^{2 \p \i F(x_j)}} \epc &&
       V^- (x,y) = \frac{K^- (x, y)}{w(x)} \epc \\
     & \frac{v^+ (x,y_k)}{2\p\i} = \frac{\res \{ w \} (y_k) K^+ (x, y_k)}
                          {\re^{2 \p \i F(y_k)} - 1} \epc &&
       V^+ (x,y) = K^+ (x, y) w(y) \epc
\end{align}
\end{subequations}
$j = 1, \dots, n_h; k = 1, \dots, n_p$, and the resolvent kernels
associated with $K_0$ and $V^\pm$ which are solutions of the linear
integral equations
\begin{equation}
     R(x - y) = K_0 (x - y) - \int_{- \p/2}^{\p/2} \rd z \: K_0 (x - z) R(z - y)
\end{equation}
and
\begin{subequations}
\begin{align}
     & R^- (x, y) = V^- (x, y) - \int_{- \p/2}^{\p/2} \rd z \: R^- (x, z) V^- (z, y) \epc \\
     & R^+ (x, y) = V^+ (x, y) - \int_{- \p/2}^{\p/2} \rd z \: V^+ (x, z) R^+ (z, y) \epp
\end{align}
\end{subequations}

Using these definitions the determinant part can be written as
\begin{align} \label{detlowtexpl}
     & D_n^{xy} (0) = P_n^{xy} \times
        \det_{\rd x, [-\p/2, \p/2]} (1 + \widehat{V}^-)
        \det_{\rd x, [-\p/2, \p/2]} (1 + \widehat{V}^+)
	\notag \\ & \mspace{54.mu} \times
	\det_{m, n = 1, \dots, n_h}
	\Bigl\{ \de_{m, n} + v^- (x_m, x_n)
	       - \int_{- \p/2}^{\p/2} \rd y \: v^- (x_m, y) R^- (y, x_n) \Bigr\}
	       \notag \\ & \mspace{54.mu} \times
	\det_{m, n = 1, \dots, n_p}
	\Bigl\{ \de_{m, n} + v^+ (y_m, y_n)
	       - \int_{- \p/2}^{\p/2} \rd y \: R^+ (y_m, y) v^+ (y, y_n) \Bigr\}
	       \notag \\ & \mspace{54.mu} \times
	\left[
	\det_{\substack{j, \ell = 1, \dots, n_p \\ k, m = 1, \dots, n_h}}
	\begin{vmatrix}
	   \de_{j \ell} + \frac{2\p\i R(y_j - y_\ell)}{\fa_n' (y_\ell|\k)} &
	   - \frac{2\p\i R(y_j - x_m)}{\fa_n' (x_m|\k)} \\
	   \frac{2\p\i R(x_k - y_\ell)}{\fa_n' (y_\ell|\k)} &
	   \de_{k m} - \frac{2\p\i R(x_k - x_m)}{\fa_n' (x_m|\k)}
        \end{vmatrix} \right]^{-1} \epc
\end{align}
where $xy = zz, -+$. Here the first two determinants on the right hand
side are Fredholm determinants of the integral operators $\widehat{V}^\pm$
with kernels $V^\pm$, contour $[-\p/2, \p/2]$ and measure $\rd x$.
It is important to note that such Fredholm determinants can be
very efficiently calculated numerically \cite{Bornemann10}.

The functions $K^\pm$ in the definition (\ref{vpm}) of the kernel
functions have to be specified as
\begin{subequations}
\label{klongi}
\begin{align}
     & K^- (x, y) = K_0 (x - y) - K_0 (\th_- - y) \epc \\
     & K^+ (x, y) = K_0 (x - y) - K_0 (x - \th_+)
\end{align}
\end{subequations}
in the longitudinal case and as
\begin{equation} \label{ktrans}
     K^\pm (x, y) = K_{\pm 1} (x - y)
\end{equation}
in the transversal case. Note that we have already set $\a = 0$
here and that in the longitudinal case the kernels depend
on two parameters $\th_\pm$.

The prefactor depends on the same parameters,
\begin{multline} \label{pzzunequal}
     P_n^{zz} =
        \frac{4 \sin^2 \bigl(\frac{\p k} 2
	      + \p \sum_{j=1}^{n_p} \bigl(p(y_j) - p(x_j)\bigr)\bigr)}
	     {(- q^2; q^2)^4 \bigl(1 - \re^{2 \p \i F(\th_-)}\bigr)
	      \bigl(1 - \re^{- 2 \p \i F(\th_+)}\bigr)} \\
        \times \prod_{k=1}^{n_p}
	   \frac{\G_{q^4} \bigl( 1 + \frac{\th_+ - y_k}{2\i} \bigr)
	         \G_{q^4} \bigl( \2 + \frac{\th_+ - x_k}{2\i} \bigr)
	         \G_{q^4} \bigl( \2 + \frac{\th_- - y_k}{2\i} \bigr)
		 \G_{q^4} \bigl( 1 + \frac{\th_- - x_k}{2\i} \bigr)}
	        {\G_{q^4} \bigl( \2 + \frac{\th_+ - y_k}{2\i} \bigr)
		 \G_{q^4} \bigl( 1 + \frac{\th_+ - x_k}{2\i} \bigr)
		 \G_{q^4} \bigl( 1 + \frac{\th_- - y_k}{2\i} \bigr)
		 \G_{q^4} \bigl( \2 + \frac{\th_- - x_k}{2\i} \bigr)} \epc
\end{multline}
in such a way that $D_n^{zz}$ is independent of $\th_\pm$
(see \cite{KKMST09a}). Choosing, for instance, $\th_+ = \th_- = \th$
the prefactor simplifies to
\begin{equation} \label{pzzequal}
     P_n^{zz} =
        \frac{\sin^2 \bigl(\frac{\p k} 2
	      + \p \sum_{j=1}^{n_p} \bigl(p(y_j) - p(x_j)\bigr)\bigr)}
	     {(- q^2; q^2)^4 \sin^2 \bigl(\p F(\th)\bigr)} \epc
\end{equation}
but e.g.\ for numerical calculations other choices may be useful.

In the transversal case the prefactor
is simply
\begin{equation}
     P_n^{-+} = \frac{1}{4 (- q^2; q^2)^4} \epp
\end{equation}
\subsection{The factorizing part}
The factorizing part is trivial in the longitudinal case,
$F_n^{zz} (\a) = 1$, since we have used the generating
function approach. In the transversal case the factoring
part is of the form
\begin{equation} \label{pmfacpart}
     F_n^{-+} (\x|\a) =
        \frac{G_+^- (\x) \overline{G}_-^+ (\x)}
	     {(q^{\a - 1} - q^{1 - \a})(q^\a - q^{- \a})} \epp
\end{equation}
Using the representation (\ref{glimitsgammaform}) derived in
Appendix~\ref{app:factorpart} and equations (\ref{asyaux}),
(\ref{lowtmeasures}), (\ref{defvminus}), (\ref{defvplus})
(\ref{kpmpmalpha}) and (\ref{ggbargammaform}) we obtain the
low-temperature limit of the functions in the numerator,
\begin{subequations}
\label{glimitslowtform}
\begin{align}
     & G_+^- (\x) = 1 - q^{- 1 - \a} w(\x)
	\notag \\ & \mspace{8.mu}
	- (q^{1 + \a} - q^{- 1 - \a}) \Biggl[
	\sum_{j=1}^{n_p} \frac{\res \{ w \} (y_j)}{\re^{2\p\i F(y_j)} - 1} G_+ (y_j, \x) +
	\int_{-\p/2}^{\p/2} \frac{\rd y}{2\p\i} w(y) G_+ (y, \x) \Biggr] \epc \\[1ex]
     & \overline{G}_-^+ (\x) = - 1 + q^{1 - \a} w^{-1} (\x)
	\notag \\ & \mspace{8.mu}
	- (q^{1 - \a} - q^{- 1 + \a}) \Biggl[
	\sum_{j=1}^{n_h} \frac{\res \{ w^{-1} \} (x_j)}{1 - \re^{2\p\i F(x_j)}}
	   \overline{G}_- (x_j, \x) +
	\int_{-\p/2}^{\p/2} \frac{\rd y}{2\p\i}
	   \frac{\overline{G}_- (y, \x)}{w (y)} \Biggr] \epp
\end{align}
\end{subequations}
Here $G_+ (\cdot, \x)$ and $\overline{G}_- (\cdot, \x)$ are the
solutions of the linear integral equations
\begin{subequations}
\label{ggbarlowtform}
\begin{align}
     & G_+ (x, \x) = - \ctg(x - \x) + q^{- 1 - \a} w(\x) \ctg(x - \x - \i \g)
	\notag \\ & \mspace{126.mu}
	- \sum_{j=1}^{n_p} v^+ (x, y_j) G_+ (y_j, \x)
	- \int_{-\p/2}^{\p/2} \rd y \: V^+ (x, y) G_+ (y, \x) \epc \\[1ex]
     & \overline{G}_- (x, \x) =
        - \ctg(x - \x) + q^{1 - \a} w^{-1} (\x) \ctg(x - \x - \i \g)
	\notag \\ & \mspace{126.mu}
	- \sum_{j=1}^{n_h} \overline{G}_- (x_j, \x) v^- (x_j, x)
	- \int_{-\p/2}^{\p/2} \rd y \: \overline{G}_- (y, \x) V^- (y,x).
\end{align}
\end{subequations}
with
\begin{subequations}
\label{vpmpmalpha}
\begin{align}
     & \frac{v^- (x_j,x)}{2\p\i} = \frac{\res \{ w^{-1} \} (x_j) K_{1 - \a} (x_j - x)}
                          {1 - \re^{2 \p \i F(x_j)}} \epc &&
       V^- (y,x) = \frac{K_{1 - \a} (y - x)}{w(y)} \epc \\
     & \frac{v^+ (x,y_j)}{2\p\i} = \frac{\res \{ w \} (y_j) K_{1 + \a} (x - y_j)}
                          {\re^{2 \p \i F(y_j)} - 1} \epc &&
       V^+ (x,y) = K_{1 + \a} (x - y) w(y) \epp
\end{align}
\end{subequations}
Though we adopt the same symbols $v^{\pm}$ and $V^{\pm}$ as in 
(\ref{vpm}), we restrict our  argument to the transversal case here.

For the physical correlation functions we have to set $\x = - \i \g/2$
and have to send $\a \rightarrow 0$ in (\ref{pmfacpart}). Recall that
this limit exists, since $\overline{G}_-^+ (\x)$ is differentiable
in $\a$ and vanishes at $\a = 0$ \cite{DGK13a}.

\section{Form factor series} \label{sec:series}
For the form factor series (\ref{ffexp}) we have to sum over all
solutions $\{x_i\}_{i=1}^{n_h}, \{y_j\}_{j=1}^{n_p}$ of the
higher-level Bethe Ansatz equations (\ref{hlbaes}) for $k = 0, 1$.
Proceeding as in our recent work \cite{DGKS15a} we use residue
calculus in several complex variables to write the sums over all
solutions of the Bethe Ansatz equations for fixed numbers of
particles and holes as multiple integrals.

Using that $\6_z \ph(x, z) = - 2 \p \i R(z - x)$ we can combine
the products over $\fa_n'$ in the denominator of (\ref{unilowtexpl})
and the last determinant in (\ref{detlowtexpl}) into
\begin{multline} \label{detoneplusa}
     \biggl[ \prod_{j=1}^{n_p} \fa_n' (y_j|\k') \biggr]
     \biggl[ \prod_{j=1}^{n_h} \fa_n' (x_j|\k') \biggr]
	\det_{\substack{j, \ell = 1, \dots, n_p \\ k, m = 1, \dots, n_h}}
	\begin{vmatrix}
	   \de_{j \ell} + \frac{2\p\i R(y_j - y_\ell)}{\fa_n' (y_\ell|\k')} &
	   - \frac{2\p\i R(y_j - x_m)}{\fa_n' (x_m|\k)'} \\[.5ex]
	   \frac{2\p\i R(x_k - y_\ell)}{\fa_n' (y_\ell|\k')} &
	   \de_{k m} - \frac{2\p\i R(x_k - x_m)}{\fa_n' (x_m|\k')}
        \end{vmatrix} \\[1ex] =
	\det_{\substack{j, \ell = 1, \dots, n_p \\ k, m = 1, \dots, n_h}}
	\begin{vmatrix}
	   \6_{v_\ell} \fa (v_j|\{u_i\}, \{v_i\}|k) &
	   \6_{u_m} \fa (v_j|\{u_i\}, \{v_i\}|k) \\[.5ex]
	   \6_{v_\ell} \fa (u_k|\{u_i\}, \{v_i\}|k) &
	   \6_{u_m} \fa (u_k|\{u_i\}, \{v_i\}|k)
        \end{vmatrix}_{\substack{\{u_i\} = \{x_i\}\\\{v_i\} = \{y_i\}}} \epp
\end{multline}
We recall that ${\cal B_\pm}$ defined in (\ref{defbpm}) are the
curves in the upper and lower half planes on which the particles
and holes condense in the low-temperature limit. Let us assume
that these curves are oriented toward the direction of growing
real part. We introduce two simple closed and positively oriented
curves going around ${\cal B_\pm}$ and enclosing all particles and
holes for small finite temperature and denote them by
$\G({\cal B}_\pm)$. Then
\begin{align} \label{multipleressum}
     & \sum_{\substack{\{x_i\}_{i=1}^{n_h}, \{y_j\}_{j=1}^{n_p}\\
           \text{solutions of HBAEs}}}
	   \frac{f(\{x_i\}_{i=1}^{n_h},  \{y_j\}_{j=1}^{n_p}|k)}{
	\displaystyle{\det_{\substack{j, \ell = 1, \dots, n_p \\ k, m = 1, \dots, n_h}}}
	\begin{vmatrix}
	   \6_{v_\ell} \fa (v_j|\{u_i\}, \{v_i\}|k) &
	   \6_{u_m} \fa (v_j|\{u_i\}, \{v_i\}|k) \\[.5ex]
	   \6_{v_\ell} \fa (u_k|\{u_i\}, \{v_i\}|k) &
	   \6_{u_m} \fa (u_k|\{u_i\}, \{v_i\}|k)
        \end{vmatrix}_{\substack{\{u_i\} = \{x_i\}\\\{v_i\} = \{y_i\}}}}
	\notag \\[1.ex] = &
     \int_{\G ({\cal B}_+)} \frac{\rd^{n_p} v}{n_p! (2\p\i)^{n_p}} \:
     \int_{\G ({\cal B}_-)} \frac{\rd^{n_h} u}{n_h! (2\p\i)^{n_h}}
     \notag \\ & \mspace{108.mu} \times
	   \frac{f(\{u_i\}_{i=1}^{n_h},  \{v_j\}_{j=1}^{n_p}|k)}{
     \bigl[ \prod_{j=1}^{n_p} (1 + \fa (v_j|\{u_i\}, \{v_i\}|k)) \bigr]
     \bigl[ \prod_{j=1}^{n_h} (1 + \fa (u_j|\{u_i\}, \{v_i\}|k)) \bigr]}
     \notag \\[1.5ex] = & (-1)^{n_p}
     \int_{{\cal B}_-} \frac{\rd^{n_h} u}{n_h! (2\p\i)^{n_h}} \:
     \int_{{\cal B}_+} \frac{\rd^{n_p} v}{n_p! (2\p\i)^{n_p}} \:
	   f(\{u_i\}_{i=1}^{n_h},  \{v_j\}_{j=1}^{n_p}|k)
	   \bigl(1 + \CO \bigl( T^\infty \bigr)\bigr) ,
\end{align}
if $f$ is holomorphic in all its variables on and inside
$\G ({\cal B}_\pm)$. In the last line we have used that
$\fa (x|\{u_i\}, \{v_i\}|k) = \CO \bigl( T^\infty \bigr)$
for $x$ slightly above ${\cal B}_+$ or slightly below
${\cal B}_-$ and that $\fa (x|\{u_i\}, \{v_i\}|k) =
\CO \bigl( T^{- \infty} \bigr)$ for $x$ slightly below
${\cal B}_+$ or slightly above ${\cal B}_-$.

Note that the terms in (\ref{detoneplusa}) are the only factors
in our expressions for the amplitudes that contain $\CO (T)$
corrections. It follows from the above consideration that these
terms combine in such a way in the form factors series that
the remaining corrections to the correlation functions are
of order $T^\infty$ in the whole antiferromagnetic massive
regime. This is in accordance with our experience with short-range
correlation functions \cite{TGK10a} which show basically
no temperature dependence at small temperatures.

\subsection{The longitudinal two-point functions}
In order to apply the above to the form factor series for
the longitudinal correlation functions we define the
form factor density
\begin{align} \label{zzdense}
     & \CA^{zz} (\{x_i\}_{i=1}^{n_p},  \{y_j\}_{j=1}^{n_p}|k) \notag \\
        & = \biggl[ \frac{2}{(1 - q^4) \G_{q^4} (\2) G_{q^4}^4 (\2)}
	            \biggr]^{2 n_p}
	    \biggl[ \prod_{j=1}^{n_p}
	            \Bigl(1 - \re^{- 2 \p \i F(x_j)}\Bigr)
	            \Bigl(1 - \re^{- 2 \p \i F(y_j)}\Bigr) \biggr] \notag \\[.5ex]
        & \qd \times
            \biggl[\prod_{j,k=1}^{n_p}
	           \re^{\ph(x_j, y_k) - \ph(y_k, x_j)} \biggr]
	  \frac{\prod_{1 \le j < k \le n_p}
	        \Ps (x_{jk}) \Ps (y_{jk})}
               {\prod_{j,k=1}^{n_p} \Ps (x_j - y_k)} \notag \\[1ex]
        & \qd \times
        \frac{\sin^2 \bigl(\frac{\p k} 2
	      + \p \sum_{j=1}^{n_p} \bigl(p(y_j) - p(x_j)\bigr)\bigr)}
	     {(- q^2; q^2)^4 \sin^2 \bigl(\p F(\th)\bigr)}
        \det_{\rd x, [-\p/2, \p/2]} (1 + \widehat{V}^-)
        \det_{\rd x, [-\p/2, \p/2]} (1 + \widehat{V}^+)
	\notag \\[.5ex] & \qd \times
	\det_{m, n = 1, \dots, n_p}
	\Bigl\{ \de_{m, n} + v^- (x_m, x_n)
	       - \int_{- \p/2}^{\p/2} \rd y \: v^- (x_m, y) R^- (y, x_n) \Bigr\}
	       \notag \\[.5ex] & \qd \times
	\det_{m, n = 1, \dots, n_p}
	\Bigl\{ \de_{m, n} + v^+ (y_m, y_n)
	       - \int_{- \p/2}^{\p/2} \rd y \: R^+ (y_m, y) v^+ (y, y_n) \Bigr\}
	       \epp
\end{align}
Here we have combined (\ref{unilowtexpl}) and (\ref{detlowtexpl})
for $n_h = n_p$ and $\a = 0$. Except for the factors in
(\ref{detoneplusa}) we have supplied a factor of $(-1)^{n_p}$
which will be absorbed by the integrals. For the longitudinal
case the integral operators are fixed by the kernels (\ref{vpm}),
(\ref{klongi}). For simplicity we have set $\th_+ = \th_- = \th$,
but since the expression is anyway independent of $\th_\pm$
we could also use (\ref{pzzunequal}) instead of (\ref{pzzequal})
here.

There is a single excited state with $n_p = n_h = 0$. For this
state $k = 1$, the products and finite determinants in
(\ref{zzdense}) are equal to one, and the remaining Fredholm
determinants can be calculated \cite{IKMT99}. This term describes
the staggered order in the antiferromagnetic massive regime
at zero temperature and is equal to the square of the staggered
magnetization first obtained by Baxter \cite{Baxter73a,Baxter76a},
\begin{equation} \label{baxtersquare}
     \CA^{zz} (\emptyset, \emptyset|1) = \frac{(q^2;q^2)^4}{(-q^2;q^2)^4} \epp
\end{equation}

Using the latter result as well as our previous result
(\ref{evarat}) for the eigenvalue ratios and the summation
formula (\ref{multipleressum}) we obtain the form factor series
\begin{multline} \label{formffserieszz}
     \<\s_1^z \s_{m+1}^z\> =
     (-1)^m \frac{(q^2;q^2)^4}{(-q^2;q^2)^4} \\ +
     \sum_{n = 1}^\infty
     \int_{{\cal B}_-} \frac{\rd^n u}{(2\p)^n n!} \:
     \int_{{\cal B}_+} \frac{\rd^n v}{(2\p)^n n!} \:
     \re^{- 2 \p \i m \sum_{j=1}^n (p(u_j) - p(v_j))} \\ \times
     \Bigl[
     \CA^{zz} (\{u_i\}_{i=1}^n,  \{v_j\}_{j=1}^n|0)
     + (-1)^m \CA^{zz} (\{u_i\}_{i=1}^n,  \{v_j\}_{j=1}^n|1) \Bigr]
\end{multline}
for the longitudinal two-point functions, holding, for every fixed $m$,
up to multiplicative corrections of the form $\bigl(1 +
\CO \bigl(T^\infty\bigr)\bigr)$. Here we have also taken into account
that the magnetization per lattice site $\<\s_1^z\> $ vanishes in the
antiferromagnetic massive regime.

Note that the form factor densities satisfy the identity
\begin{multline}
     \CA^{zz} (\{u_i\}_{i=1}^n,  \{v_j\}_{j=1}^n|0) \\ =
        \CA^{zz} (\{u_i + \p \de_{i, k} \}_{i=1}^n, \{v_j\}_{j=1}^n|1) =
	\CA^{zz} (\{u_i\}_{i=1}^n,  \{v_j + \p \de_{j, k} \}_{j=1}^n|1)
\end{multline}
for $k = 1, \dots, n$. Taking into account the quasi periodicity
of the momentum function $p$ it follows that the integrands in
(\ref{formffserieszz}) are $\p$-periodic in all variables $u_i$,
$v_j$, $i, j = 1, \dots, n$. It further follows from the definition
of $\CA^{zz}$ that the integrand is holomorphic in every $u_j$,
$j = 1, \dots, n$, inside the strip $- \g < \Im u_j < 0$ and in
every $v_j$, $j = 1, \dots, n$, inside the strip $0 < \Im v_j < \g$.
This means that the integration contours ${\cal B}_\pm$ in
(\ref{formffserieszz}) can be deformed and shifted inside their
respective strips. Since the dependence on the magnetic field entered
only through these contours, it follows that, in the low-temperature
limit, the form factor series (\ref{formffserieszz}) is independent
of the magnetic field in the full antiferromagnetic massive regime,
$0 < h < h_\ell$, or, in other words, that the magnetic field
dependence is contained in the temperature corrections of the form
$\bigl(1 + \CO \bigl(T^\infty\bigr)\bigr)$ and thus is `exponentially
small'. Hence, as claimed above, it has turned out that the dependence on the magnetic
field  of the individual form factors cancel each other out once the
summation is performed. One should keep in mind, however, that this behaviour is
not uniform in $m$. If we keep $T$ small but fixed, the magnetic field
dependence comes back for $m$ large enough.

We choose to deform the contours ${\cal B}_\pm$ into straight line
segments $[-\p/2,\p/2] \pm \i \g/2$. This choice corresponds to
the limit of ${\cal B}_\pm$ for $h \rightarrow 0$. It seems to be
particularly useful for numerical calculations. With this choice
the form factor series for the longitudinal two-point functions
finally becomes
\begin{multline} \label{formffserieszzv2}
     \<\s_1^z \s_{m+1}^z\> =
     (-1)^m \frac{(q^2;q^2)^4}{(-q^2;q^2)^4} \\[1.5ex] +
     \sum_{\substack{n \in {\mathbb N}\\k = 0, 1}} \frac{(-1)^{km}}{(n!)^2}
     \int_{-\frac\p 2 - \frac{\i \g} 2}^{\frac\p 2 - \frac{\i \g} 2}
     \frac{\rd^n u}{(2\p)^n} \:
     \int_{-\frac\p 2 + \frac{\i \g} 2}^{\frac\p 2 + \frac{\i \g} 2}
     \frac{\rd^n v}{(2\p)^n} \:
     \re^{- 2 \p \i m \sum_{j=1}^n (p(u_j) - p(v_j))} \\[-1ex] \times
     \CA^{zz} (\{u_i\}_{i=1}^n,  \{v_j\}_{j=1}^n|k) \epp
\end{multline}
This series is valid up to multiplicative temperature corrections
of the form $\bigl(1 + \CO \bigl(T^\infty\bigr)\bigr)$. Together
with the analogous result (\ref{formffseriesmpv2}) for the
transversal correlation functions below it is the main result
of this work. We would like to emphasize that it is
different from the previously known form factor series which were
obtained by means of the $q$-vertex operator approach \cite{JiMi95}
or by applying the algebraic Bethe Ansatz approach to the usual 
transfer matrix \cite{DGKS15a}. We claim that our new series
representation, based on form factors of the quantum transfer matrix,
provides a more efficient exact description of the longitudinal
two-point functions at $T = 0$, since it does neither involve
multiple-contour integrals (as the representation in \cite{JiMi95})
nor multiple-residue integrals (as the representation in
\cite{DGKS15a}). Instead we have to deal with Fredholm determinants
which, as we believe, are more efficient in numerical calculations.
\begin{remark}
In the limit $T \rightarrow 0+$ the series representation
(\ref{formffserieszzv2}) holds in the whole antiferromagnetic massive
regime $\D > 1$, $|h| < h_\ell$ and, in particular, also if the
phase boundary $h = h_\ell$ is approached from below. Hence,
when approached from below the leading asymptotic behaviour of the
longitudinal two-point function on the phase boundary is
\begin{equation} \label{frombelow}
     \<\s_1^z \s_{m+1}^z\> \sim (-1)^m \frac{(q^2;q^2)^4}{(-q^2;q^2)^4} \epp
\end{equation}
Remarkably this can be reproduced if we approach the phase
boundary from above and introduce an appropriate scaling function.
Using the the techniques developed in \cite{DGK13a} it can be
shown \cite{Dugave15} that, asymptotically for large $m$ and small
positive $h - h_\ell$,
\begin{equation}
     \<\s_1^z \s_{m+1}^z\> \sim (-1)^m
        \frac{(q^2;q^2)^4}{(-q^2;q^2)^4} \, g(m,h) \epc
\end{equation}
where
\begin{equation} \label{fromabove}
     g(m,h) = \frac{\sqrt{\rm e}\, 2^{1/6}}{A^6}
              \biggl( \frac{2 k}{1 - k^2} \biggr)^{1/4}
              \biggl( \frac{h}{h_\ell} - 1 \biggr)^{- 1/4}
	      \frac{1}{\sqrt{m}}
\end{equation}
and $A$ is the Glaisher-Kinkelin constant. Approaching the phase
boundary from above in such a way that $g(m,h) = 1$ equation
(\ref{fromabove}) reproduces (\ref{frombelow}).
\end{remark}
\subsection{The transversal two-point functions}
The form factor series for the transversal case follows as well
from our results in the previous subsections. We have to combine
(\ref{unilowtexpl}) and (\ref{detlowtexpl}) for $n_h = n_p + 2$
with (\ref{pmfacpart}) and have to send $\alpha$ to zero.
Using the summation formula (\ref{multipleressum}) we obtain
a form factor series of the form
\begin{multline} \label{formffseriesmpv2}
     \<\s_1^- \s_{m+1}^+\> = \\[1ex]
     \sum_{\substack{n \in {\mathbb N}_0\\k = 0, 1}} \frac{(-1)^{km}}{(n+2)!n!}
     \int_{-\frac\p 2 - \frac{\i \g} 2}^{\frac\p 2 - \frac{\i \g} 2}
     \frac{\rd^{n+2} u}{(2\p)^{n+2}} \:
     \int_{-\frac\p 2 + \frac{\i \g} 2}^{\frac\p 2 + \frac{\i \g} 2}
     \frac{\rd^n v}{(2\p)^n} \:
     \re^{- 2 \p \i m \sum_{j=1}^{n+2} p(u_j) + 2 \p \i m \sum_{j=1}^n p(v_j)}
     \\[-1.5ex] \times
     \CA^{-+} (\{u_i\}_{i=1}^{n + 2},  \{v_j\}_{j=1}^n|k) \epc
\end{multline}
where the amplitude densities are defined as
\begin{align} \label{mpdense}
     & \CA^{-+} (\{x_i\}_{i=1}^{n_p + 2},  \{y_j\}_{j=1}^{n_p}|k) \notag \\
        & = \lim_{\a \rightarrow 0}
	    \frac{q \, G_+^- (0) \partial_{\alpha} \overline{G}_-^+ (0)}
	         {2 \gamma (q^{- 1} - q)} \notag \\
        & \qd \biggl[ \frac{2}{(1 - q^4) \G_{q^4} (\2) G_{q^4}^4 (\2)}
	            \biggr]^{2 (n_p + 1)}
	    \biggl[ \prod_{j=1}^{n_p + 2}
	            \Bigl(1 - \re^{- 2 \p \i F(x_j)}\Bigr) \biggr]
	    \biggl[ \prod_{j=1}^{n_p}
	            \Bigl(1 - \re^{- 2 \p \i F(y_j)}\Bigr) \biggr] \notag \\[.5ex]
        & \qd \times
            \biggl[\prod_{j=1}^{n_p + 2} \prod_{k=1}^{n_p}
	           \re^{\ph(x_j, y_k) - \ph(y_k, x_j)} \biggr]
	  \frac{\bigl[ \prod_{1 \le j < k \le n_p + 2} \Ps (x_{jk}) \bigr]
	        \bigl[ \prod_{1 \le j < k \le n_p} \Ps (y_{jk}) \bigr]}
               {\prod_{j=1}^{n_p + 2} \prod_{k=1}^{n_p} \Ps (x_j - y_k)} \notag \\[1ex]
        & \qd \times
        \frac{1}{4 (- q^2; q^2)^4}
        \det_{\rd x, [-\p/2, \p/2]} (1 + \widehat{V}^-)
        \det_{\rd x, [-\p/2, \p/2]} (1 + \widehat{V}^+)
	\notag \\[.5ex] & \qd \times
	\det_{m, n = 1, \dots, n_p + 2}
	\Bigl\{ \de_{m, n} + v^- (x_m, x_n)
	       - \int_{- \p/2}^{\p/2} \rd y \: v^- (x_m, y) R^- (y, x_n) \Bigr\}
	       \notag \\[.5ex] & \qd \times
	\det_{m, n = 1, \dots, n_p}
	\Bigl\{ \de_{m, n} + v^+ (y_m, y_n)
	       - \int_{- \p/2}^{\p/2} \rd y \: R^+ (y_m, y) v^+ (y, y_n) \Bigr\}
	       \epp
\end{align}
Here we adopt the usual conventions for $n_p = 0$: the set of
hole-rapidities is the empty set, products and integrals over
empty sets of holes are replaced by 1. In the transversal case
the integral operators are fixed by the kernel functions
(\ref{vpm}) and (\ref{ktrans}).
\subsection{Discussion and numerical test cases: the longitudinal case}
It is an interesting question how our new form factor series
are related with those known previously. Both, the $q$-vertex operator
approach and the algebraic Bethe Ansatz approach applied to the
usual transfer matrix, employ different pictures of elementary
excitations. For the Hamiltonian and for the usual transfer matrix
in the antiferromagnetic massive regime these are pairs of spinons,
parameterized by pairs of real spinon rapidities. The form
factor series of the longitudinal two-point functions in this
`spinon basis' as obtained, for instance, in \cite{DGKS15a} is of
the form
\begin{multline} \label{formffserieszzspinon}
     \<\s_1^z \s_{m+1}^z\> =
     (-1)^m \frac{(q^2;q^2)^4}{(-q^2;q^2)^4} + \\[1ex] +
     \sum_{\substack{n \in {\mathbb N}\\k = 0, 1}} \frac{(-1)^{km}}{(2n)!}
     \int_{-\frac\p 2}^{\frac\p 2}  \frac{\rd^{2n} u}{(2\p)^{2n}} \:
     \re^{2 \p \i m \sum_{j=1}^{2n} p(u_j)}
     {\cal F}^{zz} (\{u_i\}_{i=1}^{2n}|k) \epp
\end{multline}
For the form factor density ${\cal F}^{zz} (\{u_i\}_{i=1}^{2n}|k)$ in
the general case see \cite{JiMi95,DGKS15a}. We do not want to reproduce
it here. Since we are unable so far to relate this $2n$-spinon
form factor density to the $n$-particle-$n$-hole form factor density%
\footnote{Henceforth denoted $n$-ph amplitude.}
$\CA^{zz} (\{u_i\}_{i=1}^n,  \{v_j\}_{j=1}^n|k)$ in the general case,
we restrict ourselves to $n = 1$. In this case we have an explicit
result \cite{DGKS16a} for ${\cal F}^{zz}$ obtained by numerical
comparison with a two-spinon form factor formula due to Lashkevich
\cite{Lashkevich02}. Namely,
\begin{multline} \label{ampfun}
    {\cal F}^{zz} (\{u_1, u_2\}|k) =
        \frac{\sin^2 \bigl(\p (p(u_1) + p(u_2) + \frac k 2) \bigr)
	\sin^2 (u_{12}) \dh_3^2 \bigl( \frac{u_{12}}2 + \frac{k \p}2, q \bigr)}
	     {\cos ((u_{12} + \i \g + k \p)/2) \cos ((u_{12} - \i \g + k \p)/2)} \\[1ex]
	\times 32 q (q^2;q^2)^2
	\prod_{\s = \pm}
        \frac{(q^4;q^4,q^4)^2}{(q^2;q^4,q^4)^2}
        \frac{(q^4 \re^{2 \i \s u_{12}};q^4,q^4)^2}
	     {(q^2 \re^{2 \i \s u_{12}};q^4,q^4)^2}
	\frac{(q^4 \re^{2 \i \s u_{12}};q^4)}
	     {(q^2 \re^{2 \i \s u_{12}};q^4)} \epc
\end{multline}
where $u_{12} = u_1 - u_2$.

When comparing (\ref{formffserieszzv2}) and (\ref{formffserieszzspinon})
a rather natural guess about the relation of the integrals
on the right hand side of both equations is that the
term corresponding to the $n$-ph contributions in
(\ref{formffserieszzv2}) is equal to the 2n-spinon term
in (\ref{formffserieszzspinon}). Since the integrals to be
compared look like Fourier integrals and since we expect
that they are pairwise equal for all $m \in {\mathbb R}$, we
expect a simple relation between the integrands. So far we
are unable to prove any relation, but we can provide at
least a conjecture supported by strong numerical evidence.
Let us consider the case $n = 1$. In this case we can calculate
$\CA^{zz} (\{u_1\}, \{u_2\}|1)$ numerically with high accuracy
and compare with (\ref{ampfun}).

A na\"\nodoti ve guess would be that ${\cal F}^{zz}
(\{u_1, u_2\}|1)$ would equal $2 \CA^{zz} (\{u_1 - \i \g\},
\{u_2\}|1)$. But this cannot be true, because of the different
properties of the two functions. While ${\cal F}^{zz}
(\{u_1, u_2\}|1)$ is symmetric in $u_1$, $u_2$ and has
a double zero at $u_1 = u_2$, neither of the two properties does hold for $2 \CA^{zz}
(\{u_1 - \i \g\}, \{u_2\}|1)$. This observation gives us a
hint which might be the true relationship between the two
functions.
\begin{conjecture}
Inside the strip $0 < \Im u_1, \Im u_2 < \g$ we have
\begin{equation}
     {\cal F}^{zz}(\{u_1, u_2\}|1) =
          \CA^{zz} (\{u_1 - \i \g\}, \{u_2\}|1)
	+ \CA^{zz} (\{u_2 - \i \g\}, \{u_1\}|1) \epp
\end{equation}
\end{conjecture}
Here both sides of the equation can be computed with several 
digits accuracy, which leaves little doubt about the
correctness of the conjecture. Comparing the combinatorial
factors in (\ref{formffserieszzv2}) and (\ref{formffserieszzspinon})
and noting that
\begin{equation}
     \frac{1}{(n!)^2} = \frac{1}{(2n)!} \begin{pmatrix} 2n \\ n \end{pmatrix}
\end{equation}
it is tempting to speculate that
\begin{equation} \label{faspec}
     {\cal F}^{zz}(\{u_j\}_{j=1}^{2n}|1) =
        \sum_{\substack{(S_1, S_2) \in p_2 (\{u_j\}_{j=1}^{2n}) \\ |S_1| = |S_2| = n}}
	\CA^{zz} (S_1 - \i \g, S_2|1) \epc
\end{equation}
where $p_2 (M)$ is the set of all ordered pairs of disjoint
subsets of $M$, and $|M|$ denotes the number of elements in $M$.
The difficulty in testing (\ref{faspec}) even numerically comes
from the fact that no efficient expressions for the higher-spinon
amplitudes ${\cal F}^{zz}$ beyond (\ref{ampfun}) are known.
So far the four-spinon amplitudes were computed only in the
isotropic limit~\cite{CaHa06}. Note, however, that explicit
expressions for the two-spinon amplitudes ${\cal F}^{-+}$ of
the transversal correlation functions are available (see
equation (\ref{twospinonffpm}) below). We have compared
these numerically with the corresponding two-hole amplitudes
of our approach. Since there are no particles involved in
this case, no symmetrization is necessary, just a proper
identification of rapidity variables. As in the longitudinal
case the numerical agreement of both types of amplitudes was
perfect.

Unlike in the case of the spinon-based approach it seems not
too hard to evaluate the first few higher-ph contributions to the
representation (\ref{formffserieszzv2}) of the longitudinal
correlation function. We denote the term in the sum on the right
hand side of (\ref{formffserieszzv2}) that involves the
$2j$-fold integrals for $k=0$ and $k=1$ by $I_{2j} (m)$ and set
\begin{equation}
     \<\s_1^z \s_{m+1}^z\>_{2n} = 
        \sum_{j = 0}^n I_{2j} (m) \epc \qqd
	   I_0(m):=(-1)^m \frac{(q^2;q^2)^4}{(-q^2;q^2)^4} \epc
\end{equation}
which includes all contributions up to $n$ particles and $n$
holes. The 2-ph contribution to the nearest-neighbour correlator, $I_4 (1)$,
for example, is a four-fold integral. For its numerical calculation
it is crucial that the integration contours are chosen as
$[-\p/2,\p/2] \pm \i \g/2$, since the factors $\re^{2\p\i p}$ are
real on these contours. Then the properties of the corresponding
amplitudes under complex conjugation guarantee that the integral
is real. We can use these properties as well as the fact that
the amplitudes are symmetric in the particle variables and in
the hole variables separately to reduce the computational cost.

We have computed the $n$-fold integrals by means of the
Gau{\ss}-Legendre quadrature rule with $N$ sampling points.
$N$ was increased until the relative change of the
result when incrementing $N$ to $N + 2$ became sufficiently
small.\footnote{If $m=1$, less than 0.01\% for $n=2$ and
$n=4$ for $\gamma \ge 0.3$. For $\gamma=0.2$ it is 0.02\%
for $n=2$ and 0.2\% for $n=4$. A similar accuracy could not
be achieved for $n=6$, as the maximum $N \sim 30$ due to
cpu time limitations. As a rule of thumb, we expect an error
of the order of 10\% in this case.}
An improvement of the numerical accuracy may be possible, but
we content ourselves to a naive approach here. Gfortran was
used to compile the programs with openmp. The computations were
mainly performed on an 8-core workstation (Xenon E5-2620, 2GHz). 
Typical runs consumed $\CO \bigl(10^2\bigr)$ seconds resp.\
minutes or hours (cpu time) for $n=2$ resp.\ $3$ or $4$.

For illustrational purposes the values of $I_{2j}(1)$, $j=1,2,3$,
for various values of $\gamma$ are listed in Table~\ref{tab:123ph}.

\begin{figure}[t]
\begin{center}
\includegraphics[width=.70\textwidth]{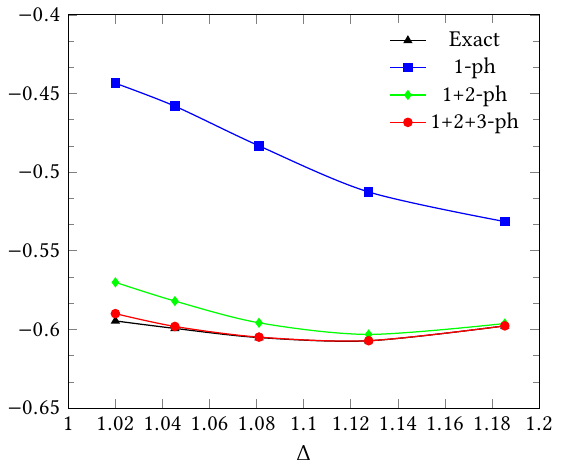} 
\end{center}
\caption{\label{fig:convergencetoexact} 
Convergence of the form factor expansion to exact exact value
of $g^{zz} (1)$, defined in (\ref{definitiongm}), for various values
of $\Delta$. The curves are added as guide for the eyes.
}
\end{figure}

\renewcommand{\arraystretch}{1.3}
\setlength{\tabcolsep}{7pt}
\begin{table}[!h]
\begin{center}
\begin{tabular}{lccccc}
\toprule
$\gamma$ & $0.2$ & $0.3$ & $0.4$ & $0.5$ & $0.6$ \\
\midrule
$I_2(1)$ & $-0.4435707$ & $-0.4581445$ & $-0.4833569$ & $-0.5126763$ &
$-0.5314993$ \\
$I_4(1)$ & $-0.1265448$ & $-0.1238852$ & $-0.1124905$ & $-0.0905522$ &
$-0.0647937$ \\
$I_6(1)$ & $-0.0142596$ & $-0.0149590$ & $-0.0089562$ & $-0.0039798$ &
$-0.0015018$ \\
\bottomrule
\end{tabular}
\caption{Explicit values of $I_{2n} (1)$ for small values of $\gamma$.}
\label{tab:123ph}
\end{center}
\end{table}%
The short-distance correlation functions $\<\s_1^z \s_{m+1}^z\>$,
$m = 1, 2, 3$, are known exactly~\cite{TKS04}. For example, the
neighbour-correlator has the representation
\begin{multline}\label{s1s2}
     \<\s_1^z \s_{2}^z\> = \\
        1 + 2 \int_{-\infty}^\infty \frac{\rd x}{\sh(\p (x + \i/2))}
	      \biggl[ \ctg (\g (x + \i/2)) \cth(\g) - 
	              \frac{x + \i/2}{\sin^2 (\g(x + \i/2))} \biggr] \epp
\end{multline}
This gives us the opportunity to test the $n$-ph approximations
$\<\s_1^z \s_{m+1}^z\>_{2n}$ obtained from our form factor expansion.
We set 
\begin{equation}
     r_n(m):= \frac{ \langle \sigma^z_1 \sigma^z_{m+1} \rangle_{2n} }
                   {\langle \sigma^z_1 \sigma^z_{m+1} \rangle  } \epp
\end{equation}
The data for these ratios in Table~\ref{tab:rn1} clearly demonstrate
that the form factor expansion converges quickly towards (\ref{s1s2}).
See also Figure~\ref{fig:convergencetoexact}.
 
\begin{table}[!h]
\begin{center}
\begin{tabular}{lccccc}
\toprule
$\gamma$ & $0.2$ & $0.3$ & $0.4$ & $0.5$ & $0.6$ \\
\midrule
$r_1(1)$ & $0.745942$ & $0.764335$ & $0.798993$ & $0.846291$ & $0.894282$ \\
$r_2(1)$ & $0.958749$ & $0.971001$ & $0.984525$ & $0.993424$ & $0.997590$ \\
$r_3(1)$ & $0.992216$ & $0.997932$ & $0.999530$ & $0.999912$ & $0.999986$ \\
\bottomrule
\end{tabular}
\caption{Comparison of $n$-ph neighbour correlators against exact results
for $n = 1, 2, 3$ and various values of $\gamma$.}
\label{tab:rn1}
\end{center}
\end{table}
The higher-ph contributions become important as the system
approaches the isotropic point. On the contrary, excitations
higher than 3-ph seem almost negligible for $\gamma>0.5$.
For $m=2, 3$ the ratios $r_3(m)$ stay closer to 1 (Table~\ref{tab:m3}).
This suggests that the contribution from higher particle-hole
excitations becomes less important.
\begin{table}[!h]
\begin{center}
\begin{tabular}{lccccc}
\toprule
$\gamma$ & $0.2$ & $0.3$ & $0.4$ & $0.5$ & $0.6$ \\
\midrule
$r_3 (2) $ & $0.999628$ & $0.999404$ & $0.999972$ & $0.999986$ & $0.999999$ \\
$r_3 (3) $ & $0.993328$ & $0.998205$ & $0.999740$ & $0.999975$ & $0.999999$ \\
\bottomrule
\end{tabular}
\caption{Ratios of third-neighbour 3-ph approximations to exact third-neighbour
correlators. The case $m=1$ appeared before in Table \ref{tab:rn1}.}
\label{tab:m3}
\end{center}
\end{table}

For $m \ge 4$ explicit formulae are not available so far. We
therefore compare our results against standard numerical methods,
the DMRG and brute force diagonalization (the Lanczos method). 
We utilized the software library ALPS ver.\ 2 \cite{Bauer_etal11}.
For the observable we chose 
\begin{equation} \label{definitiongm}
     g^{zz}(m)= (-1)^m \bigl(\langle \sigma^z_1 \sigma^z_{m+1} \rangle -I_0(m) \bigr) \epc
\end{equation}
which vanishes asymptotically and is expected to be positive
for any $m$. The above observable, measured by the DMRG and the
Lanczos method, will be compared with the expansion
$g^{zz}_{\text{ph}}(m) = (-1)^m  \sum_{j=1}^{3} I_{2j}(m)$.

For this purpose we have applied the Lanczos method to chains
of various lengths under periodic boundary conditions. For
our DMRG calculations we employed open boundary conditions%
\footnote{In order to reduce the boundary effect for DMRG, we took the
average 
$$  
     g^{zz}(m) = \frac{(-1)^m}{\ell} \sum_{n=L/2-\ell}^{L/2} 
        \Bigl(\langle \sigma^z_n \sigma^z_{m+n} \rangle - I_0(m) \Bigr).
$$
Typically we chose $\ell=10$.}
and chose parameters MAXSTATES $=50 \sim 150$.
The anisotropy parameter $\Delta$ was varied between $1.1$ and $2$.
The data were extrapolated to the thermodynamic limit assuming the form
\begin{equation} \label{assumedform}
     g^{zz}_L (m) \sim g^{zz}(m) + C(m) {\rm e}^{-\mathfrak{r}  L},
\end{equation}
where $g^{zz}_L(m)$ is a finite-$L$ datum for $g^{zz}(m)$.

\begin{figure}
\begin{center}
\begin{tabular}{@{}lr@{}}
\includegraphics[width=.48\textwidth]{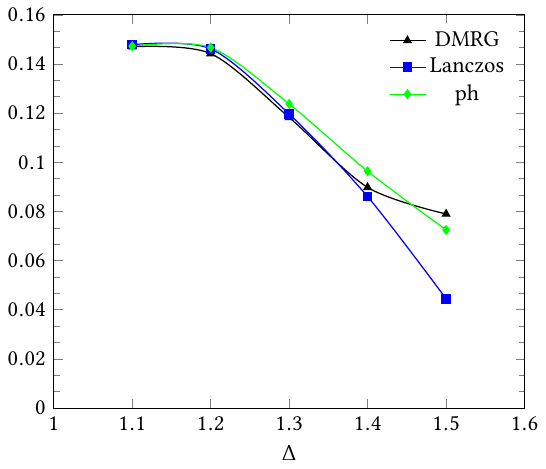} &
\includegraphics[width=.48\textwidth]{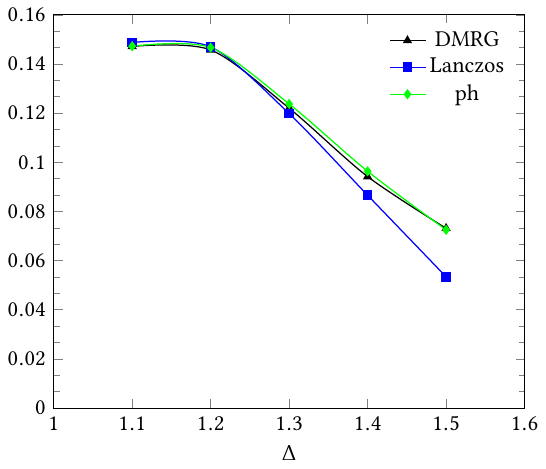} 
\end{tabular}
\end{center}
\caption{\label{fig:g5} 
Comparison of $g^{zz}(5)$ estimated by the Lanczos method (squares)
and by DMRG (triangles) against $g^{zz}(5)_{\rm ph}$. Values of $g^{zz}(5)$
extrapolated using relatively small-$L$ data (left) or large-$L$
data (right). The curves are added as guide for the eyes.
}
\end{figure}

First, we considered system sizes $12 \le L \le 24$ within the Lanczos method and
$48 \le L \le 64$ within the DMRG. Figure~\ref{fig:g5} (left) shows
the resultant values of $g^{zz}(5)$ and $g^{zz}(5)_{\rm ph}$ for $1.1 \le \Delta
\le 1.5$.
One immediately recognizes differences.
The discrepancy is partly due to the large correlation lengths
$\xi$ in the selected range of $\Delta$ (see Table~\ref{tab:correlationlength}).
\begin{table}[!h]
\begin{center}
\begin{tabular}{lcccccc}
\toprule
$\Delta$ & $1.1$ & $1.2$ & $1.3$ & $1.4$ & $1.5$ & $2.0$ \\
\midrule
$\xi$ & $8482.8$ & $347.131$ & $85.1433$ & $37.0497$ & $21.0729$ & $5.29593$ \\
\bottomrule
\end{tabular}
\caption{Correlation lengths for various values of $\Delta$.}
\label{tab:correlationlength}
\end{center}
\end{table}
We thus increased the system size up to $L = 38$ within the
Lanczos method and up to $L = 112$ within DMRG. The assumption
(\ref{assumedform}) then works well for the DMRG for $\Delta=1.4,
1.5$ with $\mathfrak{r} \propto 1/\xi$.  

\begin{figure}[!h]
\begin{center} 
\begin{tabular}{@{}lr@{}}
\includegraphics[width=.48\textwidth]{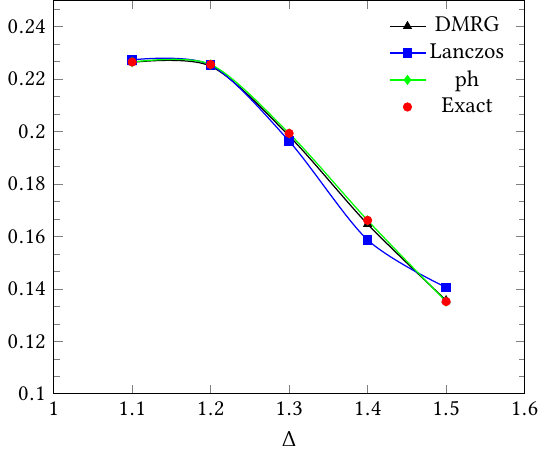} &
\includegraphics[width=.48\textwidth]{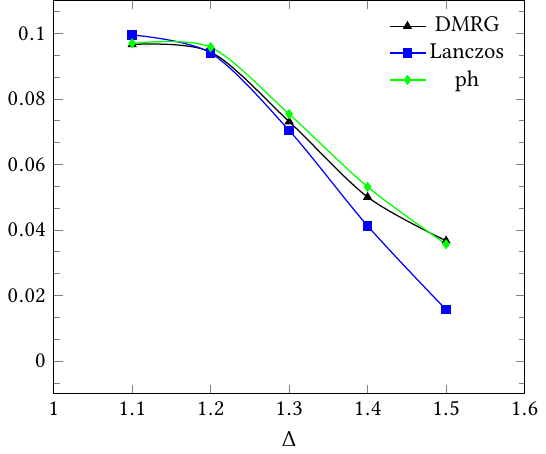} 
\end{tabular}
\end{center}
\caption{\label{fig:g3g8} 
Comparison of $g^{zz}(m)$ estimated by the Lanczos method (squares)
and by DMRG (triangles) against  $g^{zz}(m)_{\rm ph}$. The spin
distance $m$ is 3 (left panel) or 8 (right panel). The red
circles in the left panel denote the exact values.
}
\end{figure}

On the other hand,  the Lanczos data do not necessarily obey
(\ref{assumedform}) for the whole range $12\le L \le 38$.
We nevertheless fitted the data according
to (\ref{assumedform}) and the result is plotted in
Figure~\ref{fig:g5} (right). The coincidence of the DMRG with
the form factor expansion data is improved remarkably, while
it  becomes slightly better for the Lanczos method, as expected.
Probably, the agreement with the Lanczos data could be further
improved if we would consider the ground state together with
the first excited state and take the arithmetic average. Such
kind of analysis is justified as ground state and first
excited state degenerate in the thermodynamic limit, and it
is this average which corresponds to the zero-temperature
limit of the static correlation functions. Supplementary,
in Figure~\ref{fig:g3g8}, we show $g^{zz}(m)$ for some more
values of $m$ corresponding to larger system size data.

\begin{figure}[!h]
\begin{center}
\includegraphics[width=.70\textwidth]{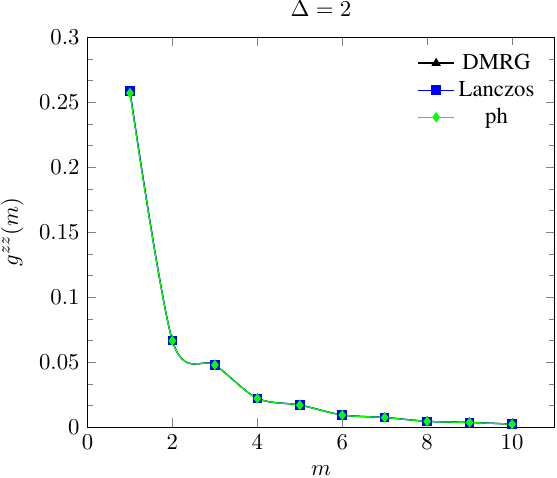}
\end{center}
\caption{\label{fig:gmLambda2} 
Plots of $g^{zz}(m)$ vs.\ $m$ for $\Delta=2$ obtained by three
different methods. The curves are almost indistinguishable.
}
\end{figure}

\begin{figure}[!h]
\begin{center}
\begin{tabular}{@{}lr@{}}
\includegraphics[width=.48\textwidth]{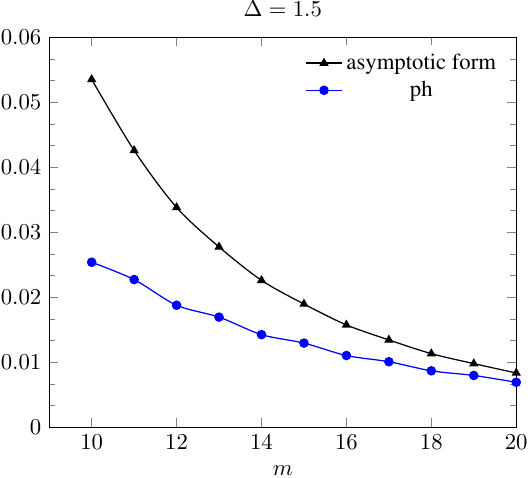} &
\includegraphics[width=.48\textwidth]{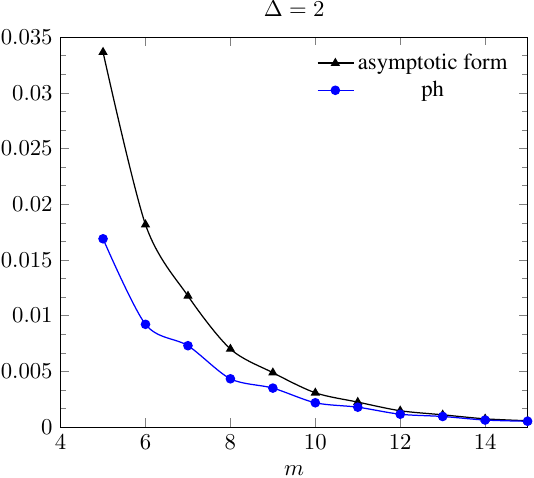} 
\end{tabular}
\end{center}
\caption{\label{fig:asymptoticgm} 
Comparison of $g^{zz}(m)$ with its asymptotic form derived in
\cite{DGKS15a}. For $\Delta=1.5$ (left), due to large $\xi$, 
$g^{zz}(m)$ still deviates considerably from its asymptotic form. 
For $\Delta=2$ (right) $g^{zz}(m)$ exhibits already a good
agreement with the asymptotic form as $\xi \sim 5.29$.
}
\end{figure}

The better agreement for larger $L$ suggests that the three
independent results eventually coincide in the limit
$\xi/L \rightarrow 0$, namely larger $L$ or larger $\Delta$
(where $\xi$ is small). This is consistent with the observation
that $g^{zz}(m)$ for $\Delta=2$ evaluated by both methods is almost
indistinguishable from $g^{zz}(m)_{\rm ph}$ (see Figure~\ref{fig:gmLambda2}).
For $m > \xi$, the form factor expansion successfully reproduces 
the known asymptotic behavior in the ground state \cite{DGKS15a} 
(see Figure~\ref{fig:asymptoticgm}).

Summarizing, we have confirmed the efficiency of the form
factor expansion for arbitrary distance and its consistency
with standard numerical methods. The numerical accuracy
reaches a  satisfactory level, except for the vicinity of
the isotropic point.  We shall discuss this problem separately
in Section~\ref{sec:isolimit}.

\subsection{Numerical test cases: the transversal case}

The numerical analysis of the previous subsection can be performed
for the transversal case in a parallel manner. We thus only briefly
summarize our results.

The form factor series of the transverse correlation function 
in the spinon basis, an analogous formula to (\ref{formffserieszzspinon}),
reads

\begin{align}
\label{ffseriesvoa}
\langle \sigma_1^- \sigma_{m+1}^+ \rangle = 
  \sum_{\substack{n \in {\mathbb N}\\k = 0, 1}}  (-1)^{mk}
 \frac{1}{(2 n)!} \int_{-\pi/2}^{\pi/2} \frac{d^{2n} u}{(2\pi)^{2n}} 
{\rm e}^{-2\pi \i m \sum_{j=1}^{2n} p(u_j)}  {\cal F}^{-+}(\{u\}|k) \epp
\end{align}

The explicit integrand for the 2 spinon case was obtained in \cite{JiMi95}, 
\begin{align} \label{twospinonffpm}
  {\cal F}^{-+} (\{u_1,u_2\} |k) &=  
   4\,  \vartheta^2_3 \Bigl( \frac{u_1+u_2+ k \pi}{2}, q \Bigr) 
   (q^2;q^4)^2    (q^4;q^4)^6   \frac{(q^4;q^4,q^4  )^4}{(q^6;q^4,q^4  )^4}   \notag \\
    & \times  \frac{\sin^2 u_{12}}{\prod_{j=1,2} \vartheta_4 (u_j - \i\gamma/2 , q^2) \vartheta_4 (u_j + \i\gamma/2 , q^2)}  \notag \\
    & \times  \prod_{\sigma=\pm} 
    \frac{ (q^4 {\rm e}^{2 \i \sigma u_{12}} ;q^4,q^4  )^2 }{(q^2 {\rm e}^{2i \sigma u_{12}} ;q^4,q^4  )^2 } 
    \,(q^2 {\rm e}^{2i \sigma u_{12}};q^4) (q^4 {\rm e}^{2i \sigma u_{12}};q^4) \epp
\end{align}
On the other hand, we have our novel form factor series in (\ref{formffseriesmpv2}).
Comparing the two leads us to
\begin{conjecture}
Inside the strip $0 < \Im u_1, \Im u_2 < \gamma$ we have
\begin{equation}
     {\cal F}^{-+}(\{u_1, u_2\}|0) =
        - \CA^{-+} (\{u_1 - \i \g,   u_2- \i \g\}|0) \epp
\end{equation}
\end{conjecture}
We have tested this conjecture numerically. The numerical evidence is
rather convincing. Since the higher-spinon contributions are not known
explicitly, we refrain from further discussion here. 

\begin{figure}
\begin{center}
\includegraphics[width=.70\textwidth]{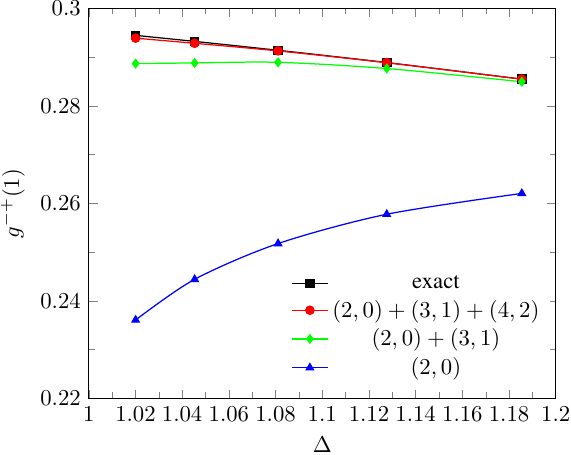}
\caption{\label{fig:gxy1convergence} 
The convergence of $g^{-+}(1)$ to its exact values for various
$\Delta$ near the isotropic point. The plots are labeled by their
quantum numbers $(n_h,n_p)$.
}
\end{center}
\end{figure}
The  formula (\ref{formffseriesmpv2}) is numerically efficient as
in the longitudinal case. Set
\[
g^{-+}(m)=(-1)^m \langle \sigma_1^- \sigma_{m+1}^+ \rangle.
\]
Figure \ref{fig:gxy1convergence} shows the convergence of $g^{-+}(1)$ to
its exact values near the isotropic point with increase in $n_h$. The
curves are indexed by $(n_h,n_p)$: $(2,0)+(3,1)$ means the sum of contributions
from the sectors $(n_h,n_p)=(2,0)$ and  $(n_h,n_p)=(3,1)$, for example.
\begin{figure}
\begin{center}
\begin{tabular}{@{}lr@{}}
\includegraphics[width=.48\textwidth]{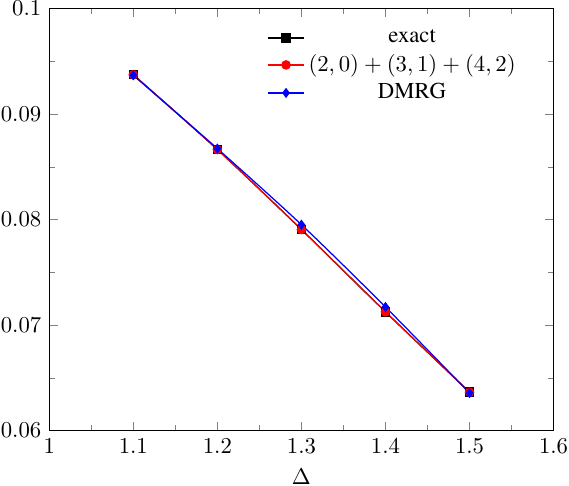} &
\includegraphics[width=.48\textwidth]{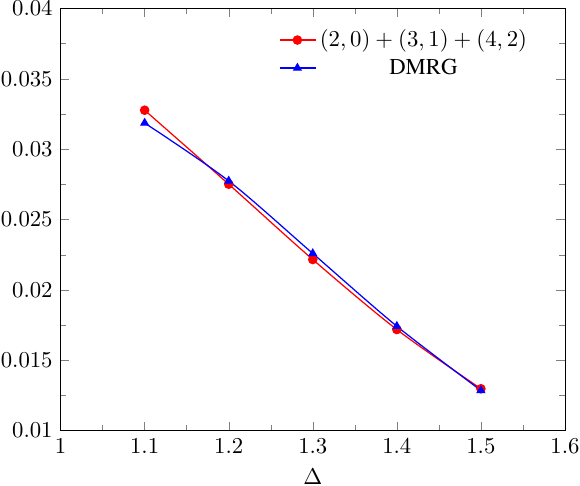}
\end{tabular}
\end{center}
\caption{\label{fig:g3g8pm} 
The  results from the form factor series agree well with those obtained by
DMRG ($L\le 112$) as demonstrated in Fig.~\ref{fig:g3g8pm} ($m=3$ left, $m=8$ right).
Comparison of $g^{-+}(m)$ as obtained by DMRG (triangles) against the form
factor series (circles) for various $\Delta$. The distance is $m=3$ (left)
and $m=8$ (right). For $m=3$ the available exact values are plotted as
black squares.
}
\end{figure}
The data obtained by the Lanczos method ($L\le 24$) deviate from these
two results for small anisotropy. When $\Delta=2$, however, the correlation
length is sufficiently small and all three results coincide with reasonable
accuracy (Table~\ref{tab:3methods}). The nice agreement supports the validity
of the form factor series (\ref{formffseriesmpv2}).
\renewcommand{\arraystretch}{1.3}
\setlength{\tabcolsep}{7pt}
\begin{table}[!h]
\begin{center}
\begin{tabular}{lcccc}
\toprule
$m$ &                      $3$ &                      $5$ &                      $7$ \\
\midrule
Lanczos&      $0.03494286$ &     $0.01136979$ & $0.00466136$ & \\
DMRG&         $0.03490542$&      $0.01136446$ & $0.00465588$ & \\
ph&               $0.03491265$ &     $0.01137057$ & $0.00465968$ & \\
\bottomrule
\end{tabular}
\caption{Explicit values of $g^{-+}(m)$, $m=3,5,7$, at $\Delta=2$ by three
different methods.}
\label{tab:3methods}
\end{center}
\end{table}%
%

\section{The isotropic limit} \label{sec:isolimit}
Within the vertex operator approach the isotropic limit was
considered, for instance, in \cite{JiMi95}. The isotropic
point $\D = 1$, $h = 0$ in the ground state phase diagram
of the XXZ chain is located at the boundary of the antiferromagnetic
massive regime (see Figure~\ref{fig:phasediagram}). In our
formulae for the ground state correlation functions, which
are independent of the magnetic field, it can be reached by
sending $\g \rightarrow 0$ and hence $q \rightarrow 1$.
As is well known this limit requires also a rescaling of
the rapidities $x, y \rightarrow \g u, \g v$ before sending
$\g \rightarrow 0$. Here we are going to perform the isotropic
limit for our form factors densities and the form factor series
for the longitudinal two-point functions, leaving the transversal
case for future study.

We remark that $\lim_{q \rightarrow 1} \G_q (u) = \G (u)$
and $\lim_{q \rightarrow 1} G_q (u) = G (u)$ (see Appendix~%
\ref{app:qfunctions}). This is enough to perform the isotropic
limit for the momentum $p$, the shift function $F$, the weight
functions $w$ and the function $\Ps$ occurring in the universal
part of the amplitudes. We shall denote the limiting functions
by hats, $\hat f (u) = \lim_{\g \rightarrow 0} f(\g u)$. Then
we obtain the momentum
\begin{equation}
     \hat p (u) = \4 + \frac{1}{2\p\i}
	\ln \Biggl(
        \frac{\ch \bigl( \frac \p 2(u + \frac \i 2) \bigr)}
             {\ch \bigl( \frac \p 2(u - \frac \i 2) \bigr)} \Biggr)
\end{equation}
in the isotropic limit. The closely related weight function
turns into
\begin{equation}
     \hat w (u) = (-1)^k \prod_{j=1}^{n_p}
                  \frac{\tgh \bigl( \frac \p 2(u - v_j) \bigr)}
		       {\tgh \bigl( \frac \p 2(u - u_j) \bigr)} \epp
\end{equation}
For the limit of the shift function we first recall the
expression of the two-spinon scattering phase \cite{FaTa81},
\begin{equation}
     \th_F (u) = \frac{1}{2\p\i}
        \ln \Biggl\{ \frac{\G \bigl(1 - \frac{\i u}{2}\bigr)
	                   \G \bigl(\2 + \frac{\i u}{2}\bigr)}
			  {\G \bigl(1 + \frac{\i u}{2}\bigr)
			   \G \bigl(\2 - \frac{\i u}{2}\bigr)} \Biggr\} \epp
\end{equation}
In terms of this scattering phase the rescaled dressed phase $\ph$ and
the rescaled shift function~$F$, 
\begin{equation}
 \hat \ph (u,v) = \lim_{ \gamma \rightarrow  0+}
                  \ph ( \gamma u, \gamma v)  \qquad \text{and}\qquad 
  \hat F(u) = \lim_{ \gamma \rightarrow  0+ } F ( \gamma u ) \epc
\end{equation}
turn into
\begin{align}
     & \hat \ph (u,v) = \frac{\i \p}{2} - 2 \p \i \th_F (u - v) \epc \\
     & \hat F (u) = \frac k 2
       + \sum_{j=1}^{n_p} \bigl( \th_F (u - u_j) - \th_F (u - v_j) \bigr) \epp
\end{align}
The isotropic limit of the function $\Ps$ is simply
\begin{equation}
     \hat \Ps (x) = \prod_{\epsilon=\pm}
                    \frac{1}{\G \bigl(\2 -\epsilon \frac{\i x}{2} \bigr)
                             \G \bigl(\epsilon \frac{\i x}{2} \bigr)} \:
                    \frac{G^4 \bigl(1 + \epsilon \frac{\i x}{2} \bigr)}
	                 {G^4 \bigl(\2 -\epsilon  \frac{\i x}{2} \bigr)} \epp
\end{equation}
With this we have gathered all what is needed to deal with
the universal part of the amplitudes.

For the determinant part we note that for our basic kernel function $K_0$
\begin{equation}
     \lim_{\g \rightarrow 0+} \rd (\g u) \: K_0 (\g u) =
          \frac{\rd u}{\p} \frac{1}{1 + u^2} = \rd u \: \hat K_0 (u) \epp
\end{equation}
In the integrals in the determinant part the rescaling
connected with the isotropic limit leads to the replacement
of the integration interval $[-\p/2,\p/2]$ by $[-\p/(2\g),\p/(2\g)]$
which in the limit $\g \rightarrow 0+$ goes to $(- \infty, \infty)$.
We set
\begin{subequations}
\label{klongiiso}
\begin{align}
     & \hat K^- (u, v) = \hat K_0 (u - v) - \hat K_0 (\th_- - v) \epc \\
     & \hat K^+ (u, v) = \hat K_0 (u - v) - \hat K_0 (u - \th_+)
\end{align}
\end{subequations}
and
\begin{subequations}
\label{vpmiso}
\begin{align}
     & \hat v^- (u_j,v) = \frac{2\p\i \res \{ \hat w^{-1} \} (u_j) \hat K^- (u_j, v)}
                          {1 - \re^{2 \p \i \hat F(u_j)}} \epc &&
       \hat V^- (u,v) = \hat w^{-1} (u) \hat K^- (u, v) \epc \\
     & \hat v^+ (u,v_k) = \frac{2\p\i \res \{ \hat w \} (v_k) \hat K^+ (u, v_k)}
                          {\re^{2 \p \i \hat F(v_k)} - 1} \epc &&
       \hat V^+ (u,v) = \hat K^+ (u, v) \hat w(v) \epp
\end{align}
\end{subequations}
We further define the corresponding resolvent kernels in the isotropic limit
as solutions of linear integral equations,
\begin{subequations}
\begin{align}
     & \hat R^- (u, v) = \hat V^- (u, v)
        - \int_{- \infty}^\infty \rd z \: \hat R^- (u, z) \hat V^- (z, v) \epc \\
     & \hat R^+ (u, v) = \hat V^+ (u, v)
        - \int_{- \infty}^\infty \rd z \: \hat V^+ (u, z) \hat R^+ (z, v) \epc
\end{align}
\end{subequations}
which completes the definitions needed in the description of the isotropic
limit of the finite determinants in (\ref{zzdense}).

In order to perform the isotropic limit of the Fredholm determinants it
is useful to distinguish the cases $k = 0$ and $k = 1$. We show in
Appendix~\ref{app:vanishingstaggered} that
\begin{equation} \label{divideoneplusr}
     \det_{\rd u, [-\p/2, \p/2]} \bigl(1 + \widehat{V}^\pm\bigr)
        = \bigl((-1)^{k+1} q^2;q^2\bigr)^2
	  \det_{\rd u, [-\p/2, \p/2]} \bigl(1 + \widehat{W}_{q, k}^\pm\bigr) \epc
\end{equation}
where $\widehat{W}_{q, k}^\pm$ are integral operators with kernels
\begin{subequations}
\begin{align}
     & W^-_{q, k} (u, v) =
        (w^{-1} (u) - (-1)^k) \bigl(R_k (u - v) - R_k (\th_- - v)\bigr) \epc \\
     & W^+_{q, k} (u, v) =
        \bigl(R_k (u - v) - R_k (u - \th_+)\bigr) (w(v) - (-1)^k)
\end{align}
\end{subequations}
defined in terms of two functions
\begin{subequations}
\label{defrfunctions}
\begin{align}
     & R_0 (u) = \frac{1}{2 \p \i} \, \6_u
                 \ln \Biggl\{ \frac{\G_{q^4} \bigl(1 + \frac{\i u}{2\g}\bigr)
	                     \G_{q^4} \bigl(\2 - \frac{\i u}{2\g}\bigr)}
		            {\G_{q^4} \bigl(1 - \frac{\i u}{2\g}\bigr)
			     \G_{q^4} \bigl(\2 + \frac{\i u}{2\g}\bigr)}
			     \Biggr\} \epc \\
     & R_1 (u) = \frac{1}{2 \p \i} \, \6_u
                 \ln \Biggl\{ \frac{\G_{q^2} \bigl(1 - \frac{\i u}{\g}\bigr)}
		                   {\G_{q^2} \bigl(1 + \frac{\i u}{\g}\bigr)}
			     \Biggr\} \epp
\end{align}
\end{subequations}
The Fredholm determinants on the right hand side of (\ref{divideoneplusr})
provide us with an alternative representation of the determinant part
of the longitudinal correlation functions from which we can easily
obtain the isotropic limit.

By virtue of the results of Appendix~\ref{app:qfunctions} the limits
\begin{equation}
     \hat W_k^\pm (u, v) =
        \lim_{\g \rightarrow 0+} \frac{W_{q, k}^\pm (\g u, \g v)}{\g}
\end{equation}
of the kernel functions exist and define integral operators acting
on the real line. The corresponding Fredholm determinants are finite.
Because of the prefactor $\bigl((-1)^{k+1} q^2;q^2\bigr)^2$ in
(\ref{divideoneplusr}), however, the Fredholm determinants 
$\det_{\rd u, [-\p/2, \p/2]} \bigl(1 + \widehat{V}^\pm\bigr)$ vanish
for $k = 1$ and with them the corresponding amplitudes,
\begin{equation}
     \hat \CA^{zz} (\{u_i\}_{i=1}^n,  \{v_j\}_{j=1}^n|1) =
        \lim_{\g \rightarrow 0+}
	(- \i \g)^{2n} \CA^{zz} (\{\g u_i\}_{i=1}^n,  \{\g v_j\}_{j=1}^n|1) = 0 \epp
\end{equation}

For $k = 0$, on the other hand, the prefactor
$\bigl((-1)^{k+1} q^2;q^2\bigr)^2$ in (\ref{divideoneplusr})
diverges, but when the Fredholm determinant is inserted into the
formula for the amplitudes is canceled by the denominator in such
a way that
\begin{equation}
     \hat \CA^{zz} (\{u_i\}_{i=1}^n,  \{v_j\}_{j=1}^n) =
        \lim_{\g \rightarrow 0}
	(- \i \g)^{2n} \CA^{zz} (\{\g u_i\}_{i=1}^n,  \{\g v_j\}_{j=1}^n|0)
\end{equation}
stays finite in the isotropic limit. Setting $\hat W^\pm (u, v) = \hat
W_0^\pm (u, v)$ we obtain the explicit expressions
\begin{subequations}
\begin{align}
     & \hat W^- (u, v) =
       (1 - \hat w^{-1} (u)) \bigl(\th_F' (u - v) - \th_F'(\th_- - v)\bigr) \epc \\
     & \hat W^+ (u, v) =
       \bigl(\th_F' (u - v) - \th_F' (u - \th_+)\bigr)(1 - \hat w(v))
\end{align}
\end{subequations}
for the remaining kernel functions in the isotropic limit.

Using all the above, the final result for the non-vanishing amplitudes is
\begin{align} \label{zzdenseiso}
     & \hat \CA^{zz} (\{u_i\}_{i=1}^n,  \{v_j\}_{j=1}^n) \notag \\
        & = \biggl[ \frac{1}{2 \G (\2) G^4 (\2)}
	            \biggr]^{2 n}
	    \biggl[ \prod_{j=1}^{n}
	            \Bigl(1 - \re^{- 2 \p \i \hat F(u_j)}\Bigr)
	            \Bigl(1 - \re^{- 2 \p \i \hat F(v_j)}\Bigr) \biggr] \notag \\[.5ex]
        & \qd \times
            \biggl[\prod_{j,k=1}^{n}
	           \re^{\hat \ph(u_j, v_k) - \hat \ph(v_k, u_j)} \biggr]
	  \frac{\prod_{1 \le j < k \le n}
	        \hat \Ps (u_{jk}) 
		\hat \Ps (v_{jk}) }
               {\prod_{j,k=1}^n \hat \Ps (u_j - v_k)} \notag  \\[1ex]
        & \qd \times
        \hat P^{zz} \:
        \det_{\rd u, {\mathbb R}} (1 + \widehat{W}^-)
        \det_{\rd u, {\mathbb R}} (1 + \widehat{W}^+)
	\notag \\[1ex] & \qd \times
	\det_{j, k = 1, \dots, n}
	\Bigl\{ \de_{j, k} + \hat v^- (u_j, u_k)
	       - \int_{- \infty}^\infty \rd v \:
	         \hat v^- (u_j, v) \hat R^- (v, u_k) \Bigr\}
	       \notag \\ & \qd \times
	\det_{j, k = 1, \dots, n}
	\Bigl\{ \de_{j, k} + \hat v^+ (v_j, v_k)
	       - \int_{- \infty}^\infty \rd v \:
	         \hat R^+ (v_j, v) \hat v^+ (v, v_k) \Bigr\} \epc
\end{align}
where
\begin{multline} \label{pzzunequaliso}
     \hat P^{zz}
      = \frac{4 \sin^2 \bigl(
	      \p \sum_{j=1}^{n_p} \bigl(\hat p(v_j) - \hat p(u_j)\bigr)\bigr)}
	     {\bigl(1 - \re^{2 \p \i \hat F(\th_-)}\bigr)
	      \bigl(1 - \re^{- 2 \p \i \hat F(\th_+)}\bigr)} \\
        \times \prod_{k=1}^n
	   \frac{\G \bigl( 1 + \frac{\th_+ - v_k}{2\i} \bigr)
	         \G \bigl( \2 + \frac{\th_+ - u_k}{2\i} \bigr)
	         \G \bigl( \2 + \frac{\th_- - v_k}{2\i} \bigr)
		 \G \bigl( 1 + \frac{\th_- - u_k}{2\i} \bigr)}
	        {\G \bigl( \2 + \frac{\th_+ - v_k}{2\i} \bigr)
		 \G \bigl( 1 + \frac{\th_+ - u_k}{2\i} \bigr)
		 \G \bigl( 1 + \frac{\th_- - v_k}{2\i} \bigr)
		 \G \bigl( \2 + \frac{\th_- - u_k}{2\i} \bigr)} \epc
\end{multline}
if we choose to keep $\th_+$ and $\th_-$ independent. This 
simplifies to
\begin{equation}
     \hat P^{zz}
        = \frac{\sin^2 \bigl(
	        \p \sum_{j=1}^{n_p} \bigl(\hat p(v_j) - \hat p(u_j)\bigr)\bigr)}
	       {\sin^2 \bigl(\hat F (\th) \bigr)}
\end{equation}
for $\th_+ = \th_- = \th$.

Finally, we end up with the following form factor series for the
longitudinal two-point functions in the isotropic limit,
\begin{multline} \label{formffserieszziso}
     \<\s_1^z \s_{m+1}^z\> =
     \sum_{n=1}^\infty \frac{1}{(n!)^2}
     \int_{{\mathbb R} - \frac{\i} 2} \frac{\rd^n u}{(2\p)^n} \:
     \int_{{\mathbb R} + \frac{\i} 2} \frac{\rd^n v}{(2\p)^n} \:
     \re^{- 2 \p \i m \sum_{j=1}^n (\hat p(u_j) - \hat p(v_j))} \\*[-1ex] \times
     \hat \CA^{zz} (\{u_i\}_{i=1}^n,  \{v_j\}_{j=1}^n) \epc
\end{multline}
where $\hat \CA^{zz} (\{u_i\}_{i=1}^n,  \{v_j\}_{j=1}^n)$ is
defined in (\ref{zzdenseiso}).

We believe that this series is a good starting point for
studying the asymptotics of the longitudinal two-point functions
at the isotropic point \cite{Affleck98}, including higher order
logarithmic corrections. As far as its numerical evaluation is
concerned, we are still struggling with technical difficulties
involved in the computation of integrals over infinite intervals.
Here we provide a numerical estimation of the ph contributions to 
$\langle \sigma_1^z \sigma_2^z \rangle$  at the isotropic limit
based on an extrapolation from $\Delta > 1$.

\begin{figure}[t]
\begin{center}
\includegraphics[width=.70\textwidth]{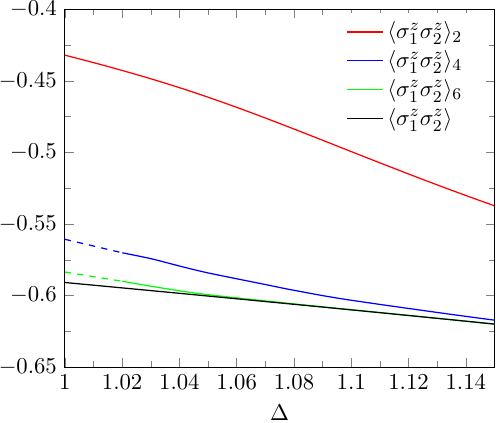}
\end{center}
\caption{\label{fig:extrapolationRational} Extrapolations to
$\Delta=1$  of 1-ph (red curve), 2-ph (blue curve) and 3-ph
(green curve) approximations compared with exact values (black
curve) of $\langle \sigma_1^z \sigma_2^z \rangle$.
}
\end{figure}

Figure~\ref{fig:extrapolationRational} shows the contributions
of the 1-, 2- and 3-ph excitations to $\langle \sigma_1^z
\sigma_2^z \rangle$ as functions of $\Delta$. When $\Delta < 1.02$
we encounter problems with numerical convergence of the 2- and 3-ph
approximations.
Extrapolating  from  $\Delta>1.02$ to the isotropic point we
obtain about 95\% (up to 2-ph)  and 98\%  (up to 3-ph) of
the exact value $\<\s_1^z \s_2^z\> = 1/3 - 4 \ln (2)/3$, which
seems consistent with the fact that the 4-spinon contribution
to the dynamic structure factor of the isotropic Heisenberg
chain saturates a frequency sum rule to 97\% \cite{CaHa06}. The
extrapolation is justified as the $n$-ph approximation
($n\le 3$)  is a continuous functions of $\D$  and the limit
$\D \rightarrow 1+$ exists and is finite as we have seen above.

For future record we supplement an estimate of $\<\s_1^- \s_2^+\>$
obtained by extrapolation of the data for $\D > 1$ to the isotropic
point: 98\% (up to $n_h=3$) and 99\% (up to $n_h=4$) of the exact
value. This seems consistent with the above result.

\section{Conclusions} \label{sec:conclusions}
We have derived novel form factor series representations
for the ground state two-point correlation function of the XXZ
chain in the antiferromagnetic massive regime and of
the XXX chain at vanishing magnetic field. These were
obtained within the algebraic Bethe Ansatz approach
applied to the quantum transfer matrix and are based on our previous
work \cite{DGKS15b} where we analysed the spectrum of
the quantum transfer matrix in the antiferromagnetic
massive regime. Our novel series are manifestly different
from the form factor series obtained within the $q$-vertex
operator approach \cite{JiMi95} or within the algebraic Bethe
Ansatz approach applied to the ordinary transfer matrix
\cite{DGKS15a}.

The novel series representations come with a different
underlying picture of elementary excitations. As we have
argued in \cite{DGKS15b} the spectrum of correlation
lengths of the quantum transfer matrix can be entirely
classified in terms of particle-hole excitations. By
contrast, the excitations of the ordinary transfer matrix
of the XXZ chain in the antiferromagnetic massive regime 
are parameterized by pairs of hole-rapidities interpreted
in terms of spinons. Within the algebraic Bethe Ansatz
approach a complete characterization of the corresponding
Bethe root patterns involve the solution of a set of
transcendental equations, the higher-level Bethe Ansatz
equations, which, for any given set of spinon rapidities,
determines a set of associated non-real Bethe roots
\cite{BVV83,ViWo84,Woynarovich82c,DGKS15a}. In the
thermodynamic limit the form factors still depend on these
roots, which makes the summation rather involved and
is the reason for the appearance of higher dimensional
residues in the description of the form factor densities
in the thermodynamic limit \cite{DGKS15a}. In this context
the form factor series derived above may be interpreted
as the result of a resummation of the contributions from
the non-real Bethe roots. To further support such interpretation
it would be important to prove our conjecture that the
spinon amplitudes can be obtained from the particle-hole
amplitudes by the symmetrization procedure suggested in
equation (\ref{faspec}).

Our preliminary attempts also suggest that the novel form
factor series may turn out to be more efficient in the
actual numerical calculation of at least the static
correlation functions at any distance.%
\footnote{It might be possible to obtain numerically more
efficient expressions for multi-spinon form factors within
the q-vertex operator approach as well (F. Smirnov, private
communication).} This seems to be an implication of our
computation of the 3-ph contribution to the two-point
functions. We further expect from the specific form of the
series that they will turn out to be useful for the
calculation of the large-distance asymptotics, in particular
also in the isotropic limit. We plan to further dwell upon
this issue in our future work.
\\[.5ex]
\noindent {\bf Acknowledgment.}
The authors would like to thank Alexander Wei{\ss}e and
Jesko Sirker for helpful discussions about the numerical
computation of correlation functions and Alexander Wei{\ss}e
in addition for providing his Lanczos data for $L = 26$-$38$.

MD and FG acknowledge financial support by the Volkswagen
Foundation and by the DFG under grant number Go 825/7-1. KKK
is supported by the CNRS. His work has been partly financed
by a Burgundy region PARI 2013-2014 FABER grant `Structures
et asymptotiques d'int\'egrales multiples' and by the ANR
`DIADEMS' SIMI 1 2010-BLAN-0120-02. JS is supported by a
JSPS Grant-in-Aid for Scientific Research (C) No.\ 15K05208.

\clearpage

\setcounter{footnote}{0}

{\appendix
\Appendix{Quantum transfer matrix and thermal form factors}%
\label{app:qtm}
In order to make this work more self-contained we review some of
our previous results on thermal form factors \cite{DGK13a} and
on the low-temperature spectrum of correlation lengths \cite{DGKS15b},
adapting the notation to the antiferromagnetic massive regime
where necessary.

\subsection{Quantum transfer matrix approach to correlation functions}
A quantum transfer matrix approach for the calculation of temperature
dependent correlation functions of Yang-Baxter integrable quantum chains
was devised in \cite{GKS04a}. Its basic input is the $R$-matrix of
the underlying vertex model. For the XXZ-chain the relevant vertex model
is the six-vertex model with $R$-matrix
\begin{equation} \label{rmatrix}
     \begin{array}{cc}
     R(x,y) = \begin{pmatrix}
                  1 & 0 & 0 & 0 \\
		  0 & b(x,y) & c(x,y) & 0 \\
		  0 & c(x,y) & b(x,y) & 0 \\
		  0 & 0 & 0 & 1
		 \end{pmatrix} \epc &
     \begin{array}{c}
     b(x, y) = \frac{\sin(y - x)}{\sin(y - x + \i \g)} \\[2ex]
     c(x, y) = \frac{\sin(\i \g)}{\sin(y - x + \i \g)}
    \end{array}
    \end{array} \epp
\end{equation}

The $R$-matrix can be used to define the statistical operator
in the canonical ensemble, $\re^{- H/T}$, which is needed to
calculate thermal expectation values. For this purpose we first
associate a staggered monodromy matrix with every site $j \in
\{- L + 1, \dots, L\}$ of the XXZ chain,
\begin{equation}
     T_j (x|\k) =
        q^{\k \s_j^z}
        R_{j \overline{N}} \bigl(x, \tst{\frac{\i \be}{N}} \bigr)
	R_{\overline{N-1} j}^{t_1}
	   \bigl(- \tst{\frac{\i \be}{N}}, x \bigr) \dots
        R_{j \overline{2}} \bigl(x, \tst{\frac{\i \be}{N}} \bigr)
	R_{\bar 1 j}^{t_1} \bigl(- \tst{\frac{\i \be}{N}}, x \bigr)
	\epp
\end{equation}
Here $N \in 2 {\mathbb N}$ is called the `Trotter number',
the indices $\bar j = \bar 1, \dots, \overline N$ refer to $N$
auxiliary sites in `Trotter direction', and `$t_1$' means transposition
with respect to the first space $R$ is acting on. The parameters
\begin{equation}
     \be = - \frac{2 J \sh(\g)}{T} \epc \qqd \k = - \frac{h}{2 \g T}
\end{equation}
are rescaled inverse temperature and magnetic field. Defining
\begin{equation}
     \r_{N, L} =
        \Tr_{\bar 1 \dots \overline N} \{T_{- L + 1} (0|\k) \dots T_L (0|\k)\}
\end{equation}
it is easy to see \cite{GKS04a} that
\begin{equation}
     \re^{- H/T} = \lim_{N \rightarrow \infty} \r_{N, L} \epp
\end{equation}

We call $\r_{N, L}$ a finite Trotter number approximant to the
statistical operator. Using $\r_{N, L}$ we can calculate
approximations to thermal expectation values which become exact in
the limit $N \rightarrow \infty$. In particular, the expectation
value of any product of local operators ${\cal O}^{(j)}
\in \End {\mathbb C}^2$, $j = 1, \dots, m+1$, $m \in {\mathbb N}$,
acting on $m+1$ consecutive sites of the \emph{infinite chain}, is
approximated by
\begin{multline} \label{opexpapr}
     \bigl\< \CO_1^{(1)} \dots \CO_{m+1}^{(m+1)} \bigr\>_N
        = \lim_{L \rightarrow \infty}
	  \frac{\Tr_{- L + 1 \dots L} \bigl\{ \r_{N, L} \CO_1^{(1)} \dots
	             \CO_{m+1}^{(m+1)}\bigr\}}
               {\Tr_{- L + 1 \dots L} \{\r_{N, L}\}} \\[1.5ex]
        = \frac{\<\k| \Tr \{\CO^{(1)} T(0|\k)\} \dots \Tr \{\CO^{(m+1)} T(0|\k)\}|\k\>}
               {\<\k|\k\> \La^{m+1} (0|\k)} \epc
\end{multline}
where $\La (0|\k) = \La_0 (0|\k)$ is the unique eigenvalue of largest
modulus of the quantum transfer matrix $t(\la|\k) = \Tr T(\la|\k)$ at
$\la = 0$, and where $|\k\> = |0;\k\>$ is the corresponding eigenvector
(see \cite{GKS04a} for more details). We call $\La (0|\k)$ the dominant
eigenvalue and $|\k\>$ the dominant eigenstate. All other states
will be called `excited states'. Below we shall be dealing with
sequences of excited states and their eigenvalues which will be
denoted somewhat unspecificly $|n;\k\>$ and $\La_n (\la|\k)$,
respectively.

\subsection{Thermal form factor expansion}
An important class of correlation functions are $\a$-twisted
two-point functions for which $\CO^{(1)}_1 = X_1$, $\CO^{(m+1)}_{m+1} = Y_{m+1}$
and $\CO^{(j)}_j = q^{\a \s^z_j}$ for $j = 2, \dots, m$. Expanding the
right hand side of (\ref{opexpapr}) in a basis of eigenstates of the
$\a$-twisted quantum transfer matrix $t(\la|\k + \a)$ we obtain the
`form factor expansion'
\begin{multline} \label{thermalformfexp}
     \bigl\< X_1 q^{\a \sum_{j=2}^m \s_j^z} Y_{m+1} \bigr\>_N = \\
        \sum_n \frac{\<\k| \Tr \{X T(0|\k)\} |n;\k'\>}
                    {\La_n (0|\k') \<\k|\k\>}
	       \frac{\<n,\k'| \Tr \{Y T(0|\k)\}|\k\>}
		    {\La (0|\k) \<n,\k'|n;\k'\>}
           \biggl( \frac{\La_n (0|\k')}{\La (0|\k)} \biggr)^m \epc
\end{multline}
where $\k' = \k + \a$. Sending $\a \rightarrow 0$ and $N \rightarrow
\infty$ we obtain the two-point functions $\< X_1 Y_{m+1} \>$.

Due to the symmetries of the Hamiltonian (\ref{ham}) there are
only two independent proper two-point functions, $\<\s_1^- \s_{m+1}^+\>$
and $\<\s_1^z \s_{m+1}^z\>$, say. For this reason we may restrict
ourselves to the cases $X = \s^-$, $Y = \s^+$ and $X = Y = \s^z$
in (\ref{thermalformfexp}). Note that
\begin{equation} \label{szconservation}
     [T_j (x|\k), \2 \s_j^z + \h^z] = 0 \epc
\end{equation}
where $\h^z$ is the pseudo spin operator $\h^z = \2 \sum_{k=1}^N
(-1)^k \s_{\overline k}^z$. Equation (\ref{szconservation})
implies that the quantum transfer matrix preserves the pseudo
spin. Hence, all eigenstates $|n;\k\>$ have definite pseudo
spin, $\h^z |n;\k\> = s |n;\k\>$, $s = - N/2, \dots, N/2$.
Furthermore, $\Tr \{X T(0|\k)\}$ changes the pseudo spin by $s$,
if $[\2 \s^z, X] = s X$. Hence, for the transversal case
$X = \s^-$, $Y = \s^+$ the non-vanishing part of the sum
over $n$ in (\ref{thermalformfexp}) is over all states with
$s = 1$, while in the longitudinal case $X = Y = \s^z$ the sum
runs over all states with $s = 0$.

For finite Trotter number $N$ the eigenvalues $\La_n (x|\k)$ and
eigenstates $|n;\k\>$ of the quantum transfer matrix are parameterized
by sets $\{x_j^r\}_{j=1}^M$, $M = N/2 - s$, of so-called Bethe roots.
These are defined with the aid of an auxiliary function
\begin{multline} \label{baaux}
     \fa(x) = \fa \bigl(x \big| \{x_k^r\}_{k=1}^M \bigr)\\ = q^{- 2 \k}
        \biggl[ \frac{\sin \bigl(x + \frac{\i \g} 2 - \frac{\i \be} N \bigr)
	              \sin \bigl(x + \frac{3 \i \g} 2 + \frac{\i \be} N \bigr)}
	             {\sin \bigl(x + \frac{\i \g} 2 + \frac{\i \be} N \bigr)
		      \sin \bigl(x - \frac{\i \g} 2 - \frac{\i \be} N \bigr)}
		      \biggr]^\frac N 2
	\prod_{k=1}^M \frac{\sin(x - x_k^r - \i \g)}{\sin(x - x_k^r + \i \g)} 
\end{multline}
as the solutions of the `Bethe Ansatz equations'
\begin{equation} \label{baes}
     \fa \bigl(x_j^r \big| \{x_k^r\}_{k=1}^M \bigr) = - 1 \epc
        \qd j = 1, \dots, M \epp
\end{equation}
Since every solution corresponds to a state label $(n, \k)$ we write
in the following $\fa_n (x|\k)$ instead of $\fa \bigl(x \big|
\{x_k^r\}_{k=1}^M \bigr)$ if $\{x_k^r\}_{k=1}^M$ satisfies (\ref{baes}).

Any auxiliary function $\fa_n (\cdot |\k) $ associated with a set
of Bethe roots satisfies a nonlinear integral equation
\cite{Kluemper93,DGKS15b}. This fact allows one to identify auxiliary
functions associated with the dominant state and the `low-lying
excited states' of the quantum transfer matrix in the Trotter limit.
Furthermore, it is known for long \cite{Kluemper93} how to write
the corresponding eigenvalues as integrals involving the auxiliary
functions. Using such type of integral representations it is easy
to obtain the eigenvalue ratios
\begin{equation}
     \r_n(x|\a) = \frac{\La_n (x + \i \g/2|\k')}{\La (x + \i \g/2|\k)}
\end{equation}
in the Trotter limit. For the XXZ chain in the antiferromagnetic
massive regime see \cite{DGKS15b}, where also the explicit expressions
(\ref{evarat}) for the eigenvalue ratios
\begin{equation}
     \r_n = \r_n (- \i \g/2|\a)
\end{equation}
in the low-temperature limit were obtained.

\subsection{Amplitudes in the Trotter limit}
In this work we study the amplitudes
\begin{equation}
     A_n^{xy} (\x|\a) = \frac{\<\k| \Tr \{X T(\x|\k)\} |n;\k'\>}
                        {\La_n (\x|\k') \<\k|\k\>}
		   \frac{\<n,\k'| \Tr \{Y T(\x|\k)\}|\k\>}
		        {\La (\x|\k) \<n,\k'|n;\k'\>}
\label{Ampleq}		        
\end{equation}
in the form factor expansion (\ref{thermalformfexp}) of the
two-point functions of the XXZ chain in the antiferromagnetic
massive regime in the Trotter limit at low temperatures. Here, we
adopt the convention that
\[
     X = \s^z\ \text{if $x = z$} \epc \qd
     X = \s^\pm\ \text{if $x = \pm$} \epc \qd 
     X = q^{\a \s^z}\ \text{if $x = \a$} \epc \qd
     X = \id\ \text{if $x = 1$}
\]
and similarly for $Y$ and $y$. We derive explicit expressions for
\begin{equation}
     A_n^{zz} = \lim_{\a \rightarrow 0} \lim_{N \rightarrow \infty}
                A_n^{zz} (0|\a) \epc \qd
     A_n^{-+} = \lim_{\a \rightarrow 0} \lim_{N \rightarrow \infty}
                A_n^{-+} (0|\a) \epp
\end{equation}
In the longitudinal case we utilize the generating function
\begin{equation} \label{defgenfun}
     A_n^{\a 1} (0|\a) = \frac{\<\k|n;\k'\> \<n,\k'|\k\>}
                              {\<\k|\k\> \<n,\k'|n;\k'\>}
\end{equation}
which seems to be more convenient than working directly with
$A_n^{zz} (0|\a)$. Setting $X = Y = q^{\alpha \s^z}$ in
(\ref{thermalformfexp}) and acting with the operator
$\2 D_m^2 \6_{\g \a}^2$, where $D_m$ is defined by $D_m f_m
= f_m - f_{m-1}$, it easy to see that
\begin{equation}
     A_n^{zz} = \lim_{N \rightarrow \infty}
                \2 \bigl(\r_n^{1/2} - \r_n^{- 1/2}\bigr)^2
                \6_{\g \a}^2 A_n^{\alpha 1} (0|\a) \Bigr|_{\a = 0} \epp
\end{equation}

In \cite{DGK13a} we considered $A_n^{\alpha 1} (\x|\a)$ and $A_n^{-+} (\x|\a)$ for
finite Trotter number and in the Trotter limit. We observed
that in both cases the amplitudes consist of three factors,
\begin{equation}
     A_n^{xy} (\x|\a) = U_{n, s} (\a) D_n^{xy} (\a) F_n^{xy} (\x|\a) \epc
\end{equation}
the universal part $U_{n, s} (\a)$, the determinant part
$D_n^{xy} (\a)$ and the factorizing part $F_n^{xy} (\x|\a)$.
The universal part $U_{n, s} (\a)$ does not depend on the
details of the operators $X$, $Y$ in \eqref{Ampleq}, but only on the spin.
Its expression in terms of Bethe roots $\{x_j^r\}_{j=1}^{N/2}$
of the dominant state and $\{y_j^r\}_{j=1}^{N/2-s}$ of an
excited state of spin $s$ takes the form
\begin{equation} \label{universalpart}
     U_{n, s} (\a) = \frac{\prod_{j=1}^{N/2} \r_n (x_j^r|\a)}
                          {\prod_{j=1}^{N/2 - s} \r_n (y_j^r|\a)} \epp
\end{equation}

The determinant part consists of four determinants,
\begin{multline} \label{detpart}
     D_n^{xy} (\a) = \\
        \frac{\det_{N/2}
	         \Bigl\{ \de^j_k + \frac{\r_n^{-1} (x_j^r|\a)}{\fa_0' (x_j^r|\k)}
		         \mathcal{U}^x (x_j^r, x_k^r) \Bigr\}}
             {\det_{N/2} \Bigl\{\de^j_k + \frac{1}{\fa_0' (x_j^r|\k)}
		                K (x_j^r - x_k^r)\Bigr\}}
        \frac{\det_{N/2-s}
	         \Bigl\{ \de^j_k + \frac{\r_n (y_j^r|\a)}{\fa_n' (y_j^r|\k')}
		         \mathcal{U}^y (y_j^r, y_k^r) \Bigr\}}
             {\det_{N/2-s} \Bigl\{ \de^j_k + \frac{1}{\fa_n' (y_j^r|\k')}
	                           K (y_j^r - y_k^r) \Bigr\}} \epp
\end{multline}
Here the primes in $\fa_0'$ and $\fa_n'$ denote the derivative with
respect to the first argument, $\fa_0$ is the auxiliary function of
the dominant state. The kernel functions in the denominator are
defined by $K(x) = 2 \p \i K_0 (x)$, where $K_\a$ was defined in
(\ref{stupidkernel}). The kernels in the numerator depend on the
operators $X$, $Y$ under consideration.
\begin{equation}
     \mathcal{U}^{\pm} (x,y) = 2 \p \i K_{\a \pm 1} (x - y) \epc
\end{equation}
while
\begin{subequations}
\begin{align}
     \mathcal{U}^{\a} (x,y) & = 2 \p \i K_\a (x - y) + \i q^{-\a} - \i q^\a \epc \\[1ex]
     \mathcal{U}^{1} (x,y) & = 2 \p \i K_\a (x - y) - \i q^{-\a} + \i q^\a \epp
\end{align}
\end{subequations}

In the longitudinal case the factorizing part is simply
\begin{equation}
     F_n^{\a 1} (\x|\a) = 1 \epp
\end{equation}
In the transversal case the factorizing part is of the form
\begin{equation} \label{facpart}
     F_n^{-+} (\x|\a) =
        \frac{G_+^- (\x) \overline{G}_-^+ (\x)}
	     {(q^{\a - 1} - q^{1 - \a})(q^\a - q^{- \a})} \epc
\end{equation}
where the functions in the numerator are determined by linear
integral equations \cite{DGK13a}. We describe these functions
below in Appendix~\ref{app:factorpart} after having introduced
some more notation that is useful for taking the zero temperature
limit.
\subsection{Low-temperature limit of auxiliary function and eigenvalue ratio}
As can be seen from the previous section we need to know the
low-temperature behaviour of the auxiliary functions $\fa_n (\cdot |\k)$
and of the eigenvalue ratios $\r_n (\cdot |\a)$ in order to
calculate the amplitudes in the form factor expansion of the
two-point function for $T \rightarrow 0+$. This low-temperature
behaviour was obtained in \cite{DGKS15b}.

After taking the Trotter limit the auxiliary functions at small
temperatures become
\begin{multline} \label{asyaux}
     \fa_n (x|\k) =
        \fa \bigl(x|\{x_i\}_{i=1}^{n_h}, \{y_j\}_{j=1}^{n_p}|k\bigr) = \\*
               (-1)^k \re^{- \frac{\e (x)} T
             + \sum_{k=1}^{n_p} \ph (x,y_k)
             - \sum_{k=1}^{n_h} \ph (x,x_k)} \epp
\end{multline}
Here $\e$ and $\ph$ are the dressed energy and the dressed phase
defined in (\ref{epsdn}) and (\ref{dressedphase}) in the main body
of the text. The number $k \in \{0,1\}$ and the two sets of `particles'
$\{y_j\}_{j=1}^{n_p}$ and `holes' $\{x_i\}_{i=1}^{n_h}$
parameterize all excited states. For given $k$ the latter are
determined by the `higher-level Bethe Ansatz equations'
\begin{equation} \label{hlbaesaform}
     \fa \bigl(x_n|\{x_i\}_{i=1}^{n_h}, \{y_j\}_{j=1}^{n_p}|k\bigr) = - 1 \epc \qd
     \fa \bigl(y_m|\{x_i\}_{i=1}^{n_h}, \{y_j\}_{j=1}^{n_p}|k\bigr) = - 1 \epc
\end{equation}
where $\Im x_n < 0$, $n = 1, \dots, n_h$, and $\Im y_m > 0$,
$m = 1, \dots, n_p$. Equations (\ref{hlbaesaform}) are equivalent
to equations (\ref{hlbaes}) in the main text. They determine the
particles and holes up to the order $T$. Corrections are of order
$T^\infty$. The auxiliary functions depend on $\k$ through $\e$ and
through the particle and hole parameters. Multiplicative temperature
corrections to (\ref{asyaux}) are uniformly of the form $1 +
\CO \bigl(T^\infty\bigr)$ inside the strip $- \g < \Im x < \g$ away
from the line $\Re \e (x) = h$, $- \g < \Im x \le - \g/2$.

Using the low-temperature formula for the eigenvalues $\La_n (x|\k)$
obtained in \cite{DGKS15b} we see that the eigenvalue ratios
behave as
\begin{equation} \label{rholowt}
     \r_n (x|\a) = \r_n^{(0)} (x|\a) \times
                      \begin{cases}
		      1 & - \g < \Im x < 0 \epc \\
		      \frac{1 + \fa_n (x|\k')}{1 + \fa_0 (x|\k)} & \Im x > 0 \epc
                      \end{cases}
\end{equation}
where
\begin{multline} \label{rhozero}
     \r_n^{(0)} (x|\a) =
        (-1)^k \exp \bigl\{ (\i \p k - \a \g) \one_{\Im x > 0} \bigr\} \\ \times
        \biggl( \frac{\cos(x + \i \g)}{\cos(x)} \biggr)^s
        \biggl[ \prod_{j=1}^{n_p} \frac{\sin(x - y_j + \i \g)}{\sin(x - y_j)} \biggr]
        \biggl[ \prod_{j=1}^{n_h} \frac{\sin(x - x_j)}{\sin(x - x_j + \i \g)}
	           \biggr] \\ \times
        \exp \biggl\{ \int_{- \frac \p 2}^{\frac \p 2} \rd y \:
	              K (x - y + \i \g/2|\g/2)
		      \Bigl[ \sum_{k=1}^{n_p} \ph (y,y_k) -
		             \sum_{k=1}^{n_h} \ph (y,x_k) \Bigr] \biggr\} \epp
\end{multline}
Here
\begin{equation}
     K(x|\de) = \frac{1}{2\p\i} \bigl( \ctg(x - \i\de) - \ctg(x + \i\de) \bigr)
\end{equation}
by definition, and
\begin{equation}
     \one_{\rm condition} = \begin{cases}
                            1 & \text{if condition is satisfied} \\
			    0 & \text{else.}
			    \end{cases}
\end{equation}
As before (\ref{rholowt}) and (\ref{rhozero}) hold up to
multiplicative corrections of the form $1 + \CO \bigl( T^\infty \bigr)$
inside the strip $- \g < \Im x < \g$ away from the line
$\Re \e (x) = h$, $- \g < \Im x \le - \g/2$.

\Appendix{Low-temperature limit of the universal part}
\label{app:unipart}\noindent
In this appendix we use (\ref{asyaux}) and (\ref{rholowt}),
(\ref{rhozero}) to calculate the universal part (\ref{universalpart})
of the amplitudes in the Trotter limit at low temperature.

Step 1. Universal part expressed by a contour integral. 

\noindent Using that the Bethe roots $\{x_j^r\}$ of the dominant state in
(\ref{universalpart}) are simple zeros of the function
$1 + \fa_0 (x|\k)$ and that the Bethe roots $\{y_j^r\}$ of the
excited states in (\ref{universalpart}) are simple zeros of
$1 + \fa_n (x|\k')$, we may rewrite (\ref{universalpart}) as
\begin{equation} \label{uniasint}
     U_{n, s} (\a) =
        \biggl[ \prod_{j=1}^{n_p} \frac{1}{\r_n (y_j|\a)} \biggr]
        \exp \biggl\{ \int_{\widetilde {\cal C}} \frac{\rd y}{2\p\i}
	              \ln \bigl( \r_n (y|\a) \bigr)
		      \6_y \ln \biggl( \frac{1 + \fa_0 (y|\k)}{1 + \fa_n (y|\k')}
		                       \biggr) \biggr\} \epp
\end{equation}
Here the contour $\widetilde {\cal C}$, sketched in Figure~\ref{fig:rhoint},
encircles all Bethe roots of the dominant state as well as all
Bethe roots with negative imaginary part of the excited state, while
the Bethe roots of the excited state which have positive imaginary
part and all other singularities of the integrand are outside
$\widetilde {\cal C}$. We assume that the temperature is low enough
for the general low-temperature picture developed in \cite{DGKS15b}
to hold true. In \cite{DGKS15a} we found that, for $T \rightarrow 0+$,
all Bethe roots condense to the curves ${\cal B}_\pm$ determined by
$\Re \e (x) = 0$, $0 < \pm \Im x < \g$. These curves are sketched
in Figure~\ref{fig:regimes} in the main text. The Bethe roots of the
dominant state all condense to ${\cal B}_-$. In the Trotter limit
the excited states have infinitely many Bethe roots located on
${\cal B}_-$ and only finitely many on ${\cal B}_+$. The latter were
called close roots or particles in \cite{DGKS15b}. We denote them by
$y_j$, $j = 1, \dots, n_p$. We define a rectangular contour ${\cal C}$
starting at $- \p/2$ and joining the points $- \p/2, - \p/2 - \i \g/2,
\p/2 - \i \g/2$, and $\p/2$ in a counterclockwise manner.\footnote{This
contour is different from the contour ${\cal C}$ in \cite{DGKS15b} as
it is only half as wide. This choice turns out to be more suited for the analysis of the $T\rightarrow 0+$ limit of the form factors.} Then ${\cal B}_-$ and hence all Bethe roots
of the dominant state and all Bethe roots with negative real part
of the excites states are located inside ${\cal C}$. The only other
singularities of the integrand inside ${\cal C}$ are a finite number
of $n_h$ zeros of $1 + \fa_n (x|\k')$ which are also zeros of
$\r_n (x|\a)$. They were called holes in \cite{DGKS15b}. We denote
them by $x_j$, $j = 1, \dots, n_h$. The holes are excluded
from $\widetilde {\cal C}$ by construction. We can achieve the exclusion
by adding contours ${\cal C}_j$ to ${\cal C}$ starting at $- \p/2$
going straight to $x_j$, going around it in a small circle and going
straight back to $- \p/2$ (see Figure \ref{fig:rhoint}).
\begin{figure}
\begin{center}
\includegraphics[width=.75\textwidth]{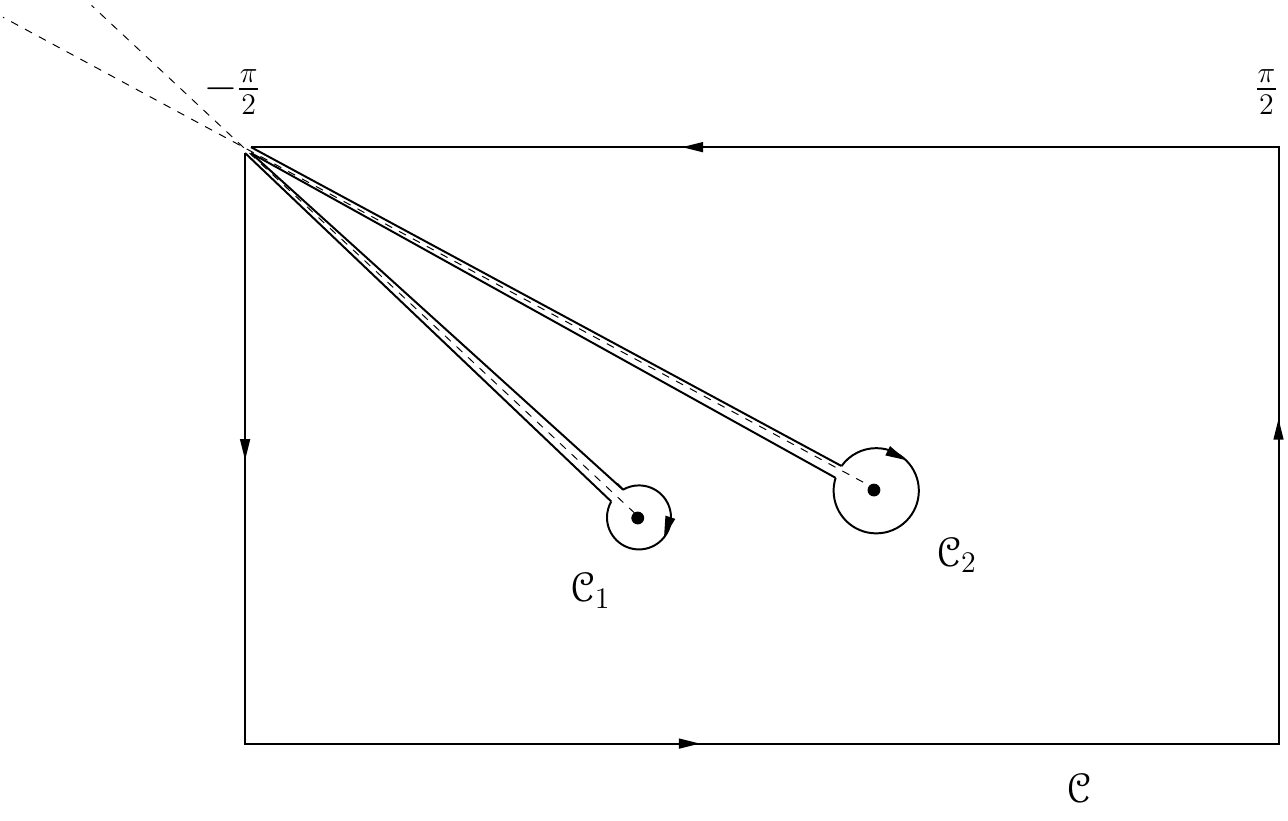}
\caption{\label{fig:rhoint}
The contour $\widetilde {\cal C} = {\cal C} + \sum_{j=1}^{n_h}  {\cal C}_j$.
Here for $n_h = 2$. The branch cuts of $\ln (\r_n (x|\a)$
originating from $x_j$, $j = 1, \dots, n_h$, leave the contour
${\cal C}$ at $- \p/2$.}
\end{center}
\end{figure}

Step 2. `Straightening the contour'.

\noindent We now perform the integrals
over the contours ${\cal C}_j$ (see Figure~\ref{fig:rhoint}) and
integrate partially in the integral over ${\cal C}$. Then some care
is necessary with the definition of the logarithms. Following 
\cite{DGKS15b} we define for any point $x$ on $\cal C$ and
$f = \fa, 1 + \fa, 1 + \fa^{-1}$, where $\fa = \fa_0 (\cdot|\k)$
or $\fa = \fa_n (\cdot|\k')$,
\begin{equation} \label{deflnc}
     \ln_{\cal C} f (x) = \int_{{\cal C}_x} \rd y \: \6_y \ln f (y) \epp
\end{equation}
Here ${\cal C}_x$ is the simple contour which starts at $- \p/2 - \i 0$
and runs along ${\cal C}$ up to the point~$x$. The function $\ln_{\cal C} f$
is holomorphic along ${\cal C}$ by construction and can be used in
partial integration.

For the integral over ${\cal C}$ a partial integration results in
\begin{multline} \label{partialintc}
     \exp \biggl\{ \int_{\cal C} \frac{\rd y}{2\p\i}
                   \ln \bigl( \r_n (y|\a) \bigr)
		   \6_y \ln \biggl( \frac{1 + \fa_0 (y|\k)}{1 + \fa_n (y|\k')}
		                    \biggr) \biggr\} = \\
     \r_n (- \p/2|\a)^{- s}
     \exp \biggl\{ - \int_{\cal C} \frac{\rd y}{2\p\i}
                   \frac{\r_n' (y|\a)}{\r_n (y|\a)}
		   \ln_{\cal C} \biggl( \frac{1 + \fa_0 (y|\k)}{1 + \fa_n (y|\k')}
		                        \biggr) \biggr\} \epp
\end{multline}
The integrals over the ${\cal C}_j$ can be calculated as follows,
\begin{align} \label{intscj}
     & \exp \biggl\{ \int_{{\cal C}_j} \frac{\rd y}{2\p\i}
                     \ln \bigl( \r_n (y|\a) \bigr)
  		     \6_y \ln \biggl( \frac{1 + \fa_0 (y|\k)}{1 + \fa_n (y|\k')}
		                      \biggr) \biggr\} = \notag \\ & \mspace{36.mu}
       \exp \biggl\{ \int_{{\cal C}_j} \frac{\rd y}{2\p\i}
                     \biggl[ \ln \biggl( \frac{\r_n (y|\a)}{y - x_j} \biggr)
		             + \ln(y - x_j) \biggr]
                     \notag \\ & \mspace{162.mu} \times
  		     \biggl[ \6_y \ln \biggl( \frac{(1 + \fa_0 (y|\k))(y - x_j)}
		                                   {1 + \fa_n (y|\k')} \biggr)
                             - \frac{1}{y - x_j} \biggr] \biggr\} =
	             \notag \\ & \mspace{36.mu}
		     - \frac{1 + \fa_n (- \p/2|\k')}{1 + \fa_0 (- \p/2|\k)}
		       \frac{\r_n' (x_j|\a) \bigl(1 + \fa_0 (x_j|\k)\bigr)}
		            {\fa_n' (x_j|\k')} \epp
\end{align}
Here the first logarithms in the square brackets under the second
integral are holomorphic inside ${\cal C}_j$. The second logarithm in
the first square bracket is defined with a branch cut originating
from $x_j$ and going through $- \p/2$. For the second equation
see Appendix~\ref{app:fourierandintegrals}. Inserting (\ref{partialintc})
and (\ref{intscj}) into (\ref{uniasint}) we obtain
\begin{multline} \label{uniasintlowt}
     U_{n, s} (\a) = (-1)^{n_p}
        \biggl[ \prod_{j=1}^{n_p} \frac{1}{\r_n (y_j|\a)} \biggr]
        \biggl[ \prod_{j=1}^{n_h} 
	        \frac{\r_n' (x_j|\a) \bigl(1 + \fa_0 (x_j|\k)\bigr)}
	             {\fa_n' (x_j|\k')} \biggr] \r_n (- \p/2|\a)^{- s} \\
        \biggl( \frac{1 + \fa_n (- \p/2|\k')}{1 + \fa_0 (- \p/2|\k)} \biggr)^{n_h}
        \exp \biggl\{ - \int_{\cal C} \frac{\rd y}{2\p\i}
                      \frac{\r_n' (y|\a)}{\r_n (y|\a)}
		      \ln_{\cal C} \biggl( \frac{1 + \fa_0 (y|\k)}{1 + \fa_n (y|\k')}
		                           \biggr) \biggr\} \epp
\end{multline}
The next steps now consist of inserting the low-temperature expressions
(\ref{asyaux}) and (\ref{rholowt}) into the various terms on the right
hand side of this equation.

Step 3. Low-temperature limit of the integral term and replacing
$\r_n (\cdot|\a)$ by its low-temperature limit. 

\noindent Following essentially
the same reasoning as in equations (31)-(35) of our paper \cite{DGKS15b}
and using that
\begin{equation} \label{semidefect}
     \int_{{\cal C}_+} \frac{\rd y}{2\p\i} \6_y \ln \bigl( \r_n^{(0)} (y|\a) \bigr) =
     \int_{{\cal C}_-} \frac{\rd y}{2\p\i} \6_y \ln \bigl( \r_n^{(0)} (y|\a) \bigr) =
     n_p + s \epc
\end{equation}
where ${\cal C}_+$ and ${\cal C}_-$ are the straight directed contours
connecting $\p/2$ with $-\p/2$ and $-\p/2 - \i \g/2$ with $\p/2 - \i \g/2$,
respectively, we obtain
\begin{multline} \label{expintlowt}
        \biggl( \frac{1 + \fa_n (- \p/2|\k')}{1 + \fa_0 (- \p/2|\k)} \biggr)^{n_h}
        \exp \biggl\{ - \int_{\cal C} \frac{\rd y}{2\p\i}
                      \frac{\r_n' (y|\a)}{\r_n (y|\a)}
		      \ln_{\cal C} \biggl( \frac{1 + \fa_0 (y|\k)}{1 + \fa_n (y|\k')}
		                           \biggr) \biggr\} = \\
        \exp \biggl\{ \int_{- \p/2}^{\p/2} \frac{\rd y}{2\p\i}
                      \frac{\r_n^{(0) \prime} (y|\a)}{\r_n^{(0)} (y|\a)}
		      \bigl[ \ln \bigl( \fa_0 (y|\k) \bigr)
		           - \ln \bigl( \fa_n (y|\k') \bigr) \bigr] \biggr\} \epc
\end{multline}
up to multiplicative corrections of the form $1 + \CO \bigl( T^\infty \bigr)$.
Using (\ref{rholowt}) we further see that
\begin{equation}
     \r_n (y_j|\a) = \lim_{x \rightarrow y_j} \r_n^{(0)} (x|\a)
                     \frac{1 + \fa_n (x|\k')}{1 + \fa_0 (x|\k)} =
		     \res \bigl\{ \r_n^{(0)} (y_j|\a) \bigr\}
                     \frac{\fa_n' (y_j|\k')}{1 + \fa_0 (y_j|\k)} \epc
\end{equation}
up to multiplicative corrections of the form $1 + \CO \bigl( T^\infty \bigr)$.
Equation (\ref{semidefect}) also implies the identity
\begin{multline} \label{expintdlogrhoy}
     \int_{-\p/2}^{\p/2} \frac{\rd y}{2\p\i}
        \bigl( \6_y \ln \bigl( \r_n^{(0)} (y|\a) \bigr) \bigr) 2 \i y s = \\
	s \ln \bigl( \r_n^{(0)} (-\p/2|\a) \bigr) - \i \p s (n_p + s)
	- s \int_{-\p/2}^{\p/2} \frac{\rd y}{\p} \ln \bigl( \r_n^{(0)} (y|\a) \bigr) \epp 
\end{multline}
Inserting (\ref{expintlowt})-(\ref{expintdlogrhoy}) and (\ref{hlbaesaform})
into (\ref{uniasintlowt}) we obtain the following low-temperature
expression for the universal part of the amplitudes,
\begin{multline} \label{unilowt}
     U_{n, s} (\a) =
	\Biggl[ \prod_{j=1}^{n_h}
        \frac{1 - \frac{\fa_0 (x_j|\k)}{\fa_n (x_j|\k')}}{\fa_n' (x_j|\a)} \Biggr]
	\Biggl[ \prod_{j=1}^{n_p}
        \frac{1 - \frac{\fa_0 (y_j|\k)}{\fa_n (y_j|\k')}}{\fa_n' (y_j|\a)} \Biggr]
	\frac{\prod_{j=1}^{n_h} \r_n^{(0)\prime} (x_j|\a) }
	     {\prod_{j=1}^{n_p} \res \bigl\{ \r_n^{(0)} (y_j|\a) \bigr\}} \\
	(-1)^{n_p + s n_p + s}
	\exp \biggl\{
	- s \int_{-\p/2}^{\p/2} \frac{\rd y}{\p} \ln \bigl( \r_n^{(0)} (y|\a) \bigr)
	\biggr\} \\
        \exp \biggl\{ \int_{- \p/2}^{\p/2} \frac{\rd y}{2\p\i}
	              \bigl( \6_y \ln \bigl( \r_n^{(0)} (y|\a) \bigr) \bigr)
		      \bigl[ \ln \bigl( \fa_0 (y|\k) \bigr)
		           - \ln \bigl( \fa_n (y|\k') \bigr) - 2 \i ys \bigr] \biggr\}
\end{multline}
which is again valid up to multiplicative corrections of the form
$1 + \CO \bigl( T^\infty \bigr)$.

Step 4. Inserting explicit expressions, evaluating remaining integrals.

\noindent 
If we insert (\ref{asyaux}) and (\ref{rhozero}) into equation
(\ref{unilowt}) we obtain an expression containing explicit functions
and integrals over explicit functions. The only slightly cumbersome
task that remains is to calculate these integrals. This can be done
in various ways. One way is to use Fourier series representations
and the convolution theorem for Fourier series. We have gathered some
formulae needed in that case in Appendix~\ref{app:fourierandintegrals}.
Before presenting the final formulae we give a few intermediate
results.

First of all, using (\ref{deflnsin}),
\begin{multline}
     \exp \biggl\{ - s \int_{-\p/2}^{\p/2} \frac{\rd y}{\p}
                          \ln \bigl( \r_n^{(0)} (y|\a) \bigr) \biggr\} = \\
     (-1)^{sk + s n_p +s} q^{- s^2}
        \exp \biggl\{ - \i s \Bigl[ \sum_{j=1}^{n_h} x_j
	                          + \sum_{j=1}^{n_p} y_j \Bigr] \biggr\} \epp
\end{multline}
Next, replacing $\ln \bigl( \fa_0 (y|\k) \bigr)
- \ln \bigl( \fa_n (y|\k') \bigr) - 2 \i ys$ by its low-$T$ limit
(\ref{asyaux}), introducing the `periodic form of the dressed
phase'
\begin{equation} \label{defphip}
     \ph_p (x, z) = \ph (x, z) - \i (\p/2 + x - z)
\end{equation}
and using (\ref{semidefect}) we obtain
\begin{multline}
     \exp \biggl\{ \int_{- \p/2}^{\p/2} \frac{\rd y}{2\p\i}
	           \bigl( \6_y \ln \bigl( \r_n^{(0)} (y|\a) \bigr) \bigr)
		   \bigl[ \ln \bigl( \fa_0 (y|\k) \bigr)
		         - \ln \bigl( \fa_n (y|\k') \bigr) - 2 \i ys \bigr] \biggr\} = \\
        (-1)^{(k + s)(n_p + s)} q^{- \a (n_p + s)}
        \exp \biggl\{ \i(n_p + s) \Bigl[ \sum_{j=1}^{n_h} x_j
	                               - \sum_{j=1}^{n_p} y_j \Bigr] \biggr\} \\ \times
     \exp \biggl\{ \int_{- \p/2}^{\p/2} \frac{\rd y}{2\p\i}
	           \bigl( \6_y \ln \bigl( \r_n^{(0)} (y|\a) \bigr) \bigr)
                   \Bigl[ \sum_{j=1}^{n_h} \ph_p (y, x_j)
		        - \sum_{j=1}^{n_p} \ph_p (y, y_j) \Bigr] \biggr\} \epp
\end{multline}
Then we insert (\ref{rholowt}) and (\ref{rhozero}) into the
third factor on the right hand side of (\ref{unilowt}), implying that
\begin{align}
     & \frac{\prod_{j=1}^{n_h} \r_n^{(0)\prime} (x_j|\a) }
            {\prod_{j=1}^{n_p} \res \bigl\{ \r_n^{(0)} (y_j|\a) \bigr\}} = \notag \\
     & \mspace{36.mu} (-1)^{(k + s) n_p} q^{- \a n_p - 2 s^2}
       \exp \biggl\{ \i n_p \sum_{j=1}^{n_h} x_j
	             - \i n_h \sum_{j=1}^{n_p} y_j \biggr\}
       \biggl( \frac{1}{\sin(\i \g)} \biggr)^{2(n_p + s)} \notag \\
     & \mspace{36.mu}
       \biggl[ \prod_{\substack{j, k = 1 \\ j \ne k}}^{n_h}
                  \frac{\sin(x_j - x_k)}{\sin(x_j - x_k + \i \g)} \biggr]
       \biggl[ \prod_{\substack{j, k = 1 \\ j \ne k}}^{n_p}
                  \frac{\sin(y_j - y_k)}{\sin(y_j - y_k + \i \g)} \biggr] \notag \\
     & \mspace{36.mu}
       \biggl[ \prod_{j = 1}^{n_h} \prod_{k = 1}^{n_p}
                  \frac{\sin(x_j - y_k + \i \g)}{\sin(x_j - y_k)}
                  \frac{\sin(y_k - x_j + \i \g)}{\sin(y_k - x_j)} \biggr] \notag \\
     & \mspace{36.mu}
       \exp \biggl\{ \int_{- \p/2}^{\p/2} \rd y \:
                     \Bigl[ \sum_{j=1}^{n_h} K (x_j - y + \i \g/2|\g/2)
		          - \sum_{j=1}^{n_p} K (y_j - y + \i \g/2|\g/2) \Bigr]
			  \notag \\ & \mspace{144.mu} \times
                     \Bigl[ \sum_{j=1}^{n_h} \ph_p (y, x_j)
		          - \sum_{j=1}^{n_p} \ph_p (y, y_j) \Bigr]
			  \biggr\} \epp
\end{align}
Moreover,
\begin{multline}
     \frac{1}{2\p\i} \6_y \ln \bigl( \r_n^{(0)} (y|\a) \bigr) =
        - \frac{s}{\p}
	+ \sum_{j=1}^{n_h} K (y - x_j + \i \g/2|\g/2)
	- \sum_{j=1}^{n_p} K (y - y_j + \i \g/2|\g/2) \\
     + \int_{- \p/2}^{\p/2} \frac{\rd z}{2 \p \i}
        K (y - z + \i \g/2|\g/2) \6_z
	\Bigl[ \sum_{j=1}^{n_p} \ph_p (z, y_j) - 
	       \sum_{j=1}^{n_h} \ph_p (z, x_j) \Bigr] \epp
\end{multline}

All the remaining integrals can now be calculated e.g.\ by means of
equations (\ref{convolution})-(\ref{reduction}) in
Appendix~\ref{app:fourierandintegrals}. This leads to the following
expression for the universal part of the amplitudes expressed
in terms of $q$-multi factorials.
\begin{align} \label{unilowtexplmfform}
     U_{n, s} (\a) & =
	    \biggl[ \prod_{j=1}^{n_h}
	            \frac{1 - \re^{- 2 \p \i F(x_j)}}{\fa_n' (x_j|\k')} \biggr]
	    \biggl[ \prod_{j=1}^{n_p}
	            \frac{1 - \re^{- 2 \p \i F(y_j)}}{\fa_n' (y_j|\k')} \biggr] \notag \\
        & \qd \times
	  (-1)^{n_p + s } q^{- \a (2 n_p + s) + s^2} 2^{4 s^2}
	  \exp\biggl\{ 2 \i n_p \sum_{k=1}^{n_h} x_k 
	               - 2 \i n_h \sum_{k=1}^{n_p} y_k \biggr\} \notag \\
        & \qd \times
	  \frac{\Bigl[\prod_{j, k = 1 \atop j \ne k}^{n_h} \sin(x_{jk})\Bigr]
	        \Bigl[\prod_{j, k = 1 \atop j \ne k}^{n_p} \sin(y_{jk})\Bigr]}
               {\prod_{j=1}^{n_h} \prod_{k=1}^{n_p} \sin(x_j - y_k)
	        \sin(y_k - x_j)} \notag \\
        & \qd \times
          \biggl[\prod_{j, k = 1}^{n_h} (q^2 \re^{2 \i x_{jk}}; q^4)
	            (q^4 \re^{2 \i x_{jk}}; q^4)
		    \frac{(q^4 \re^{2 \i x_{jk}}; q^4,q^4)^4}
		         {(q^2 \re^{2 \i x_{jk}}; q^4,q^4)^4} \biggr] \notag \\
        & \qd \times
          \biggl[\prod_{j, k = 1}^{n_p} (q^2 \re^{2 \i y_{jk}}; q^4)
	            (q^4 \re^{2 \i y_{jk}}; q^4)
		    \frac{(q^4 \re^{2 \i y_{jk}}; q^4,q^4)^4}
		         {(q^2 \re^{2 \i y_{jk}}; q^4,q^4)^4} \biggr] \notag \\
        & \qd \times
          \biggl[\prod_{j=1}^{n_h} \prod_{k=1}^{n_p}
		    \frac{(q^2 \re^{- 2 \i (x_j - y_k)}; q^4)}
		         {(q^2 \re^{2 \i (x_j - y_k)}; q^4)^3}
		    \frac{(q^4 \re^{2 \i (x_j - y_k)}; q^4)}
		         {(q^4 \re^{- 2 \i (x_j - y_k)}; q^4)^3}
			 \biggr] \notag \\
        & \qd \times
          \biggl[\prod_{j=1}^{n_h} \prod_{k=1}^{n_p} \prod_{\s = \pm 1}
		    \frac{(q^2 \re^{2 \i \s (x_j - y_k)}; q^4,q^4)^4}
		         {(q^4 \re^{2 \i \s (x_j - y_k)}; q^4,q^4)^4}
			 \biggr] \epp
\end{align}

From here we arrive at equation (\ref{unilowtexpl}) in the
main body of the text if we replace systematically the $q$-multi
factorials by $q$-gamma and $q$-Barnes function and the
sine functions by $q$-numbers using the formulae collected
in Appendix~\ref{app:qfunctions}.

\Appendix{Low-temperature limit of the determinant part}
\label{app:detpart}\noindent
In the Trotter limit the determinants in (\ref{detpart}) turn
into Fredholm determinants of linear integral operators
defined by their kernels and by certain contours and `measures'
(or `weight functions'). All of this was described in some detail
in \cite{DGK13a}. Here we have to adapt the notation to the
antiferromagnetic massive regime.

\subsection{Determinants in the numerator}
We begin our discussion with the determinants in the numerator
in (\ref{detpart}). Let
\begin{equation}
     \rd M^\pm (x) =
        \frac{\rd x \: \r_n^{\pm 1} (x|\a)}
	     {1 - \Bigl(\frac{\fa_0 (x|\k)}{\fa_n (x|\k')}\Bigr)^{\pm 1}}
        \biggl( \frac{1 + \fa_0 (x|\k)}{1 + \fa_n (x|\k')} \biggr)^{\pm 1}
	\epp
\end{equation}
\begin{figure}
\begin{center}
\includegraphics[width=.85\textwidth]{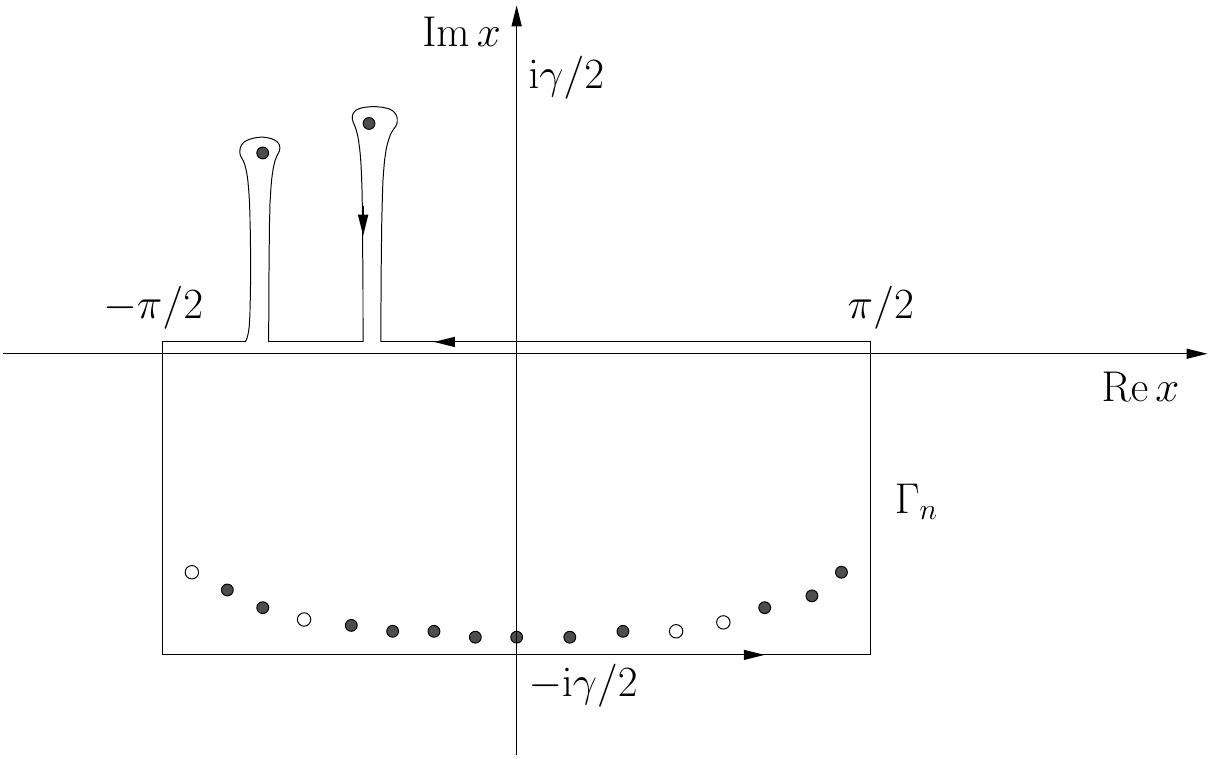}
\caption{\label{fig:gammacontour} Sketch of the contour $\G_n$.
Black dots depict Bethe roots, white dots depict holes.
}
\end{center}
\end{figure}
The function $\r_n^{\pm 1} (x|\a)\bigl(1 + \fa_0 (x|\k)\bigr)/
\bigl(1 + \fa_n (x|\k')\bigr)$ is meromorphic inside the strip
$- \g/2 < \Im x < \g/2$, where the Bethe roots are located for
low enough temperature. Its only poles and zeros inside this
strip are simple poles at the Bethe roots of the excited state
$|n; \k'\>$ and simple zeros at the Bethe roots of the dominant
state $|\k\>$. This becomes clear when we write this function
explicitly in terms of products over Bethe roots. Let us now
define a simple closed contour $\G_n$ located inside the strip
$- \g/2 < \Im x < \g/2$ and encircling all Bethe roots of the
states $|n; \k'\>$ and $|\k\>$ but none of the possible zeros of
$1 - \fa_0 (x|\k)/\fa_n (x|\k')$. Then, for any function $f$
which is holomorphic inside the strip $- \g/2 < \Im x < \g/2$
\begin{equation} \label{capitalmeasures}
     \int_{\G_n} \frac{\rd M^\s (x)}{2 \p \i} f(x) =
        \begin{cases}
	  \displaystyle
          \sum_{j=1}^{N/2} \frac{f(x_j^r)}{\r_n (x_j^r|\a) \fa_0' (x_j^r|\k)} &
	  \text{if $\s = -$,} \\[3ex]
	  \displaystyle
          \sum_{j=1}^{N/2 - s} \frac{\r_n (y_j^r|\a) f(y_j^r)}{\fa_n' (y_j^r|\k')} &
	  \text{if $\s = +$.}
        \end{cases}
\end{equation}
With these formulae we can interpret the determinants in the
numerator of (\ref{detpart}) as Fredholm determinants with
kernels $\mathcal{U}^{x, y}$, contour $\G_n$ and measures $\rd M^\pm$
(cf.\ \cite{DGK13a}), and we can take the Trotter limit.

The temperature dependence of the Fredholm determinants comes
from the measures $\rd M^\pm$. We will consider these measures
in the low-temperature limit. The contour $\G_n$ sketched in
Figure~\ref{fig:gammacontour} consists of an upper part
with $\Im x > 0$, a lower part with $\Im x = - \g/2$ and
a left and a right part connecting $\mp \p/2$ with $\mp \p/2
- \i \g/2$. The contributions to the Fredholm determinants
stemming from the left and right parts of the contour cancel
each other due to the $\p$-periodicity of the integrand. For
this reason it suffices to consider the measures on the
upper and lower part of the contour.

We start by calculating the explicit form of the function
$\r_n^{(0)}$ for $- \g < \Im x <~0$. Inserting $\ph_p$,
equation (\ref{defphip}), into (\ref{rhozero}) and
calculating the integrals by means of the formulae in
Appendix~\ref{app:fourierandintegrals}, we obtain
\begin{align}
     & 
        \r_n^{(0)} (x|\a) =
	(-1)^{k+s} q^{-s} \exp \biggl\{- 2 \i s x + \i \sum_{k=1}^{n_h} x_k
	                           - \i \sum_{k=1}^{n_p} y_k \biggr\} \notag \\
     & \mspace{36.mu}
       \biggl[ \prod_{k = 1}^{n_h}
               \frac{\sin(x - x_k)}{\sin(x - x_k + \i \g)} \biggr]
       \biggl[ \prod_{k = 1}^{n_p}
               \frac{\sin(x - y_k + \i \g)}{\sin(x - y_k)} \biggr] \notag \\
     & \mspace{36.mu}
       \exp\biggl\{
       \int_{- \p/2}^{\p/2} \frac{\rd y}{2 \p \i}
        K (x - y + \i \g/2|\g/2)
	\Bigl[ \sum_{k=1}^{n_p} \ph_p (y, y_k) - 
	       \sum_{k=1}^{n_h} \ph_p (y, x_k) \Bigr] \biggr\}  = \notag \\
     & \mspace{36.mu}
       (-1)^k (4 q)^s
       \biggl[\prod_{k = 1}^{n_h} \sin(x - x_k)
              \prod_{\s = \pm 1} \frac{(q^2 \re^{2 \i \s (x - x_k)};q^4)}
	                              {(q^4 \re^{2 \i \s (x - x_k)};q^4)}
              \biggr] \notag \\
     & \mspace{138.mu}
       \biggl[\prod_{k = 1}^{n_p} \frac{1}{\sin(x - y_k)}
              \prod_{\s = \pm 1} \frac{(q^2 \re^{2 \i \s (x - y_k)};q^4)}
	                              {(q^4 \re^{2 \i \s (x - y_k)};q^4)} \biggr]
       = w(x) \epp
\end{align}
The last equation defines the function $w(x)$ in the entire
complex plane, which will be needed below. It can be nicely
expressed in terms of Jacobian theta functions or in terms
of the dressed momentum,
\begin{subequations}
\begin{align} \label{wforms}
     w(x) & = (-1)^k \biggl[\prod_{k = 1}^{n_h}
                            \frac{\dh_1 (x - x_k, q^2)}{\dh_4 (x - x_k, q^2)} \biggr]
                     \biggl[\prod_{k = 1}^{n_p}
                            \frac{\dh_4 (x - y_k, q^2)}{\dh_1 (x - y_k, q^2)} \biggr] \\
          & = (-1)^k \exp \biggl\{ 2\p \i \Bigl[
	               \sum_{k=1}^{n_h} p(x - x_k + \i \g/2)
	               - \sum_{k=1}^{n_p} p(x - y_k + \i \g/2) \Bigr] \biggr\} \epp
\end{align}
\end{subequations}
Replacing $x$ by $- \i \g/2$ this reproduces equation (\ref{evarat})
as it should be.

The function $\r_n^{(0)}$ has a jump discontinuity across
the real axis which comes from the explicit prefactor in
(\ref{rhozero}) and from the pole at $y = x$ in the integration
kernel. It can be calculated e.g.\ by `pulling $x$ across the
real axis'. It follows that
\begin{equation} \label{rhojump}
     \r_n^{(0)} (x|\a) = \begin{cases}
                            w(x) & \text{for $- \g < \Im x < 0$,} \\
                           w(x) \re^{- 2 \p \i F(x)}
			       & \text{for $0 < \Im x < \g$.} \\
                         \end{cases}
\end{equation}
Combining now (\ref{rholowt}) and (\ref{rhojump}) and using the
explicit low-temperature form of the auxiliary functions (\ref{asyaux})
we see that, up to multiplicative corrections of the form
$1 + \CO \bigl(T^\infty\bigr)$,
\begin{equation} \label{lowtmeasures}
     \rd M^\pm (x) = \begin{cases}
                        \displaystyle
                        \frac{\rd x\: w^{\pm 1} (x) \re^{\mp 2 \p \i F(x)}}
			     {1 - \re^{\mp 2 \p \i F(x)}}
			& \text{for $x$ on the upper part of $\G_n$,} \\[2ex]
                        \displaystyle
                        \frac{\rd x\: w^{\pm 1} (x)}
			     {1 - \re^{\mp 2 \p \i F(x)}}
			& \text{for $x$ on the lower part of $\G_n$.}
                         \end{cases}
\end{equation}

With this we can simplify Fredholm determinants of the form
$\det_{\rd M^\pm, \G_n} (1 + \widehat{\mathcal{U}}^\pm)$, where
$\widehat{\mathcal{U}}^\pm$ are integral operators with regular kernels
of the form $\mathcal{U}^\pm (x,y)$. First note that for any function
$f$ holomorphic on and inside $\G_n$
\begin{multline} \label{leftactionkminus}
     [f (1 + \widehat{\mathcal{U}}^-)] (y) =
        f(y) + \int_{\G_n} \rd M^- (x) \: f(x) \mathcal{U}^- (x, y)  \\
	= f(y) + \sum_{j=1}^{n_h} f(x_j) v^- (x_j,y)
	     + \int_{- \p/2}^{\p/2} \rd x \: f(x) V^- (x, y)
	     + \CO \bigl( T^\infty \bigr) \epc
\end{multline}
where
\begin{equation} \label{defvminus}
     v^- (x_j,y) = \frac{2\p\i \res \{ w^{-1} \} (x_j) \mathcal{U}^- (x_j, y)}
                        {1 - \re^{2 \p \i F(x_j)}} \epc \qd
     V^- (x,y) = w^{-1} (x) \mathcal{U}^- (x, y) \epp
\end{equation}
In (\ref{leftactionkminus}) we have pushed the upper part of
$\G_n$ down and the lower part of $\G_n$ up to the interval
$[- \p/2, \p/2]$. Pushing up the lower part produces the
sum over holes which are simple poles of $w^{-1}$ (see
(\ref{wforms})).

Equation (\ref{leftactionkminus}) shows that we may interpret
$1 + \widehat{\mathcal{U}}^-$ as an integral operator acting on functions
supported on $[- \p/2, \p/2] \cup \{x_1, \dots, x_{n_h}\}$.
Its determinant is
\[
     \det \begin{vmatrix}
             \de (x - y) + V^- (x, y) & V^- (x, x_1) & \dots & \dots & 
	     V^- (x, x_{n_h}) \\
	     v^- (x_1, y) & 1 + v^- (x_1, x_1) & v^- (x_1, x_2) & \dots &
	     v^- (x_1, x_{n_h}) \\
	     \vdots & \vdots & \dots & \dots & \vdots \\
	     v^- (x_{n_h}, y) & v^- (x_{n_h}, x_1) & v^- (x_{n_h}, x_2) & \dots &
	     1 + v^- (x_{n_h}, x_{n_h})
          \end{vmatrix} \epp
\]
Here we would like to extract the Fredholm determinant
corresponding to the upper left block $1 + \widehat{V}^-$,
using the identity
\begin{equation} \label{detpseudodet}
     \det \begin{pmatrix} A & B \\ C & D \end{pmatrix} =
        \det (A) \det (D - C A^{-1} B)
\end{equation}
for block matrices.

The kernel of the inverse of $1 + \widehat{V}^-$ may be expressed
with the aid of the resolvent defined by
\begin{equation}
     R^- (x, y) = V^- (x, y) - \int_{- \p/2}^{\p/2} \rd z \: R^- (x, z) V^- (z, y) \epp
\end{equation}
Then
\begin{multline} \label{detnumminus}
     \det_{\rd M^-, \G_n} (1 + \widehat{\mathcal{U}}^-) = 
        \det_{\rd x, [-\p/2, \p/2]} (1 + \widehat{V}^-) \\ \times
	\det_{m, n = 1, \dots, n_h}
	\Bigl\{ \de_{m, n} + v^- (x_m, x_n)
	       - \int_{- \p/2}^{\p/2} \rd y \: v^- (x_m, y) R^- (y, x_n) \Bigr\} \epc
\end{multline}
up to multiplicative corrections of the form $1 +
\CO \bigl( T^\infty \bigr)$.

A very similar reasoning may be applied to the other Fredholm determinant
in the numerator, $\det_{\rd M^+, \G_n} (1 + \widehat{\mathcal{U}}^+)$. Setting
\begin{equation} \label{defvplus}
     v^+ (x,y_j) = \frac{2\p\i \res \{ w \} (y_j) \mathcal{U}^+ (x, y_j)}
                        {\re^{2 \p \i F(y_j)} - 1} \epc \qd
     V^+ (x,y) = \mathcal{U}^+ (x, y) w(y) \epc
\end{equation}
and introducing the resolvent kernel $R^+$ as the solution of
the linear integral equation
\begin{equation}
     R^+ (x, y) = V^+ (x, y) - \int_{- \p/2}^{\p/2} \rd z \: V^+ (x, z) R^+ (z, y)
\end{equation}
we obtain
\begin{multline} \label{detnumplus}
     \det_{\rd M^+, \G_n} (1 + \widehat{\mathcal{U}}^+) = 
        \det_{\rd x, [-\p/2, \p/2]} (1 + \widehat{V}^+) \\ \times
	\det_{m, n = 1, \dots, n_p}
	\Bigl\{ \de_{m, n} + v^+ (y_m, y_n)
	       - \int_{- \p/2}^{\p/2} \rd y \: R^+ (y_m, y) v^+ (y, y_n) \Bigr\} \epc
\end{multline}
which is again valid up to multiplicative corrections of the form $1 +
\CO \bigl( T^\infty \bigr)$.

Equations (\ref{detnumminus}) and (\ref{detnumplus}) provide
computable and efficient expressions for the determinants in the
numerator of (\ref{detpart}) in the Trotter limit and for low
temperatures. In the longitudinal case we have to substitute 
\begin{equation}
    \mathcal{U}^\pm (x, y)   =
        K_\a (x - y) \pm \frac{q^\a - q^{-\a}}{2\p} \epc
\end{equation}
into (\ref{defvminus}) and (\ref{defvplus}), while, in the transversal
case, 
\begin{equation} \label{kpmpmalpha}
     \mathcal{U}^\pm (x, y) = K_{\a \pm 1} (x - y) \epp
\end{equation}
Since we are working with the generating function in the
longitudinal case, we still have to explain how to perform the
derivative with respect to $\a$. We can proceed the same
way as in our work on the massless regime \cite{DGK13a} using
an idea going back to \cite{KKMST09a}. The idea is to extract a
factor linear in $\a$ from each of the determinants in the
numerator.

\subsection{\boldmath Extraction of $\a$ in the longitudinal case}
We define
\begin{subequations}
\begin{align}
     & U_\th^+ (x, y) = K_\a (x - y) - K_\a (x - \th) \epc \\
     & U_\th^- (x, y) = K_\a (x - y) - K_\a (\th - y) \epp
\end{align}
\end{subequations}
Then
\begin{equation}
     \lim_{\th \rightarrow \i \infty} U_\th^\pm (x, y)
        = \mathcal{U}^\pm (x, y) \epp
\end{equation}
Thus, in the longitudinal case we may substitute $U_\th^\pm (x, y)$
into (\ref{defvminus}) and (\ref{defvplus}) and then send
$\th \rightarrow \i \infty$.

We further define a function
\begin{equation}
     f(x) = \exp \biggl\{ \int_{{\cal C}_n} \frac{\rd y}{2 \p \i}
                          \ctg(x - y) \ln_{{\cal C}_n}
			  \biggl( \frac{1 + \fa_n (y|\k')}{1 + \fa_0 (y|\k)} \biggr)
			  \biggr\} \epc
\end{equation}
where ${\cal C}_n$ is a contour enclosing all Bethe roots of the
dominant state and of the excited state as well as the point
$- \i \g/2$, but none of the holes. The logarithm is defined
along the contour as in (\ref{deflnc}). Following the reasoning
of Appendix A.3 of \cite{KKMST09a} one can show that the ratios
\begin{equation}
     \frac{\det_{\rd M^\pm, \G_n} (1 + \widehat{U}_\th^\pm)}
          {q^\a f^{\mp 1} (\th \mp \i \g) - q^{-\a} f^{\mp 1} (\th \pm \i \g)}
\end{equation}
are independent of $\th$.

As is clear from its definition (\ref{defgenfun}), the function
$A_n (\a)$ has a double zero in $\a$ at $\a = 0$. It originates
from the determinants in the numerator of the determinant part. Since
$\lim_{\th \rightarrow \i \infty} f(\th)$ exists, we may conclude that
\begin{multline} \label{azzupm}
     \2 \6_{\g \a}^2
     \det_{\rd M^-, \G_n} \bigl(1 + \widehat{\mathcal{U}}^-\bigr)
     \det_{\rd M^+, \G_n} \bigl(1 + \widehat{\mathcal{U}}^+\bigr) \Bigr|_{\a = 0} =\\[1ex]
        \lim_{\a \rightarrow 0}
	\frac{4 \det_{\rd M^-, \G_n} \bigl(1 + \widehat{\mathcal{U}}^-\bigr)
	        \det_{\rd M^+, \G_n} \bigl(1 + \widehat{\mathcal{U}}^+\bigr)}
             {(q^\a - q^{-\a})^2} = \\[1ex]
     4 \lim_{\a \rightarrow 0} \lim_{\th_\pm \rightarrow \i \infty}
     \frac{\det_{\rd M^-, \G_n} (1 + \widehat{U}_{\th_-}^-)}
          {q^\a f (\th_- + \i \g) - q^{-\a} f (\th_- - \i \g)} \:
     \frac{\det_{\rd M^+, \G_n} (1 + \widehat{U}_{\th_+}^+)}
          {q^\a/f (\th_+ - \i \g) - q^{-\a}/f (\th_+ + \i \g)} = \\[1ex]
     \frac{4 \det_{\rd M^-, \G_n} (1 + \widehat{U}_{\th_-}^-)
           \det_{\rd M^+, \G_n} (1 + \widehat{U}_{\th_+}^+)}
          {\Bigl(1 - \frac{f (\th_- - \i \g)}{f (\th_- + \i \g)}\Bigr)
           \Bigl(\frac{f (\th_+ + \i \g)}{f (\th_+ - \i \g)} - 1\Bigr)} \:
           \frac{f (\th_+ + \i \g)}{f (\th_- + \i \g)} \Bigg|_{\a = 0} \epp
\end{multline}
Here we can insert
\begin{equation} \label{ffaa}
     \frac{f (\th - \i \g)}{f (\th + \i \g)} =
        \frac{\fa_n (\th|\k')}{\fa_0 (\th|\k)} \epc
\end{equation}
which follows from the nonlinear integral equations satisfied by
the auxiliary functions \cite{DGKS15b}.

We want to perform the low temperature limit in (\ref{azzupm}).
This is now easy for the first fraction on the right hand side.
The numerator is of a form such that we can apply the formulae
of the previous subsection, and for the denominator we can use
(\ref{ffaa}) and (\ref{asyaux}). For the second fraction on
the right hand side we can utilize the fact that $\th_+$ and
$\th_-$ are free parameters. Choosing $\th_+ = \th_- = \th$
this fraction equals one, and in the low-temperature limit we
end up with
\begin{multline} \label{nomiderivative}
     \2 \6_{\g \a}^2
     \det_{\rd M^-, \G_n} \bigl(1 + \widehat{\mathcal{U}}^-\bigr)
     \det_{\rd M^+, \G_n} \bigl(1 + \widehat{\mathcal{U}}^+\bigr) \Bigr|_{\a = 0} =\\[1ex]
     - \frac{\det_{\rd M^-, \G_n} (1 + \widehat{U}_{\th}^-)
             \det_{\rd M^+, \G_n} (1 + \widehat{U}_{\th}^+)}
            {\sin^2 (\p F(\th))} \biggr|_{\a = 0} \epc
\end{multline}
being valid up to multiplicative corrections of the form
$1 + \CO \bigl( T^\infty \bigr)$. For the determinants in the
numerator we have to substitute (\ref{defvminus}), (\ref{detnumminus})
and (\ref{defvplus}), (\ref{detnumplus}) with $\mathcal{U}^\pm (x, y) =
U^\pm_\th (x,y)$.

Alternatively it is possible to keep the two free parameters
$\th_+$ and $\th_-$ in the low-temperature limit. In that case
the resulting expression looks more involved, since the
factors $f(\th_+ + \i\g)$ and $1/f(\th_- + \i\g)$ do not cancel
each other anymore and have to be calculated using similar techniques 
as in Appendix~\ref{app:unipart}. Here we only give the final
result,
\begin{multline}
     f(\th + \i \g) = \\
        \exp \biggl\{ \2 \biggl( \g \a - \i \p k
	              + \i \sum_{j=1}^{n_p} (y_j - x_j) \biggr) \biggr\}
        \prod_{j=1}^{n_p}
	\frac{(q^2 \re^{2\i(\th - y_j)}; q^4)(q^4 \re^{2\i(\th - x_j)}; q^4)}
	     {(q^4 \re^{2\i(\th - y_j)}; q^4)(q^2 \re^{2\i(\th - x_j)}; q^4)} \epc
\end{multline}
and leave the details to the reader. Choosing $\th_+$ and $\th_-$
independently is sometimes advantageous, e.g.\ in numerical
calculations.

\subsection{Determinants in the denominator}
With the determinants in the denominator we can proceed in a similar
way as with the determinants in the numerator. We introduce a measure
\begin{equation}
     \rd m (x) = \frac{\rd x}{1 + \fa_n (x|\k')} \epp
\end{equation}
In the Trotter limit the second determinant in the denominator of
(\ref{detpart}) becomes a Fredholm determinant $\det_{\rd m, {\cal C}_n}
(1 + \widehat{K})$ with measure $\rd m$ and with respect to a
contour ${\cal C}_n$ which includes all Bethe roots but excludes
the holes of the state $|n;\k'\>$ (cf.\ \cite{DGK13a}).

For any $\p$-periodic function, holomorphic on and inside ${\cal C}_n$,
\begin{multline} \label{rightactionk}
     [(1 + \widehat{K}) f] (x) =
        f(x) + \int_{{\cal C}_n} \rd m(y) \: K_0 (x - y) f(y) = \\
	f(x) + \sum_{k=1}^{n_p} \frac{2\p\i K_0 (x - y_k) f(y_k)}{\fa_n' (y_k|\k')}
	     - \sum_{k=1}^{n_h} \frac{2\p\i K_0 (x - x_k) f(x_k)}{\fa_n' (x_k|\k')} \\
	     + \int_{- \p/2}^{\p/2} \rd y \: K_0 (x - y) f(y)
	     + \CO \bigl( T^\infty \bigr) \epp
\end{multline}
This shows that we may interpret $1 + \widehat{K}$ as an integral
operator acting on functions supported on $[- \p/2, \p/2] \cup
\{y_1, \dots, y_{n_p}; x_1, \dots, x_{n_h}\}$. The determinant
of this integral operator is
\[
     \det \begin{vmatrix}
             \de (x - y) + K_0 (x - y) & \frac{2\p\i K_0 (x - y_1)}{\fa_n' (y_1|\k')} &
	     \dots & \dots & - \frac{2\p\i K_0 (x - x_{n_h})}{\fa_n' (x_{n_h}|\k')} \\
	     K_0 (y_1 - y) & 1 + \frac{2\p\i K_0 (0)}{\fa_n' (y_1|\k')} &
	     \frac{2\p\i K_0 (y_1 - y_2)}{\fa_n' (y_2|\k')} & \dots &
	     - \frac{2\p\i K_0 (y_1 - x_{n_h})}{\fa_n' (x_{n_h}|\k')} \\
	     \vdots & \vdots & \dots & \dots & \vdots \\
	     K_0 (x_{n_h} - y) & \frac{2\p\i K_0 (x_{n_h} - y_1)}{\fa_n' (y_1|\k')} &
	     \frac{2\p\i K_0 (x_{n_h} - y_2)}{\fa_n' (y_2|\k')} & \dots &
	     1 - \frac{2\p\i K_0 (0)}{\fa_n' (x_{n_h}|\k')}
          \end{vmatrix} \epp
\]

Now we can proceed as above. We introduce the resolvent kernel $R$
as the solution of the linear integral equation
\begin{equation} \label{resolventkernel}
     R(x - y) = K_0 (x - y) - \int_{- \p/2}^{\p/2} \rd z \: K_0 (x - z) R(z - y) \epp
\end{equation}
Then, using (\ref{detpseudodet}), we end up with
\begin{multline}
     \det_{\rd m, {\cal C}_n} (1 + \widehat{K}) = \\
        \det_{\rd x, [- \p/2, \p/2]} (1 + \widehat{K}_0)
	\det_{\substack{j, \ell = 1, \dots, n_p \\ k, m = 1, \dots, n_h}}
	\begin{vmatrix}
	   \de_{j \ell} + \frac{2\p\i R(y_j - y_\ell)}{\fa_n' (y_\ell|\k')} &
	   - \frac{2\p\i R(y_j - x_m)}{\fa_n' (x_m|\k')} \\
	   \frac{2\p\i R(x_k - y_\ell)}{\fa_n' (y_\ell|\k')} &
	   \de_{k m} - \frac{2\p\i R(x_k - x_m)}{\fa_n' (x_m|\k')}
        \end{vmatrix} \epc
\end{multline}
where we have neglected multiplicative corrections of the form
$ 1 + \CO \bigl( T^\infty \bigr)$. It follows from equation
(\ref{asyaux}) that the finite determinant on the right hand side
goes to one as $T \rightarrow 0+$. But here the multiplicative
corrections are of the form $ 1 + \CO (T)$. Therefore we keep the
finite determinant. It nicely combines with the $\CO (T)$ contributions
of the universal part.

With the first determinant in the denominator of (\ref{detpart}) we can
proceed in a very similar way as above. We define
\begin{equation}
     \rd m_0 (x) = \frac{\rd x}{1 + \fa_0 (x|\k)} \epp
\end{equation}
In the Trotter limit the first determinant in the denominator of
(\ref{detpart}) then becomes the Fredholm determinant
$\det_{\rd m_0, {\cal C}_0} (1 + \widehat{K})$ where the 
contour ${\cal C}_0$ includes all Bethe roots of the dominant
state $|\k\>$. Following the same steps as above we obtain the
low-temperature asymptotic value
\begin{equation}
     \det_{\rd m_0, {\cal C}_0} (1 + \widehat{K}) =
        \det_{\rd x, [- \p/2, \p/2]} (1 + \widehat{K}_0)
\end{equation}
valid up to multiplicative corrections of the form
$ 1 + \CO \bigl( T^\infty \bigr)$. Alternatively, the Fredholm
determinant on the right hand side can be expressed in terms
of the integral operator $\widehat{R}$ connected with the resolvent
kernel (\ref{resolventkernel}) or in terms of $q$-factorials
\cite{IKMT99},
\begin{equation} \label{detresolve}
     \det_{\rd x, [- \p/2, \p/2]} (1 + \widehat{K}_0) = 
        \frac{1}{\det_{\rd x, [- \p/2, \p/2]} (1 - \widehat{R})} =
	2 (- q^2; q^2)^2 \epp
\end{equation}

\Appendix{Low-temperature limit of the factorizing part}
\label{app:factorpart}\noindent
In \cite{DGK13a} we introduced two functions $G_+$ and
$\overline{G}_-$ which determine the factorizing part
of the transversal correlation functions as solutions
of linear integral equations. Here we need these equations
in a form which respects the notational conventions
for the antiferromagnetic massive regime and is at
the same time appropriate for taking the low-temperature
limit. Such a form can be obtained e.g.\ from the linear
integral equations in \cite{DGK13a} by first going back
to finite Trotter number and sums over Bethe roots instead
of integrals, then switching to the conventions of the
antiferromagnetic massive regime and finally using
(\ref{capitalmeasures}) to obtain integrals more
appropriate for the low-temperature limit. This way we obtain
the linear integral equations
\begin{subequations}
\label{ggbargammaform}
\begin{align}
     & G_+ (x, \x) = - \ctg(x - \x)
        + \frac{q^{- 1 - \a} \r_n (\x|\a) \ctg(x - \x - \i \g)}{1 + \fa_n (\x|\k')}
	\notag \\ & \mspace{36.mu}
        + \frac{q^{1 + \a} \r_n (\x|\a) \ctg(x - \x + \i \g)}{1 + \fa_n^{-1} (\x|\k')}
	- \int_{\G_n} \rd M^+ (y) K_{1+\a} (x - y) G_+ (y, \x) \epc \\
     & \overline{G}_- (x, \x) = - \ctg(x - \x)
        + \frac{q^{1 - \a} \ctg(x - \x - \i \g)}
	       {\r_n (\x|\a) \bigl(1 + \fa_0 (\x|\k)\bigr)}
	\notag \\ & \mspace{36.mu}
        + \frac{q^{- 1 + \a} \ctg(x - \x + \i \g)}
	       {\r_n (\x|\a) \bigl(1 + \fa_0^{-1} (\x|\k)\bigr)}
	- \int_{\G_n} \rd M^- (y) K_{\a - 1} (y - x) \overline{G}_- (y, \x) \epc
\end{align}
\end{subequations}
where, for low enough temperature, the contour $\G_n$ is the same as
in Figure~\ref{fig:gammacontour} and where $\x$ is outside $\G_n$ with
$\Im \x < - \g/2$.

The functions $G_+^- (\x)$ and $\overline{G}_-^+ (\x)$ in equation
(\ref{facpart}) can be represented by means of integrals involving
$G_+ (\cdot, \x)$ and $\overline{G}_- (\cdot, \x)$. Starting again
from the corresponding equations in \cite{DGK13a} and proceeding in
a similar way as above we obtain
\begin{subequations}
\label{glimitsgammaform}
\begin{align}
     & G_+^- (\x) =
        1 - \frac{q^{- 1 - \a} \r_n (\x|\a)}{1 + \fa_n (\x|\k')}
	\notag \\ & \mspace{36.mu}
        - \frac{q^{1 + \a} \r_n (\x|\a)}{1 + \fa_n^{-1} (\x|\k')}
	- (q^{1 + \a} - q^{- 1 - \a})
	  \int_{\G_n} \frac{\rd M^+ (y)}{2\p\i} G_+ (y, \x) \epc \\[1ex]
     & \overline{G}_-^+ (\x) = - 1
        + \frac{q^{1 - \a}}{\r_n (\x|\a) \bigl(1 + \fa_0 (\x|\k)\bigr)}
	\notag \\ & \mspace{36.mu}
        + \frac{q^{- 1 + \a}}{\r_n (\x|\a) \bigl(1 + \fa_0^{-1} (\x|\k)\bigr)}
	- (q^{1 - \a} - q^{- 1 + \a})
	\int_{\G_n} \frac{\rd M^- (y)}{2\p\i} \overline{G}_- (y, \x) \epp
\end{align}
\end{subequations}
Using (\ref{lowtmeasures}) and arguments similar to those in
Appendix~\ref{app:detpart} in order to perform the low-temperature
limit we obtain equations (\ref{glimitslowtform}) of the main text.

\Appendix{Some Fourier series and integrals}
\label{app:fourierandintegrals}\noindent
Many of the integrals occurring in Appendix~\ref{app:unipart} can
be calculated using the convolution theorem for Fourier series
combined with resummation.
\begin{lemma}
Convolution of Fourier series. Given the Fourier series representations
of two functions $f, g: [-\p/2, \p/2] \rightarrow {\mathbb C}$,
\begin{equation}
     f(x) = \sum_{n \in {\mathbb Z}} f_n \re^{2 \i n x} \epc \qd
     g(x) = \sum_{n \in {\mathbb Z}} g_n \re^{2 \i n x} \epc
\end{equation}
their convolution has the Fourier series representation
\begin{equation} \label{convolution}
     \int_{- \p/2}^{\p/2} \rd y \: f(x - y) g(y) = 
        \p \sum_{n \in {\mathbb Z}} f_n g_n \re^{2 \i n x} \epp
\end{equation}
\end{lemma}
In Appendix~\ref{app:unipart} one may use the Fourier series
\begin{align}
     \frac{1}{2 \p \i} \ctg (x) & =
        \begin{cases} \displaystyle
	   - \frac{1}{2\p} - \frac{1}{\p} \sum_{n=1}^\infty \re^{2 \i nx} &
	   \text{if $\Im x > 0$,} \\[1ex] \displaystyle
	   \frac{1}{2\p} + \frac{1}{\p} \sum_{n=-\infty}^{-1} \re^{2 \i nx} &
	   \text{if $\Im x < 0$.}
	\end{cases} \\[2ex]
     \ph_p (x,z) & =
        \sum_{n=1}^\infty \frac{1}{n} \frac{\re^{2 \i n(x - z)}}{1 + q^{- 2n}} +
        \sum_{n=-\infty}^{-1} \frac{1}{n} \frac{\re^{2 \i n(x - z)}}{1 + q^{2n}}
        \qd \text{for $|\Im z| < \g$}
\end{align}
and the `resummation formulae'
\begin{subequations}
\begin{align}
     \sum_{n=1}^\infty \frac{1}{n} \frac{x^n}{1 + q^{2n}} & =
        \ln \biggl( \frac{(x q^2; q^4)}{(x; q^4)} \biggr) \epc \\
     \sum_{n=1}^\infty \frac{1}{n} \frac{x^n}{(1 + q^{2n})^2} & =
        \ln \biggl( \frac{(x q^2; q^4, q^4)^2}
	                 {(x; q^4, q^4) (xq^4; q^4, q^4)} \biggr) \epc
\end{align}
\end{subequations}
$|x| < 1$, as well as the reduction formulae
\begin{equation} \label{reduction}
     (xq; q, q) = \frac{(x; q, q)}{(x; q)} \epc \qd
     (xq; q) = \frac{(x; q)}{1 - x}
\end{equation}
in order to calculate the remaining integrals and to simplify the
result.

In Appendix~\ref{app:unipart} we frequently encountered certain
elementary integrals involving logarithms along lines parallel to
the real axis. To make it easier to verify the results of
Appendix~\ref{app:unipart} we briefly discuss these integrals here.
In all cases involving integrals over $\ln \sin (x)$ or its derivatives,
the following definition is very helpful
\begin{equation} \label{deflnsin}
     \ln \sin (x) = \begin{cases}
		       \frac{\i \p}{2} - \ln (2) - \i x + \Ln (1 - \re^{2 \i x})
		       & \text{if $\Im x > 0$,} \\
		       - \frac{\i \p}{2} - \ln (2) + \i x + \Ln (1 - \re^{- 2 \i x})
		       & \text{if $\Im x < 0$}.
                    \end{cases}
\end{equation}
Here $\Ln$ denotes the principal branch of the logarithm. Note
that $\Ln (1 - \re^{2 \i x})$ is holomorphic and $\p$-periodic in
the upper half plane, while $\Ln (1 - \re^{- 2 \i x})$ is holomorphic
and $\p$-periodic in the lower half plane. Using (\ref{deflnsin})
we obtain, for instance, for $- \g < \Im x < 0$ that
\begin{equation}
     \int_{- \p/2}^{\p/2} \frac{\rd y}{2\p\i} \bigl\{
        \ln \bigl( \sin (y - x - \i \g) \bigr) - \ln \bigl( \sin (y -x) \bigr\}
	= - \frac{\p}{2} - x - \i \g \epp
\end{equation}
\begin{figure}
\begin{center}
\includegraphics[width=.75\textwidth]{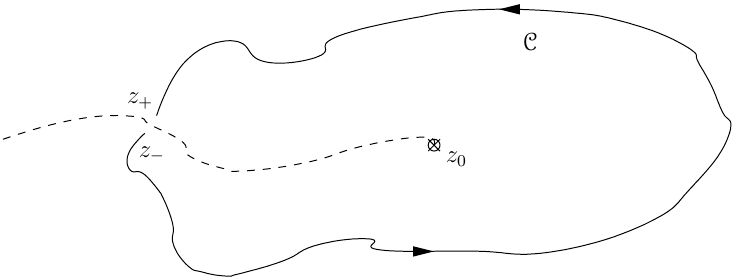}
\caption{\label{fig:intaroundcut} Integration contour for an integral
around the cut of a logarithm with branch point at $z_0$.}
\end{center}
\end{figure}

We also encountered the integral
\begin{equation}
     I_1 = \int_{z_-}^{z_+} \frac{\rd z}{2 \p \i} \: \frac{\ln (z - z_0)}{z - z_0} \epc
\end{equation}
where $z_\pm$ denote the boundary values from above and below
the cut of a point on the cut and where the integration contour
is a simple closed contour from $z_-$ to $z_+$ that goes around
$z_0$ (see Figure~\ref{fig:intaroundcut}). Then $\ln (z - z_0)$
is holomorphic and single-valued on the contour. Hence,
\begin{equation}
     I_2 = \frac{1}{2\p\i} \frac{\ln^2 (z - z_0)}{2}\biggr|_{z_-}^{z_+}
         = \ln(z_+ - z_0) - \p\i = \ln(z_- - z_0) + \p\i \epp
\end{equation}

Lastly, we needed to know integrals over the same type of contour of
the form
\begin{equation}
     I_2 = \int_{z_-}^{z_+} \frac{\rd z}{2 \p \i} \: \ln (z - z_0) f'(z) \epc
\end{equation}
where $f$ is holomorphic and single valued on the contour with
$f(z_+) = f(z_-)$. For these integrals we use partial integration
to obtain
\begin{equation}
     I_2 = f(z_+) - f(z_0) = f(z_-) - f(z_0) \epp
\end{equation}

\Appendix{\boldmath The $q$-gamma family}
\label{app:qfunctions}\noindent
In this appendix we collect some basic facts about $q$-gamma
and $q$-Barnes functions. They belong to the $q$-analogue of
the family of multiple-gamma functions. With the definition
\begin{equation}
     [x]_q = \frac{1 - q^x}{1 - q}
\end{equation}
of a `$q$-number' we have the following
\begin{theorem*}
\cite{Nishizawa96}
Let $q \in {\mathbb C}$, $|q| < 1$. The sequence of
functional equations
\begin{equation} \label{qgammafun}
     g_r (x + 1) = g_{r-1} (x) g_r (x) \epc \qd r \in {\mathbb N}
\end{equation}
with boundary conditions
\begin{equation}
     g_r (1) = 1 \epc \qd g_0 (x) = [x]_q
\end{equation}
and
\begin{equation}
     \6_x^{r+1} \ln g_r (x+1) \ge 0 \qd \text{for $x \ge 0$}
\end{equation}
uniquely determines a sequence of meromorphic functions $g_r$.
\end{theorem*}
The function $\G_q = g_1$ is the $q$-gamma function and
$G_q = g_2$ is the $q$-Barnes function. For the whole
sequence of multiple $q$-gamma functions infinite product
representations exist \cite{Nishizawa96}. Alternatively
they can be expressed in terms of $q$-multi factorials.
Here we give only the $q$-multi factorial representations
of $\G_q$ and $G_q$,
\begin{equation} \label{defgammaqgq}
     \G_q (x) = (1 - q)^{1 - x} \frac{(q;q)}{(q^x;q)} \epc \qd
     G_q (x) = (1 - q)^{- \2 (1 - x)(2 - x)} (q;q)^{x - 1}
               \frac{(q^x;q,q)}{(q;q,q)} \epp
\end{equation}
These definitions together with the functional equations
(\ref{qgammafun}) were used to obtain the expression
(\ref{unilowtexpl}) for the universal part of the amplitudes
in the main text.

For the isotropic limit we have used that the $q$-gamma and
$q$-Barnes functions turn into their classical counterparts
as $q \rightarrow 1$. Here we include a short proof of this
fact which is in the spirit of our treatment of the nonlinear
integral equations in \cite{DGKS15a}.

For $|q| < 1$ we obtain the following series expansion for
$\ln \G_q$ directly from (\ref{defgammaqgq}),
\begin{equation} \label{lngammaqseries}
     \ln \G_q (x) =
        \sum_{p \ge 1} \frac{1}{p} \biggl\{
	   \frac{q^{xp} - q^p}{1 - q^p} - q^p (1 - x) \biggr\} \epp
\end{equation}
Upon setting $q = \re^{-t}$ and introducing
\begin{equation}
     f_\G (s) =
        \frac{1}{s} \biggl\{ \frac{\re^{- xs} - \re^{-s}}{1 - \re^{-s}}
	                     - \re^{-s}(x -1) \biggr\}
\end{equation}
we can rewrite (\ref{lngammaqseries}) as
\begin{equation} \label{lngammaqintrep}
     \ln \G_q (x) =
        \sum_{p \ge 1} t f_\G (tp) =
	\int_{\widehat{\cal C}_t}
	   \rd s \: \frac{f_\G (s)}{\re^{\frac{2\p\i s}{t}} - 1} \epc
\end{equation}
where the contour $\widehat{\cal C}_t$ consists of three
straight line segments,
\begin{equation}
     \widehat{\cal C}_t =
        \bigl\{ ] + \infty; \tst{\frac t 2}] + \i \de \bigr\}
	\cup \bigl\{ \tst{\frac 1 2} + [\i \de; - \i \de] \bigr\}
	\cup \bigl\{ [\tst{\frac t 2}; + \infty [ - \i \de \bigr\} \epp
\end{equation}
The contour integral can be decomposed as
\begin{equation}
     \ln \G_q (x) =
	\int_0^{+ \infty} \rd s \: f_\G (s) - \int_0^\frac{t}{2} \rd s \: f_\G (s)
	- \sum_{\eps = \pm} \eps
	\int_{\widehat{\cal C}_t^{(\eps)}}
	   \rd s \: \frac{f_\G (s)}{\re^{- \frac{\eps 2\p\i s}{t}} - 1} \epc
\end{equation}
where $\widehat{\cal C}_t^{(\epsilon)} = \widehat{\cal C}_t
\cap {\mathbb H}^\epsilon$. Since $f_\G$ is smooth, decays
exponentially fast as $\Re s \rightarrow + \infty$ and
$f_\G (s) = \CO (s)$ for $s \rightarrow 0$, the last two
terms on the right hand side produce $\CO (t^2)$ contributions,
while the first term is a known integral representation of
$\ln \G (x)$ (see e.g. \cite{BaEr53gamma}).

The calculations are similar for the $q$-Barnes function.
The logarithms of the $q$-multi factorials in (\ref{defgammaqgq}) 
may be expanded into the series
\begin{equation}
     \ln G_q (x) =
        \sum_{p \ge 1} \frac 1 p \biggl\{ \2 (1 - x)(2 - x)q^p +
	   \frac{(1 - x) q^p}{1 - q^p} -
	   \frac{q^{xp} - q^p}{(1 - q^p)^2} \biggr\} \epp
\end{equation}
The function
\begin{equation}
     f_G (s) = \frac 1 s \biggl\{ \2 (1 - x)(2 - x) \re^{-s} +
        \frac{(1 - x) \re^{-s}}{1 - \re^{-s}} -
        \frac{\re^{-xs} - \re^{-s}}{(1 - \re^{-s})^2} \biggr\} \epc
\end{equation}
which satisfies $f_G (s) = \CO (1)$ for $s \rightarrow 0+$,
allows one to recast $\ln G_q (x)$ into the form
\begin{multline}
     \ln G_q (x) =
        \sum_{p \ge 1} t f_G (tp) =
	\int_{\widehat{\cal C}_t}
	   \rd s \: \frac{f_G (s)}{\re^{\frac{2\p\i s}{t}} - 1} \\
	= \int_0^{+ \infty} \rd s \: f_G (s) - \int_0^\frac{t}{2} \rd s \: f_G (s)
	- \sum_{\epsilon = \pm} \epsilon
	\int_{\widehat{\cal C}_t^{(\epsilon)}}
	   \rd s \: \frac{f_G (s)}{\re^{- \frac{\epsilon 2\p\i s}{t}} - 1} \epp
\end{multline}
The last two terms are $\CO (t)$ in the limit $t \rightarrow 0+$.
This establishes that
\begin{equation}
     G_q (x) \rightarrow \widetilde{G} (x) =
        \exp \biggl\{ \int_0^{+ \infty} \rd s \: f_G (s) \biggr\} \epc
\end{equation}
pointwise for $\Re x > 0$. Performing such a pointwise limit
in the functional equation (\ref{qgammafun}) for $r = 2$, we
conclude that $\widetilde{G} (x + 1) = \G(x) \widetilde{G} (x)$.
Since $G_q (1) = 1$ and $\6_x^3 \ln G_q (x) \ge 0$ for $x \ge 0$
it follows that the same properties hold for $\widetilde{G}$.
Thus, $\widetilde{G}$ must be equal to the Barnes $G$ function
owing to the uniqueness theorem of Vign\'eras \cite{Vigneras79}.

\Appendix{Fredholm determinants in the isotropic limit}
\label{app:vanishingstaggered}\noindent
In this appendix we provide some details of the derivation of
equation (\ref{divideoneplusr}) in the main text.
The kernels $K^\pm$, equation (\ref{klongi}), define integral operators
$\widehat{K}^\pm$ acting on $L^2 ([-\p/2, \p/2])$. Let us further
define an operator $\widehat w$ which acts on $L^2 ([-\p/2, \p/2])$
by pointwise multiplication with the values of the function $w$,
equation (\ref{defw}).

The integral operators $\widehat{V}^\pm$
appearing in the Fredholm determinant contributions to the
longitudinal correlation functions can then be written as
\begin{equation}
     \widehat{V}^- = \widehat{w}^{-1} \widehat{K}^- \epc \qd
     \widehat{V}^+ = \widehat{K}^+ \widehat{w} \epp
\end{equation}
It is not difficult to see that, for $k = 0, 1$, the resolvents
$\widehat{L}_k^\pm$, defined by
\begin{subequations}
\begin{align}
     & (1 + (-1)^k \widehat{K}^-)(1 - (-1)^k \widehat{L}_k^-) = 1 \epc \\[1ex]
     & (1 - (-1)^k \widehat{L}_k^+)(1 + (-1)^k \widehat{K}^+) = 1 \epc
\end{align}
\end{subequations}
exist as operators on $L^2 ([-\p/2, \p/2])$ and can represented as
integral operators with kernels
\begin{subequations}
\begin{align}
     & L_k^- (x,y) = R_k (x - y) - R_k (\th_- - y) \epc \\[1ex]
     & L_k^+ (x,y) = R_k (x - y) - R_k (x - \th_+) \epc
\end{align}
\end{subequations}
where the functions $R_k$, $k = 0, 1$, were introduced in
(\ref{defrfunctions}). It is further known that
\begin{equation}
     \det_{\rd u, [-\p/2, \p/2]} (1 \mp \widehat{K}^-) =
     \det_{\rd u, [-\p/2, \p/2]} (1 \mp \widehat{K}^+) =
        (\pm q^2, q^2)^2 \epp
\end{equation}

The above equations imply that
\begin{align}
     & \det_{\rd u, [-\p/2, \p/2]} \bigl(1 + \widehat{V}^-\bigr) =
       \det_{\rd u, [-\p/2, \p/2]} \bigl(1 + \widehat{w}^{-1} \widehat{K}^-\bigr)
       \notag \\[1ex]
     & \qd =
       \det_{\rd u, [-\p/2, \p/2]} \bigl(1 + (-1)^k \widehat{K}^-\bigr) \notag \\
     & \qqd \times
       \det_{\rd u, [-\p/2, \p/2]}
          \bigl(1 + (-1)^k \widehat{K}^-
	          + (\widehat{w}^{-1} - (-1)^k) \widehat{K}^-\bigr)
       \det_{\rd u, [-\p/2, \p/2]} \bigl(1 - (-1)^k \widehat{L}_k^-\bigr)
       \notag \\[1ex]
     & \qd =
       \bigl((-1)^{k+1} q^2, q^2\bigr)^2 \det_{\rd u, [-\p/2, \p/2]}
       \bigl(1 + (\widehat{w}^{-1} - (-1)^k) \widehat{L}_k^-\bigr) \epp
\end{align}
Similarly
\begin{equation}
     \det_{\rd u, [-\p/2, \p/2]} \bigl(1 + \widehat{V}^+\bigr) =
       \bigl((-1)^{k+1} q^2, q^2\bigr)^2 \det_{\rd u, [-\p/2, \p/2]}
       \bigl(1 + \widehat{L}_k^+ (\widehat{w} - (-1)^k)\bigr) \epp
\end{equation}
The latter two equations are equivalent to (\ref{divideoneplusr})
in the main text.
}



\providecommand{\bysame}{\leavevmode\hbox to3em{\hrulefill}\thinspace}
\providecommand{\MR}{\relax\ifhmode\unskip\space\fi MR }
\providecommand{\MRhref}[2]{%
  \href{http://www.ams.org/mathscinet-getitem?mr=#1}{#2}
}
\providecommand{\href}[2]{#2}

\end{document}